\documentclass[a4paper,traditabstract,longauth]{aa}

\def\setsymbol#1#2{\expandafter\def\csname #1\endcsname{#2}}
\def\getsymbol#1{\csname #1\endcsname}

\def\Planck{{\it Planck\/}}

\newbox\tablebox    \newdimen\tablewidth
\def\leaderfil{\leaders\hbox to 5pt{\hss.\hss}\hfil}
\def\tablenote#1 #2\par{\begingroup \parindent=0.8em
    \abovedisplayshortskip=0pt\belowdisplayshortskip=0pt
    \noindent
    $$\hss\vbox{\hsize\tablewidth \hangindent=\parindent \hangafter=1 \noindent
    \hbox to \parindent{$^#1$\hss}\strut#2\strut\par}\hss$$
    \endgroup}
\def\doubleline{\vskip 3pt\hrule \vskip 1.5pt \hrule \vskip 5pt}

\def\L2{\ifmmode L_2\else $L_2$\fi}

\def\DeltaT{\ifmmode \Delta T\else $\Delta T$\fi}
\def\deltat{\ifmmode \Delta t\else $\Delta t$\fi}
\def\fknee{\ifmmode f_{\rm knee}\else $f_{\rm knee}$\fi}
\def\Fmax{\ifmmode F_{\rm max}\else $F_{\rm max}$\fi}
\def\solar{\ifmmode{\rm M}_{\mathord\odot}\else${\rm M}_{\mathord\odot}$\fi}
\def\Msolar{\ifmmode{\rm M}_{\mathord\odot}\else${\rm M}_{\mathord\odot}$\fi}
\def\Lsolar{\ifmmode{\rm L}_{\mathord\odot}\else${\rm L}_{\mathord\odot}$\fi}

\def\inv{\ifmmode^{-1}\else$^{-1}$\fi}
\def\mo{\ifmmode^{-1}\else$^{-1}$\fi}
\def\sup#1{\ifmmode ^{\rm #1}\else $^{\rm #1}$\fi}
\def\expo#1{\ifmmode \times 10^{#1}\else $\times 10^{#1}$\fi}
\def\,{\thinspace}
\def\lsim{\mathrel{\raise .4ex\hbox{\rlap{$<$}\lower 1.2ex\hbox{$\sim$}}}}
\def\gsim{\mathrel{\raise .4ex\hbox{\rlap{$>$}\lower 1.2ex\hbox{$\sim$}}}}

\def\simprop{\mathrel{\raise .4ex\hbox{\rlap{$\propto$}\lower 1.2ex\hbox{$\sim$}}}}
\def\deg{\ifmmode^\circ\else$^\circ$\fi}
\def\pdeg{\ifmmode $\setbox0=\hbox{$^{\circ}$}\rlap{\hskip.11\wd0 .}$^{\circ}
          \else \setbox0=\hbox{$^{\circ}$}\rlap{\hskip.11\wd0 .}$^{\circ}$\fi}
\def\arcs{\ifmmode {^{\scriptstyle\prime\prime}}
          \else $^{\scriptstyle\prime\prime}$\fi}
\def\arcm{\ifmmode {^{\scriptstyle\prime}}
          \else $^{\scriptstyle\prime}$\fi}
\newdimen\sa  \newdimen\sb
\def\parcs{\sa=.07em \sb=.03em
     \ifmmode \hbox{\rlap{.}}^{\scriptstyle\prime\kern -\sb\prime}\hbox{\kern -\sa}
     \else \rlap{.}$^{\scriptstyle\prime\kern -\sb\prime}$\kern -\sa\fi}
\def\parcm{\sa=.08em \sb=.03em
     \ifmmode \hbox{\rlap{.}\kern\sa}^{\scriptstyle\prime}\hbox{\kern-\sb}
     \else \rlap{.}\kern\sa$^{\scriptstyle\prime}$\kern-\sb\fi}
\def\ra[#1 #2 #3.#4]{#1\sup{h}#2\sup{m}#3\sup{s}\llap.#4}
\def\dec[#1 #2 #3.#4]{#1\deg#2\arcm#3\arcs\llap.#4}
\def\deco[#1 #2 #3]{#1\deg#2\arcm#3\arcs}
\def\rra[#1 #2]{#1\sup{h}#2\sup{m}}

\def\dots{\relax\ifmmode \ldots\else $\ldots$\fi}
\def\WHzsr{\ifmmode $W\,Hz\mo\,sr\mo$\else W\,Hz\mo\,sr\mo\fi}
\def\mHz{\ifmmode $\,mHz$\else \,mHz\fi}
\def\GHz{\ifmmode $\,GHz$\else \,GHz\fi}
\def\mKs{\ifmmode $\,mK\,s$^{1/2}\else \,mK\,s$^{1/2}$\fi}
\def\muKs{\ifmmode \,\mu$K\,s$^{1/2}\else \,$\mu$K\,s$^{1/2}$\fi}
\def\muKRJs{\ifmmode \,\mu$K$_{\rm RJ}$\,s$^{1/2}\else \,$\mu$K$_{\rm RJ}$\,s$^{1/2}$\fi}
\def\muKHz{\ifmmode \,\mu$K\,Hz$^{-1/2}\else \,$\mu$K\,Hz$^{-1/2}$\fi}
\def\MJysr{\ifmmode \,$MJy\,sr\mo$\else \,MJy\,sr\mo\fi}
\def\MJysrmK{\ifmmode \,$MJy\,sr\mo$\,mK$_{\rm CMB}\mo\else \,MJy\,sr\mo\,mK$_{\rm CMB}\mo$\fi}
\def\microns{\ifmmode \,\mu$m$\else \,$\mu$m\fi}

\def\muK{\ifmmode \,\mu$K$\else \,$\mu$\hbox{K}\fi}
\def\microK{\ifmmode \,\mu$K$\else \,$\mu$\hbox{K}\fi}
\def\muW{\ifmmode \,\mu$W$\else \,$\mu$\hbox{W}\fi}
\def\kms{\ifmmode $\,km\,s$^{-1}\else \,km\,s$^{-1}$\fi}
\def\kmsMpc{\ifmmode $\,\kms\,Mpc\mo$\else \,\kms\,Mpc\mo\fi}
\setsymbol{LFI:center:frequency:70GHz:units}{70.3\,GHz}
\setsymbol{LFI:center:frequency:44GHz:units}{44.1\,GHz}
\setsymbol{LFI:center:frequency:30GHz:units}{28.5\,GHz}

\setsymbol{LFI:center:frequency:70GHz}{70.3}
\setsymbol{LFI:center:frequency:44GHz}{44.1}
\setsymbol{LFI:center:frequency:30GHz}{28.5}

\setsymbol{LFI:center:frequency:LFI18:Rad:M:units}{71.7\GHz}
\setsymbol{LFI:center:frequency:LFI19:Rad:M:units}{67.5\GHz}
\setsymbol{LFI:center:frequency:LFI20:Rad:M:units}{69.2\GHz}
\setsymbol{LFI:center:frequency:LFI21:Rad:M:units}{70.4\GHz}
\setsymbol{LFI:center:frequency:LFI22:Rad:M:units}{71.5\GHz}
\setsymbol{LFI:center:frequency:LFI23:Rad:M:units}{70.8\GHz}
\setsymbol{LFI:center:frequency:LFI24:Rad:M:units}{44.4\GHz}
\setsymbol{LFI:center:frequency:LFI25:Rad:M:units}{44.0\GHz}
\setsymbol{LFI:center:frequency:LFI26:Rad:M:units}{43.9\GHz}
\setsymbol{LFI:center:frequency:LFI27:Rad:M:units}{28.3\GHz}
\setsymbol{LFI:center:frequency:LFI28:Rad:M:units}{28.8\GHz}
\setsymbol{LFI:center:frequency:LFI18:Rad:S:units}{70.1\GHz}
\setsymbol{LFI:center:frequency:LFI19:Rad:S:units}{69.6\GHz}
\setsymbol{LFI:center:frequency:LFI20:Rad:S:units}{69.5\GHz}
\setsymbol{LFI:center:frequency:LFI21:Rad:S:units}{69.5\GHz}
\setsymbol{LFI:center:frequency:LFI22:Rad:S:units}{72.8\GHz}
\setsymbol{LFI:center:frequency:LFI23:Rad:S:units}{71.3\GHz}
\setsymbol{LFI:center:frequency:LFI24:Rad:S:units}{44.1\GHz}
\setsymbol{LFI:center:frequency:LFI25:Rad:S:units}{44.1\GHz}
\setsymbol{LFI:center:frequency:LFI26:Rad:S:units}{44.1\GHz}
\setsymbol{LFI:center:frequency:LFI27:Rad:S:units}{28.5\GHz}
\setsymbol{LFI:center:frequency:LFI28:Rad:S:units}{28.2\GHz}

\setsymbol{LFI:center:frequency:LFI18:Rad:M}{71.7}
\setsymbol{LFI:center:frequency:LFI19:Rad:M}{67.5}
\setsymbol{LFI:center:frequency:LFI20:Rad:M}{69.2}
\setsymbol{LFI:center:frequency:LFI21:Rad:M}{70.4}
\setsymbol{LFI:center:frequency:LFI22:Rad:M}{71.5}
\setsymbol{LFI:center:frequency:LFI23:Rad:M}{70.8}
\setsymbol{LFI:center:frequency:LFI24:Rad:M}{44.4}
\setsymbol{LFI:center:frequency:LFI25:Rad:M}{44.0}
\setsymbol{LFI:center:frequency:LFI26:Rad:M}{43.9}
\setsymbol{LFI:center:frequency:LFI27:Rad:M}{28.3}
\setsymbol{LFI:center:frequency:LFI28:Rad:M}{28.8}
\setsymbol{LFI:center:frequency:LFI18:Rad:S}{70.1}
\setsymbol{LFI:center:frequency:LFI19:Rad:S}{69.6}
\setsymbol{LFI:center:frequency:LFI20:Rad:S}{69.5}
\setsymbol{LFI:center:frequency:LFI21:Rad:S}{69.5}
\setsymbol{LFI:center:frequency:LFI22:Rad:S}{72.8}
\setsymbol{LFI:center:frequency:LFI23:Rad:S}{71.3}
\setsymbol{LFI:center:frequency:LFI24:Rad:S}{44.1}
\setsymbol{LFI:center:frequency:LFI25:Rad:S}{44.1}
\setsymbol{LFI:center:frequency:LFI26:Rad:S}{44.1}
\setsymbol{LFI:center:frequency:LFI27:Rad:S}{28.5}
\setsymbol{LFI:center:frequency:LFI28:Rad:S}{28.2}

\setsymbol{LFI:white:noise:sensitivity:70GHz:units}{134.7\muKs}
\setsymbol{LFI:white:noise:sensitivity:44GHz:units}{164.7\muKs}
\setsymbol{LFI:white:noise:sensitivity:30GHz:units}{143.4\muKs}

\setsymbol{LFI:white:noise:sensitivity:70GHz}{134.7}
\setsymbol{LFI:white:noise:sensitivity:44GHz}{164.7}
\setsymbol{LFI:white:noise:sensitivity:30GHz}{143.4}

\setsymbol{LFI:white:noise:sensitivity:LFI18:Rad:M:units}{512.0\muKs}
\setsymbol{LFI:white:noise:sensitivity:LFI19:Rad:M:units}{581.4\muKs}
\setsymbol{LFI:white:noise:sensitivity:LFI20:Rad:M:units}{590.8\muKs}
\setsymbol{LFI:white:noise:sensitivity:LFI21:Rad:M:units}{455.2\muKs}
\setsymbol{LFI:white:noise:sensitivity:LFI22:Rad:M:units}{492.0\muKs}
\setsymbol{LFI:white:noise:sensitivity:LFI23:Rad:M:units}{507.7\muKs}
\setsymbol{LFI:white:noise:sensitivity:LFI24:Rad:M:units}{462.2\muKs}
\setsymbol{LFI:white:noise:sensitivity:LFI25:Rad:M:units}{413.6\muKs}
\setsymbol{LFI:white:noise:sensitivity:LFI26:Rad:M:units}{478.6\muKs}
\setsymbol{LFI:white:noise:sensitivity:LFI27:Rad:M:units}{277.7\muKs}
\setsymbol{LFI:white:noise:sensitivity:LFI28:Rad:M:units}{312.3\muKs}
\setsymbol{LFI:white:noise:sensitivity:LFI18:Rad:S:units}{465.7\muKs}
\setsymbol{LFI:white:noise:sensitivity:LFI19:Rad:S:units}{555.6\muKs}
\setsymbol{LFI:white:noise:sensitivity:LFI20:Rad:S:units}{623.2\muKs}
\setsymbol{LFI:white:noise:sensitivity:LFI21:Rad:S:units}{564.1\muKs}
\setsymbol{LFI:white:noise:sensitivity:LFI22:Rad:S:units}{534.4\muKs}
\setsymbol{LFI:white:noise:sensitivity:LFI23:Rad:S:units}{542.4\muKs}
\setsymbol{LFI:white:noise:sensitivity:LFI24:Rad:S:units}{399.2\muKs}
\setsymbol{LFI:white:noise:sensitivity:LFI25:Rad:S:units}{392.6\muKs}
\setsymbol{LFI:white:noise:sensitivity:LFI26:Rad:S:units}{418.6\muKs}
\setsymbol{LFI:white:noise:sensitivity:LFI27:Rad:S:units}{302.9\muKs}
\setsymbol{LFI:white:noise:sensitivity:LFI28:Rad:S:units}{285.3\muKs}

\setsymbol{LFI:white:noise:sensitivity:LFI18:Rad:M}{512.0}
\setsymbol{LFI:white:noise:sensitivity:LFI19:Rad:M}{581.4}
\setsymbol{LFI:white:noise:sensitivity:LFI20:Rad:M}{590.8}
\setsymbol{LFI:white:noise:sensitivity:LFI21:Rad:M}{455.2}
\setsymbol{LFI:white:noise:sensitivity:LFI22:Rad:M}{492.0}
\setsymbol{LFI:white:noise:sensitivity:LFI23:Rad:M}{507.7}
\setsymbol{LFI:white:noise:sensitivity:LFI24:Rad:M}{462.2}
\setsymbol{LFI:white:noise:sensitivity:LFI25:Rad:M}{413.6}
\setsymbol{LFI:white:noise:sensitivity:LFI26:Rad:M}{478.6}
\setsymbol{LFI:white:noise:sensitivity:LFI27:Rad:M}{277.7}
\setsymbol{LFI:white:noise:sensitivity:LFI28:Rad:M}{312.3}
\setsymbol{LFI:white:noise:sensitivity:LFI18:Rad:S}{465.7}
\setsymbol{LFI:white:noise:sensitivity:LFI19:Rad:S}{555.6}
\setsymbol{LFI:white:noise:sensitivity:LFI20:Rad:S}{623.2}
\setsymbol{LFI:white:noise:sensitivity:LFI21:Rad:S}{564.1}
\setsymbol{LFI:white:noise:sensitivity:LFI22:Rad:S}{534.4}
\setsymbol{LFI:white:noise:sensitivity:LFI23:Rad:S}{542.4}
\setsymbol{LFI:white:noise:sensitivity:LFI24:Rad:S}{399.2}
\setsymbol{LFI:white:noise:sensitivity:LFI25:Rad:S}{392.6}
\setsymbol{LFI:white:noise:sensitivity:LFI26:Rad:S}{418.6}
\setsymbol{LFI:white:noise:sensitivity:LFI27:Rad:S}{302.9}
\setsymbol{LFI:white:noise:sensitivity:LFI28:Rad:S}{285.3}

\setsymbol{LFI:knee:frequency:70GHz:units}{29.5\mHz}
\setsymbol{LFI:knee:frequency:44GHz:units}{56.2\mHz}
\setsymbol{LFI:knee:frequency:30GHz:units}{113.7\mHz}

\setsymbol{LFI:knee:frequency:70GHz}{29.5}
\setsymbol{LFI:knee:frequency:44GHz}{56.2}
\setsymbol{LFI:knee:frequency:30GHz}{113.7}

\setsymbol{LFI:knee:frequency:LFI18:Rad:M:units}{16.3\mHz}
\setsymbol{LFI:knee:frequency:LFI19:Rad:M:units}{15.1\mHz}
\setsymbol{LFI:knee:frequency:LFI20:Rad:M:units}{18.7\mHz}
\setsymbol{LFI:knee:frequency:LFI21:Rad:M:units}{37.2\mHz}
\setsymbol{LFI:knee:frequency:LFI22:Rad:M:units}{12.7\mHz}
\setsymbol{LFI:knee:frequency:LFI23:Rad:M:units}{34.6\mHz}
\setsymbol{LFI:knee:frequency:LFI24:Rad:M:units}{46.2\mHz}
\setsymbol{LFI:knee:frequency:LFI25:Rad:M:units}{24.9\mHz}
\setsymbol{LFI:knee:frequency:LFI26:Rad:M:units}{67.6\mHz}
\setsymbol{LFI:knee:frequency:LFI27:Rad:M:units}{187.4\mHz}
\setsymbol{LFI:knee:frequency:LFI28:Rad:M:units}{122.2\mHz}
\setsymbol{LFI:knee:frequency:LFI18:Rad:S:units}{17.7\mHz}
\setsymbol{LFI:knee:frequency:LFI19:Rad:S:units}{22.0\mHz}
\setsymbol{LFI:knee:frequency:LFI20:Rad:S:units}{8.7\mHz}
\setsymbol{LFI:knee:frequency:LFI21:Rad:S:units}{25.9\mHz}
\setsymbol{LFI:knee:frequency:LFI22:Rad:S:units}{15.8\mHz}
\setsymbol{LFI:knee:frequency:LFI23:Rad:S:units}{129.8\mHz}
\setsymbol{LFI:knee:frequency:LFI24:Rad:S:units}{100.9\mHz}
\setsymbol{LFI:knee:frequency:LFI25:Rad:S:units}{38.9\mHz}
\setsymbol{LFI:knee:frequency:LFI26:Rad:S:units}{58.9\mHz}
\setsymbol{LFI:knee:frequency:LFI27:Rad:S:units}{104.4\mHz}
\setsymbol{LFI:knee:frequency:LFI28:Rad:S:units}{40.7\mHz}

\setsymbol{LFI:knee:frequency:LFI18:Rad:M}{16.3}
\setsymbol{LFI:knee:frequency:LFI19:Rad:M}{15.1}
\setsymbol{LFI:knee:frequency:LFI20:Rad:M}{18.7}
\setsymbol{LFI:knee:frequency:LFI21:Rad:M}{37.2}
\setsymbol{LFI:knee:frequency:LFI22:Rad:M}{12.7}
\setsymbol{LFI:knee:frequency:LFI23:Rad:M}{34.6}
\setsymbol{LFI:knee:frequency:LFI24:Rad:M}{46.2}
\setsymbol{LFI:knee:frequency:LFI25:Rad:M}{24.9}
\setsymbol{LFI:knee:frequency:LFI26:Rad:M}{67.6}
\setsymbol{LFI:knee:frequency:LFI27:Rad:M}{187.4}
\setsymbol{LFI:knee:frequency:LFI28:Rad:M}{122.2}
\setsymbol{LFI:knee:frequency:LFI18:Rad:S}{17.7}
\setsymbol{LFI:knee:frequency:LFI19:Rad:S}{22.0}
\setsymbol{LFI:knee:frequency:LFI20:Rad:S}{8.7}
\setsymbol{LFI:knee:frequency:LFI21:Rad:S}{25.9}
\setsymbol{LFI:knee:frequency:LFI22:Rad:S}{15.8}
\setsymbol{LFI:knee:frequency:LFI23:Rad:S}{129.8}
\setsymbol{LFI:knee:frequency:LFI24:Rad:S}{100.9}
\setsymbol{LFI:knee:frequency:LFI25:Rad:S}{38.9}
\setsymbol{LFI:knee:frequency:LFI26:Rad:S}{58.9}
\setsymbol{LFI:knee:frequency:LFI27:Rad:S}{104.4}
\setsymbol{LFI:knee:frequency:LFI28:Rad:S}{40.7}

\setsymbol{LFI:slope:70GHz:units}{$-1.03$\mHz}
\setsymbol{LFI:slope:44GHz:units}{$-0.89$\mHz}
\setsymbol{LFI:slope:30GHz:units}{$-0.87$\mHz}

\setsymbol{LFI:slope:70GHz}{$-1.03$}
\setsymbol{LFI:slope:44GHz}{$-0.89$}
\setsymbol{LFI:slope:30GHz}{$-0.87$}

\setsymbol{LFI:slope:LFI18:Rad:M:units}{$-1.04$\mHz}
\setsymbol{LFI:slope:LFI19:Rad:M:units}{$-1.09$\mHz}
\setsymbol{LFI:slope:LFI20:Rad:M:units}{$-0.69$\mHz}
\setsymbol{LFI:slope:LFI21:Rad:M:units}{$-1.56$\mHz}
\setsymbol{LFI:slope:LFI22:Rad:M:units}{$-1.01$\mHz}
\setsymbol{LFI:slope:LFI23:Rad:M:units}{$-0.96$\mHz}
\setsymbol{LFI:slope:LFI24:Rad:M:units}{$-0.83$\mHz}
\setsymbol{LFI:slope:LFI25:Rad:M:units}{$-0.91$\mHz}
\setsymbol{LFI:slope:LFI26:Rad:M:units}{$-0.95$\mHz}
\setsymbol{LFI:slope:LFI27:Rad:M:units}{$-0.87$\mHz}
\setsymbol{LFI:slope:LFI28:Rad:M:units}{$-0.88$\mHz}
\setsymbol{LFI:slope:LFI18:Rad:S:units}{$-1.15$\mHz}
\setsymbol{LFI:slope:LFI19:Rad:S:units}{$-1.00$\mHz}
\setsymbol{LFI:slope:LFI20:Rad:S:units}{$-0.95$\mHz}
\setsymbol{LFI:slope:LFI21:Rad:S:units}{$-0.92$\mHz}
\setsymbol{LFI:slope:LFI22:Rad:S:units}{$-1.01$\mHz}
\setsymbol{LFI:slope:LFI23:Rad:S:units}{$-0.95$\mHz}
\setsymbol{LFI:slope:LFI24:Rad:S:units}{$-0.73$\mHz}
\setsymbol{LFI:slope:LFI25:Rad:S:units}{$-1.16$\mHz}
\setsymbol{LFI:slope:LFI26:Rad:S:units}{$-0.79$\mHz}
\setsymbol{LFI:slope:LFI27:Rad:S:units}{$-0.82$\mHz}
\setsymbol{LFI:slope:LFI28:Rad:S:units}{$-0.91$\mHz}

\setsymbol{LFI:slope:LFI18:Rad:M}{$-1.04$}
\setsymbol{LFI:slope:LFI19:Rad:M}{$-1.09$}
\setsymbol{LFI:slope:LFI20:Rad:M}{$-0.69$}
\setsymbol{LFI:slope:LFI21:Rad:M}{$-1.56$}
\setsymbol{LFI:slope:LFI22:Rad:M}{$-1.01$}
\setsymbol{LFI:slope:LFI23:Rad:M}{$-0.96$}
\setsymbol{LFI:slope:LFI24:Rad:M}{$-0.83$}
\setsymbol{LFI:slope:LFI25:Rad:M}{$-0.91$}
\setsymbol{LFI:slope:LFI26:Rad:M}{$-0.95$}
\setsymbol{LFI:slope:LFI27:Rad:M}{$-0.87$}
\setsymbol{LFI:slope:LFI28:Rad:M}{$-0.88$}
\setsymbol{LFI:slope:LFI18:Rad:S}{$-1.15$}
\setsymbol{LFI:slope:LFI19:Rad:S}{$-1.00$}
\setsymbol{LFI:slope:LFI20:Rad:S}{$-0.95$}
\setsymbol{LFI:slope:LFI21:Rad:S}{$-0.92$}
\setsymbol{LFI:slope:LFI22:Rad:S}{$-1.01$}
\setsymbol{LFI:slope:LFI23:Rad:S}{$-0.95$}
\setsymbol{LFI:slope:LFI24:Rad:S}{$-0.73$}
\setsymbol{LFI:slope:LFI25:Rad:S}{$-1.16$}
\setsymbol{LFI:slope:LFI26:Rad:S}{$-0.79$}
\setsymbol{LFI:slope:LFI27:Rad:S}{$-0.82$}
\setsymbol{LFI:slope:LFI28:Rad:S}{$-0.91$}

\setsymbol{LFI:FWHM:70GHz:units}{13\parcm01}
\setsymbol{LFI:FWHM:44GHz:units}{27\parcm92}
\setsymbol{LFI:FWHM:30GHz:units}{32\parcm65}

\setsymbol{LFI:FWHM:70GHz}{13.01}
\setsymbol{LFI:FWHM:44GHz}{27.92}
\setsymbol{LFI:FWHM:30GHz}{32.65}

\setsymbol{LFI:FWHM:LFI18:units}{13\parcm39}
\setsymbol{LFI:FWHM:LFI19:units}{13\parcm01}
\setsymbol{LFI:FWHM:LFI20:units}{12\parcm75}
\setsymbol{LFI:FWHM:LFI21:units}{12\parcm74}
\setsymbol{LFI:FWHM:LFI22:units}{12\parcm87}
\setsymbol{LFI:FWHM:LFI23:units}{13\parcm27}
\setsymbol{LFI:FWHM:LFI24:units}{22\parcm98}
\setsymbol{LFI:FWHM:LFI25:units}{30\parcm46}
\setsymbol{LFI:FWHM:LFI26:units}{30\parcm31}
\setsymbol{LFI:FWHM:LFI27:units}{32\parcm65}
\setsymbol{LFI:FWHM:LFI28:units}{32\parcm66}

\setsymbol{LFI:FWHM:LFI18}{13.39}
\setsymbol{LFI:FWHM:LFI19}{13.01}
\setsymbol{LFI:FWHM:LFI20}{12.75}
\setsymbol{LFI:FWHM:LFI21}{12.74}
\setsymbol{LFI:FWHM:LFI22}{12.87}
\setsymbol{LFI:FWHM:LFI23}{13.27}
\setsymbol{LFI:FWHM:LFI24}{22.98}
\setsymbol{LFI:FWHM:LFI25}{30.46}
\setsymbol{LFI:FWHM:LFI26}{30.31}
\setsymbol{LFI:FWHM:LFI27}{32.65}
\setsymbol{LFI:FWHM:LFI28}{32.66}

\setsymbol{LFI:FWHM:uncertainty:LFI18:units}{0.170\arcm}
\setsymbol{LFI:FWHM:uncertainty:LFI19:units}{0.174\arcm}
\setsymbol{LFI:FWHM:uncertainty:LFI20:units}{0.170\arcm}
\setsymbol{LFI:FWHM:uncertainty:LFI21:units}{0.156\arcm}
\setsymbol{LFI:FWHM:uncertainty:LFI22:units}{0.164\arcm}
\setsymbol{LFI:FWHM:uncertainty:LFI23:units}{0.171\arcm}
\setsymbol{LFI:FWHM:uncertainty:LFI24:units}{0.652\arcm}
\setsymbol{LFI:FWHM:uncertainty:LFI25:units}{1.075\arcm}
\setsymbol{LFI:FWHM:uncertainty:LFI26:units}{1.131\arcm}
\setsymbol{LFI:FWHM:uncertainty:LFI27:units}{1.266\arcm}
\setsymbol{LFI:FWHM:uncertainty:LFI28:units}{1.287\arcm}

\setsymbol{LFI:FWHM:uncertainty:LFI18}{0.170}
\setsymbol{LFI:FWHM:uncertainty:LFI19}{0.174}
\setsymbol{LFI:FWHM:uncertainty:LFI20}{0.170}
\setsymbol{LFI:FWHM:uncertainty:LFI21}{0.156}
\setsymbol{LFI:FWHM:uncertainty:LFI22}{0.164}
\setsymbol{LFI:FWHM:uncertainty:LFI23}{0.171}
\setsymbol{LFI:FWHM:uncertainty:LFI24}{0.652}
\setsymbol{LFI:FWHM:uncertainty:LFI25}{1.075}
\setsymbol{LFI:FWHM:uncertainty:LFI26}{1.131}
\setsymbol{LFI:FWHM:uncertainty:LFI27}{1.266}
\setsymbol{LFI:FWHM:uncertainty:LFI28}{1.287}

\setsymbol{HFI:center:frequency:100GHz:units}{100\,GHz}
\setsymbol{HFI:center:frequency:143GHz:units}{143\,GHz}
\setsymbol{HFI:center:frequency:217GHz:units}{217\,GHz}
\setsymbol{HFI:center:frequency:353GHz:units}{353\,GHz}
\setsymbol{HFI:center:frequency:545GHz:units}{545\,GHz}
\setsymbol{HFI:center:frequency:857GHz:units}{857\,GHz}

\setsymbol{HFI:center:frequency:100GHz}{100}
\setsymbol{HFI:center:frequency:143GHz}{143}
\setsymbol{HFI:center:frequency:217GHz}{217}
\setsymbol{HFI:center:frequency:353GHz}{353}
\setsymbol{HFI:center:frequency:545GHz}{545}
\setsymbol{HFI:center:frequency:857GHz}{857}

\setsymbol{HFI:Ndetectors:100GHz}{8}
\setsymbol{HFI:Ndetectors:143GHz}{11}
\setsymbol{HFI:Ndetectors:217GHz}{12}
\setsymbol{HFI:Ndetectors:353GHz}{12}
\setsymbol{HFI:Ndetectors:545GHz}{3}
\setsymbol{HFI:Ndetectors:857GHz}{4}

\setsymbol{HFI:FWHM:Maps:100GHz:units}{9\parcm88}
\setsymbol{HFI:FWHM:Maps:143GHz:units}{7\parcm18}
\setsymbol{HFI:FWHM:Maps:217GHz:units}{4\parcm87}
\setsymbol{HFI:FWHM:Maps:353GHz:units}{4\parcm65}
\setsymbol{HFI:FWHM:Maps:545GHz:units}{4\parcm72}
\setsymbol{HFI:FWHM:Maps:857GHz:units}{4\parcm39}
\setsymbol{HFI:FWHM:Maps:100GHz}{9.88}
\setsymbol{HFI:FWHM:Maps:143GHz}{7.18}
\setsymbol{HFI:FWHM:Maps:217GHz}{4.87}
\setsymbol{HFI:FWHM:Maps:353GHz}{4.65}
\setsymbol{HFI:FWHM:Maps:545GHz}{4.72}
\setsymbol{HFI:FWHM:Maps:857GHz}{4.39}

\setsymbol{HFI:beam:ellipticity:Maps:100GHz}{1.15}
\setsymbol{HFI:beam:ellipticity:Maps:143GHz}{1.01}
\setsymbol{HFI:beam:ellipticity:Maps:217GHz}{1.06}
\setsymbol{HFI:beam:ellipticity:Maps:353GHz}{1.05}
\setsymbol{HFI:beam:ellipticity:Maps:545GHz}{1.14}
\setsymbol{HFI:beam:ellipticity:Maps:857GHz}{1.19}

\setsymbol{HFI:FWHM:Mars:100GHz:units}{9\parcm37}
\setsymbol{HFI:FWHM:Mars:143GHz:units}{7\parcm04}
\setsymbol{HFI:FWHM:Mars:217GHz:units}{4\parcm68}
\setsymbol{HFI:FWHM:Mars:353GHz:units}{4\parcm43}
\setsymbol{HFI:FWHM:Mars:545GHz:units}{3\parcm80}
\setsymbol{HFI:FWHM:Mars:857GHz:units}{3\parcm67}

\setsymbol{HFI:FWHM:Mars:100GHz}{9.37}
\setsymbol{HFI:FWHM:Mars:143GHz}{7.04}
\setsymbol{HFI:FWHM:Mars:217GHz}{4.68}
\setsymbol{HFI:FWHM:Mars:353GHz}{4.43}
\setsymbol{HFI:FWHM:Mars:545GHz}{3.80}
\setsymbol{HFI:FWHM:Mars:857GHz}{3.67}

\setsymbol{HFI:beam:ellipticity:Mars:100GHz}{1.18}
\setsymbol{HFI:beam:ellipticity:Mars:143GHz}{1.03}
\setsymbol{HFI:beam:ellipticity:Mars:217GHz}{1.14}
\setsymbol{HFI:beam:ellipticity:Mars:353GHz}{1.09}
\setsymbol{HFI:beam:ellipticity:Mars:545GHz}{1.25}
\setsymbol{HFI:beam:ellipticity:Mars:857GHz}{1.03}

\setsymbol{HFI:CMB:relative:calibration:100GHz}{$\lsim 1\%$}
\setsymbol{HFI:CMB:relative:calibration:143GHz}{$\lsim 1\%$}
\setsymbol{HFI:CMB:relative:calibration:217GHz}{$\lsim 1\%$}
\setsymbol{HFI:CMB:relative:calibration:353GHz}{$\lsim 1\%$}
\setsymbol{HFI:CMB:relative:calibration:545GHz}{}
\setsymbol{HFI:CMB:relative:calibration:857GHz}{}

\setsymbol{HFI:CMB:absolute:calibration:100GHz}{$\lsim 2\%$}
\setsymbol{HFI:CMB:absolute:calibration:143GHz}{$\lsim 2\%$}
\setsymbol{HFI:CMB:absolute:calibration:217GHz}{$\lsim 2\%$}
\setsymbol{HFI:CMB:absolute:calibration:353GHz}{$\lsim 2\%$}
\setsymbol{HFI:CMB:absolute:calibration:545GHz}{}
\setsymbol{HFI:CMB:absolute:calibration:857GHz}{}

\setsymbol{HFI:FIRAS:gain:calibration:accuracy:statistical:100GHz}{}
\setsymbol{HFI:FIRAS:gain:calibration:accuracy:statistical:143GHz}{}
\setsymbol{HFI:FIRAS:gain:calibration:accuracy:statistical:217GHz}{}
\setsymbol{HFI:FIRAS:gain:calibration:accuracy:statistical:353GHz}{2.5\%}
\setsymbol{HFI:FIRAS:gain:calibration:accuracy:statistical:545GHz}{1\%}
\setsymbol{HFI:FIRAS:gain:calibration:accuracy:statistical:857GHz}{0.5\%}

\setsymbol{HFI:FIRAS:gain:calibration:accuracy:systematic:100GHz}{}
\setsymbol{HFI:FIRAS:gain:calibration:accuracy:systematic:143GHz}{}
\setsymbol{HFI:FIRAS:gain:calibration:accuracy:systematic:217GHz}{}
\setsymbol{HFI:FIRAS:gain:calibration:accuracy:systematic:353GHz}{}
\setsymbol{HFI:FIRAS:gain:calibration:accuracy:systematic:545GHz}{7\%}
\setsymbol{HFI:FIRAS:gain:calibration:accuracy:systematic:857GHz}{7\%}

\setsymbol{HFI:FIRAS:zero:point:accuracy:100GHz:units}{0.8\MJysr}
\setsymbol{HFI:FIRAS:zero:point:accuracy:143GHz:units}{}
\setsymbol{HFI:FIRAS:zero:point:accuracy:217GHz:units}{}
\setsymbol{HFI:FIRAS:zero:point:accuracy:353GHz:units}{1.4\MJysr}
\setsymbol{HFI:FIRAS:zero:point:accuracy:545GHz:units}{2.2\MJysr}
\setsymbol{HFI:FIRAS:zero:point:accuracy:857GHz:units}{1.7\MJysr}

\setsymbol{HFI:FIRAS:zero:point:accuracy:100GHz}{0.8}
\setsymbol{HFI:FIRAS:zero:point:accuracy:143GHz}{}
\setsymbol{HFI:FIRAS:zero:point:accuracy:217GHz}{}
\setsymbol{HFI:FIRAS:zero:point:accuracy:353GHz}{1.4}
\setsymbol{HFI:FIRAS:zero:point:accuracy:545GHz}{2.2}
\setsymbol{HFI:FIRAS:zero:point:accuracy:857GHz}{1.7}

\setsymbol{HFI:unit:conversion:100GHz:units}{0.2415\MJysrmK}
\setsymbol{HFI:unit:conversion:143GHz:units}{0.3694\MJysrmK}
\setsymbol{HFI:unit:conversion:217GHz:units}{0.4811\MJysrmK}
\setsymbol{HFI:unit:conversion:353GHz:units}{0.2883\MJysrmK}
\setsymbol{HFI:unit:conversion:545GHz:units}{0.05826\MJysrmK}
\setsymbol{HFI:unit:conversion:857GHz:units}{0.002238\MJysrmK}

\setsymbol{HFI:unit:conversion:100GHz}{0.2415}
\setsymbol{HFI:unit:conversion:143GHz}{0.3694}
\setsymbol{HFI:unit:conversion:217GHz}{0.4811}
\setsymbol{HFI:unit:conversion:353GHz}{0.2883}
\setsymbol{HFI:unit:conversion:545GHz}{0.05826}
\setsymbol{HFI:unit:conversion:857GHz}{0.002238}

\setsymbol{HFI:colour:correction:alpha=-2:V1.01:100GHz}{0.9893}
\setsymbol{HFI:colour:correction:alpha=-2:V1.01:143GHz}{0.9759}
\setsymbol{HFI:colour:correction:alpha=-2:V1.01:217GHz}{1.0007}
\setsymbol{HFI:colour:correction:alpha=-2:V1.01:353GHz}{1.0028}
\setsymbol{HFI:colour:correction:alpha=-2:V1.01:545GHz}{1.0019}
\setsymbol{HFI:colour:correction:alpha=-2:V1.01:857GHz}{0.9889}

\setsymbol{HFI:colour:correction:alpha=0:V1.01:100GHz}{1.0008}
\setsymbol{HFI:colour:correction:alpha=0:V1.01:143GHz}{1.0148}
\setsymbol{HFI:colour:correction:alpha=0:V1.01:217GHz}{0.9909}
\setsymbol{HFI:colour:correction:alpha=0:V1.01:353GHz}{0.9888}
\setsymbol{HFI:colour:correction:alpha=0:V1.01:545GHz}{0.9878}
\setsymbol{HFI:colour:correction:alpha=0:V1.01:857GHz}{1.0014}

\usepackage{graphicx}
\usepackage{lscape}
\usepackage{color}
\definecolor{AliceBlue}{rgb}{0.94,0.97,1.00}
\definecolor{AntiqueWhite1}{rgb}{1.00,0.94,0.86}
\definecolor{AntiqueWhite2}{rgb}{0.93,0.87,0.80}
\definecolor{AntiqueWhite3}{rgb}{0.80,0.75,0.69}
\definecolor{AntiqueWhite4}{rgb}{0.55,0.51,0.47}
\definecolor{AntiqueWhite}{rgb}{0.98,0.92,0.84}
\definecolor{BlanchedAlmond}{rgb}{1.00,0.92,0.80}
\definecolor{BlueViolet}{rgb}{0.54,0.17,0.89}
\definecolor{CadetBlue1}{rgb}{0.60,0.96,1.00}
\definecolor{CadetBlue2}{rgb}{0.56,0.90,0.93}
\definecolor{CadetBlue3}{rgb}{0.48,0.77,0.80}
\definecolor{CadetBlue4}{rgb}{0.33,0.53,0.55}
\definecolor{CadetBlue}{rgb}{0.37,0.62,0.63}
\definecolor{CornflowerBlue}{rgb}{0.39,0.58,0.93}
\definecolor{DarkBlue}{rgb}{0.00,0.00,0.55}
\definecolor{DarkCyan}{rgb}{0.00,0.55,0.55}
\definecolor{DarkGoldenrod1}{rgb}{1.00,0.73,0.06}
\definecolor{DarkGoldenrod2}{rgb}{0.93,0.68,0.05}
\definecolor{DarkGoldenrod3}{rgb}{0.80,0.58,0.05}
\definecolor{DarkGoldenrod4}{rgb}{0.55,0.40,0.03}
\definecolor{DarkGoldenrod}{rgb}{0.72,0.53,0.04}
\definecolor{DarkGray}{rgb}{0.66,0.66,0.66}
\definecolor{DarkGreen}{rgb}{0.00,0.39,0.00}
\definecolor{DarkGrey}{rgb}{0.66,0.66,0.66}
\definecolor{DarkKhaki}{rgb}{0.74,0.72,0.42}
\definecolor{DarkMagenta}{rgb}{0.55,0.00,0.55}
\definecolor{DarkOliveGreen1}{rgb}{0.79,1.00,0.44}
\definecolor{DarkOliveGreen2}{rgb}{0.74,0.93,0.41}
\definecolor{DarkOliveGreen3}{rgb}{0.64,0.80,0.35}
\definecolor{DarkOliveGreen4}{rgb}{0.43,0.55,0.24}
\definecolor{DarkOliveGreen}{rgb}{0.33,0.42,0.18}
\definecolor{DarkOrange1}{rgb}{1.00,0.50,0.00}
\definecolor{DarkOrange2}{rgb}{0.93,0.46,0.00}
\definecolor{DarkOrange3}{rgb}{0.80,0.40,0.00}
\definecolor{DarkOrange4}{rgb}{0.55,0.27,0.00}
\definecolor{DarkOrange}{rgb}{1.00,0.55,0.00}
\definecolor{DarkOrchid1}{rgb}{0.75,0.24,1.00}
\definecolor{DarkOrchid2}{rgb}{0.70,0.23,0.93}
\definecolor{DarkOrchid3}{rgb}{0.60,0.20,0.80}
\definecolor{DarkOrchid4}{rgb}{0.41,0.13,0.55}
\definecolor{DarkOrchid}{rgb}{0.60,0.20,0.80}
\definecolor{DarkRed}{rgb}{0.55,0.00,0.00}
\definecolor{DarkSalmon}{rgb}{0.91,0.59,0.48}
\definecolor{DarkSeaGreen1}{rgb}{0.76,1.00,0.76}
\definecolor{DarkSeaGreen2}{rgb}{0.71,0.93,0.71}
\definecolor{DarkSeaGreen3}{rgb}{0.61,0.80,0.61}
\definecolor{DarkSeaGreen4}{rgb}{0.41,0.55,0.41}
\definecolor{DarkSeaGreen}{rgb}{0.56,0.74,0.56}
\definecolor{DarkSlateBlue}{rgb}{0.28,0.24,0.55}
\definecolor{DarkSlateGray1}{rgb}{0.59,1.00,1.00}
\definecolor{DarkSlateGray2}{rgb}{0.55,0.93,0.93}
\definecolor{DarkSlateGray3}{rgb}{0.47,0.80,0.80}
\definecolor{DarkSlateGray4}{rgb}{0.32,0.55,0.55}
\definecolor{DarkSlateGray}{rgb}{0.18,0.31,0.31}
\definecolor{DarkSlateGrey}{rgb}{0.18,0.31,0.31}
\definecolor{DarkTurquoise}{rgb}{0.00,0.81,0.82}
\definecolor{DarkViolet}{rgb}{0.58,0.00,0.83}
\definecolor{DeepPink1}{rgb}{1.00,0.08,0.58}
\definecolor{DeepPink2}{rgb}{0.93,0.07,0.54}
\definecolor{DeepPink3}{rgb}{0.80,0.06,0.46}
\definecolor{DeepPink4}{rgb}{0.55,0.04,0.31}
\definecolor{DeepPink}{rgb}{1.00,0.08,0.58}
\definecolor{DeepSkyBlue1}{rgb}{0.00,0.75,1.00}
\definecolor{DeepSkyBlue2}{rgb}{0.00,0.70,0.93}
\definecolor{DeepSkyBlue3}{rgb}{0.00,0.60,0.80}
\definecolor{DeepSkyBlue4}{rgb}{0.00,0.41,0.55}
\definecolor{DeepSkyBlue}{rgb}{0.00,0.75,1.00}
\definecolor{DimGray}{rgb}{0.41,0.41,0.41}
\definecolor{DimGrey}{rgb}{0.41,0.41,0.41}
\definecolor{DodgerBlue1}{rgb}{0.12,0.56,1.00}
\definecolor{DodgerBlue2}{rgb}{0.11,0.53,0.93}
\definecolor{DodgerBlue3}{rgb}{0.09,0.45,0.80}
\definecolor{DodgerBlue4}{rgb}{0.06,0.31,0.55}
\definecolor{DodgerBlue}{rgb}{0.12,0.56,1.00}
\definecolor{FloralWhite}{rgb}{1.00,0.98,0.94}
\definecolor{ForestGreen}{rgb}{0.13,0.55,0.13}
\definecolor{GhostWhite}{rgb}{0.97,0.97,1.00}
\definecolor{GreenYellow}{rgb}{0.68,1.00,0.18}
\definecolor{HotPink1}{rgb}{1.00,0.43,0.71}
\definecolor{HotPink2}{rgb}{0.93,0.42,0.65}
\definecolor{HotPink3}{rgb}{0.80,0.38,0.56}
\definecolor{HotPink4}{rgb}{0.55,0.23,0.38}
\definecolor{HotPink}{rgb}{1.00,0.41,0.71}
\definecolor{IndianRed1}{rgb}{1.00,0.42,0.42}
\definecolor{IndianRed2}{rgb}{0.93,0.39,0.39}
\definecolor{IndianRed3}{rgb}{0.80,0.33,0.33}
\definecolor{IndianRed4}{rgb}{0.55,0.23,0.23}
\definecolor{IndianRed}{rgb}{0.80,0.36,0.36}
\definecolor{LavenderBlush1}{rgb}{1.00,0.94,0.96}
\definecolor{LavenderBlush2}{rgb}{0.93,0.88,0.90}
\definecolor{LavenderBlush3}{rgb}{0.80,0.76,0.77}
\definecolor{LavenderBlush4}{rgb}{0.55,0.51,0.53}
\definecolor{LavenderBlush}{rgb}{1.00,0.94,0.96}
\definecolor{LawnGreen}{rgb}{0.49,0.99,0.00}
\definecolor{LemonChiffon1}{rgb}{1.00,0.98,0.80}
\definecolor{LemonChiffon2}{rgb}{0.93,0.91,0.75}
\definecolor{LemonChiffon3}{rgb}{0.80,0.79,0.65}
\definecolor{LemonChiffon4}{rgb}{0.55,0.54,0.44}
\definecolor{LemonChiffon}{rgb}{1.00,0.98,0.80}
\definecolor{LightBlue1}{rgb}{0.75,0.94,1.00}
\definecolor{LightBlue2}{rgb}{0.70,0.87,0.93}
\definecolor{LightBlue3}{rgb}{0.60,0.75,0.80}
\definecolor{LightBlue4}{rgb}{0.41,0.51,0.55}
\definecolor{LightBlue}{rgb}{0.68,0.85,0.90}
\definecolor{LightCoral}{rgb}{0.94,0.50,0.50}
\definecolor{LightCyan1}{rgb}{0.88,1.00,1.00}
\definecolor{LightCyan2}{rgb}{0.82,0.93,0.93}
\definecolor{LightCyan3}{rgb}{0.71,0.80,0.80}
\definecolor{LightCyan4}{rgb}{0.48,0.55,0.55}
\definecolor{LightCyan}{rgb}{0.88,1.00,1.00}
\definecolor{LightGoldenrod1}{rgb}{1.00,0.93,0.55}
\definecolor{LightGoldenrod2}{rgb}{0.93,0.86,0.51}
\definecolor{LightGoldenrod3}{rgb}{0.80,0.75,0.44}
\definecolor{LightGoldenrod4}{rgb}{0.55,0.51,0.30}
\definecolor{LightGoldenrodYellow}{rgb}{0.98,0.98,0.82}
\definecolor{LightGoldenrod}{rgb}{0.93,0.87,0.51}
\definecolor{LightGray}{rgb}{0.83,0.83,0.83}
\definecolor{LightGreen}{rgb}{0.56,0.93,0.56}
\definecolor{LightGrey}{rgb}{0.83,0.83,0.83}
\definecolor{LightPink1}{rgb}{1.00,0.68,0.73}
\definecolor{LightPink2}{rgb}{0.93,0.64,0.68}
\definecolor{LightPink3}{rgb}{0.80,0.55,0.58}
\definecolor{LightPink4}{rgb}{0.55,0.37,0.40}
\definecolor{LightPink}{rgb}{1.00,0.71,0.76}
\definecolor{LightSalmon1}{rgb}{1.00,0.63,0.48}
\definecolor{LightSalmon2}{rgb}{0.93,0.58,0.45}
\definecolor{LightSalmon3}{rgb}{0.80,0.51,0.38}
\definecolor{LightSalmon4}{rgb}{0.55,0.34,0.26}
\definecolor{LightSalmon}{rgb}{1.00,0.63,0.48}
\definecolor{LightSeaGreen}{rgb}{0.13,0.70,0.67}
\definecolor{LightSkyBlue1}{rgb}{0.69,0.89,1.00}
\definecolor{LightSkyBlue2}{rgb}{0.64,0.83,0.93}
\definecolor{LightSkyBlue3}{rgb}{0.55,0.71,0.80}
\definecolor{LightSkyBlue4}{rgb}{0.38,0.48,0.55}
\definecolor{LightSkyBlue}{rgb}{0.53,0.81,0.98}
\definecolor{LightSlateBlue}{rgb}{0.52,0.44,1.00}
\definecolor{LightSlateGray}{rgb}{0.47,0.53,0.60}
\definecolor{LightSlateGrey}{rgb}{0.47,0.53,0.60}
\definecolor{LightSteelBlue1}{rgb}{0.79,0.88,1.00}
\definecolor{LightSteelBlue2}{rgb}{0.74,0.82,0.93}
\definecolor{LightSteelBlue3}{rgb}{0.64,0.71,0.80}
\definecolor{LightSteelBlue4}{rgb}{0.43,0.48,0.55}
\definecolor{LightSteelBlue}{rgb}{0.69,0.77,0.87}
\definecolor{LightYellow1}{rgb}{1.00,1.00,0.88}
\definecolor{LightYellow2}{rgb}{0.93,0.93,0.82}
\definecolor{LightYellow3}{rgb}{0.80,0.80,0.71}
\definecolor{LightYellow4}{rgb}{0.55,0.55,0.48}
\definecolor{LightYellow}{rgb}{1.00,1.00,0.88}
\definecolor{LimeGreen}{rgb}{0.20,0.80,0.20}
\definecolor{MediumAquamarine}{rgb}{0.40,0.80,0.67}
\definecolor{MediumBlue}{rgb}{0.00,0.00,0.80}
\definecolor{MediumOrchid1}{rgb}{0.88,0.40,1.00}
\definecolor{MediumOrchid2}{rgb}{0.82,0.37,0.93}
\definecolor{MediumOrchid3}{rgb}{0.71,0.32,0.80}
\definecolor{MediumOrchid4}{rgb}{0.48,0.22,0.55}
\definecolor{MediumOrchid}{rgb}{0.73,0.33,0.83}
\definecolor{MediumPurple1}{rgb}{0.67,0.51,1.00}
\definecolor{MediumPurple2}{rgb}{0.62,0.47,0.93}
\definecolor{MediumPurple3}{rgb}{0.54,0.41,0.80}
\definecolor{MediumPurple4}{rgb}{0.36,0.28,0.55}
\definecolor{MediumPurple}{rgb}{0.58,0.44,0.86}
\definecolor{MediumSeaGreen}{rgb}{0.24,0.70,0.44}
\definecolor{MediumSlateBlue}{rgb}{0.48,0.41,0.93}
\definecolor{MediumSpringGreen}{rgb}{0.00,0.98,0.60}
\definecolor{MediumTurquoise}{rgb}{0.28,0.82,0.80}
\definecolor{MediumVioletRed}{rgb}{0.78,0.08,0.52}
\definecolor{MidnightBlue}{rgb}{0.10,0.10,0.44}
\definecolor{MintCream}{rgb}{0.96,1.00,0.98}
\definecolor{MistyRose1}{rgb}{1.00,0.89,0.88}
\definecolor{MistyRose2}{rgb}{0.93,0.84,0.82}
\definecolor{MistyRose3}{rgb}{0.80,0.72,0.71}
\definecolor{MistyRose4}{rgb}{0.55,0.49,0.48}
\definecolor{MistyRose}{rgb}{1.00,0.89,0.88}
\definecolor{NavajoWhite1}{rgb}{1.00,0.87,0.68}
\definecolor{NavajoWhite2}{rgb}{0.93,0.81,0.63}
\definecolor{NavajoWhite3}{rgb}{0.80,0.70,0.55}
\definecolor{NavajoWhite4}{rgb}{0.55,0.47,0.37}
\definecolor{NavajoWhite}{rgb}{1.00,0.87,0.68}
\definecolor{NavyBlue}{rgb}{0.00,0.00,0.50}
\definecolor{OldLace}{rgb}{0.99,0.96,0.90}
\definecolor{OliveDrab1}{rgb}{0.75,1.00,0.24}
\definecolor{OliveDrab2}{rgb}{0.70,0.93,0.23}
\definecolor{OliveDrab3}{rgb}{0.60,0.80,0.20}
\definecolor{OliveDrab4}{rgb}{0.41,0.55,0.13}
\definecolor{OliveDrab}{rgb}{0.42,0.56,0.14}
\definecolor{OrangeRed1}{rgb}{1.00,0.27,0.00}
\definecolor{OrangeRed2}{rgb}{0.93,0.25,0.00}
\definecolor{OrangeRed3}{rgb}{0.80,0.22,0.00}
\definecolor{OrangeRed4}{rgb}{0.55,0.15,0.00}
\definecolor{OrangeRed}{rgb}{1.00,0.27,0.00}
\definecolor{PaleGoldenrod}{rgb}{0.93,0.91,0.67}
\definecolor{PaleGreen1}{rgb}{0.60,1.00,0.60}
\definecolor{PaleGreen2}{rgb}{0.56,0.93,0.56}
\definecolor{PaleGreen3}{rgb}{0.49,0.80,0.49}
\definecolor{PaleGreen4}{rgb}{0.33,0.55,0.33}
\definecolor{PaleGreen}{rgb}{0.60,0.98,0.60}
\definecolor{PaleTurquoise1}{rgb}{0.73,1.00,1.00}
\definecolor{PaleTurquoise2}{rgb}{0.68,0.93,0.93}
\definecolor{PaleTurquoise3}{rgb}{0.59,0.80,0.80}
\definecolor{PaleTurquoise4}{rgb}{0.40,0.55,0.55}
\definecolor{PaleTurquoise}{rgb}{0.69,0.93,0.93}
\definecolor{PaleVioletRed1}{rgb}{1.00,0.51,0.67}
\definecolor{PaleVioletRed2}{rgb}{0.93,0.47,0.62}
\definecolor{PaleVioletRed3}{rgb}{0.80,0.41,0.54}
\definecolor{PaleVioletRed4}{rgb}{0.55,0.28,0.36}
\definecolor{PaleVioletRed}{rgb}{0.86,0.44,0.58}
\definecolor{PapayaWhip}{rgb}{1.00,0.94,0.84}
\definecolor{PeachPuff1}{rgb}{1.00,0.85,0.73}
\definecolor{PeachPuff2}{rgb}{0.93,0.80,0.68}
\definecolor{PeachPuff3}{rgb}{0.80,0.69,0.58}
\definecolor{PeachPuff4}{rgb}{0.55,0.47,0.40}
\definecolor{PeachPuff}{rgb}{1.00,0.85,0.73}
\definecolor{PowderBlue}{rgb}{0.69,0.88,0.90}
\definecolor{RosyBrown1}{rgb}{1.00,0.76,0.76}
\definecolor{RosyBrown2}{rgb}{0.93,0.71,0.71}
\definecolor{RosyBrown3}{rgb}{0.80,0.61,0.61}
\definecolor{RosyBrown4}{rgb}{0.55,0.41,0.41}
\definecolor{RosyBrown}{rgb}{0.74,0.56,0.56}
\definecolor{RoyalBlue1}{rgb}{0.28,0.46,1.00}
\definecolor{RoyalBlue2}{rgb}{0.26,0.43,0.93}
\definecolor{RoyalBlue3}{rgb}{0.23,0.37,0.80}
\definecolor{RoyalBlue4}{rgb}{0.15,0.25,0.55}
\definecolor{RoyalBlue}{rgb}{0.25,0.41,0.88}
\definecolor{SaddleBrown}{rgb}{0.55,0.27,0.07}
\definecolor{SandyBrown}{rgb}{0.96,0.64,0.38}
\definecolor{SeaGreen1}{rgb}{0.33,1.00,0.62}
\definecolor{SeaGreen2}{rgb}{0.31,0.93,0.58}
\definecolor{SeaGreen3}{rgb}{0.26,0.80,0.50}
\definecolor{SeaGreen4}{rgb}{0.18,0.55,0.34}
\definecolor{SeaGreen}{rgb}{0.18,0.55,0.34}
\definecolor{SkyBlue1}{rgb}{0.53,0.81,1.00}
\definecolor{SkyBlue2}{rgb}{0.49,0.75,0.93}
\definecolor{SkyBlue3}{rgb}{0.42,0.65,0.80}
\definecolor{SkyBlue4}{rgb}{0.29,0.44,0.55}
\definecolor{SkyBlue}{rgb}{0.53,0.81,0.92}
\definecolor{SlateBlue1}{rgb}{0.51,0.44,1.00}
\definecolor{SlateBlue2}{rgb}{0.48,0.40,0.93}
\definecolor{SlateBlue3}{rgb}{0.41,0.35,0.80}
\definecolor{SlateBlue4}{rgb}{0.28,0.24,0.55}
\definecolor{SlateBlue}{rgb}{0.42,0.35,0.80}
\definecolor{SlateGray1}{rgb}{0.78,0.89,1.00}
\definecolor{SlateGray2}{rgb}{0.73,0.83,0.93}
\definecolor{SlateGray3}{rgb}{0.62,0.71,0.80}
\definecolor{SlateGray4}{rgb}{0.42,0.48,0.55}
\definecolor{SlateGray}{rgb}{0.44,0.50,0.56}
\definecolor{SlateGrey}{rgb}{0.44,0.50,0.56}
\definecolor{SpringGreen1}{rgb}{0.00,1.00,0.50}
\definecolor{SpringGreen2}{rgb}{0.00,0.93,0.46}
\definecolor{SpringGreen3}{rgb}{0.00,0.80,0.40}
\definecolor{SpringGreen4}{rgb}{0.00,0.55,0.27}
\definecolor{SpringGreen}{rgb}{0.00,1.00,0.50}
\definecolor{SteelBlue1}{rgb}{0.39,0.72,1.00}
\definecolor{SteelBlue2}{rgb}{0.36,0.67,0.93}
\definecolor{SteelBlue3}{rgb}{0.31,0.58,0.80}
\definecolor{SteelBlue4}{rgb}{0.21,0.39,0.55}
\definecolor{SteelBlue}{rgb}{0.27,0.51,0.71}
\definecolor{VioletRed1}{rgb}{1.00,0.24,0.59}
\definecolor{VioletRed2}{rgb}{0.93,0.23,0.55}
\definecolor{VioletRed3}{rgb}{0.80,0.20,0.47}
\definecolor{VioletRed4}{rgb}{0.55,0.13,0.32}
\definecolor{VioletRed}{rgb}{0.82,0.13,0.56}
\definecolor{WhiteSmoke}{rgb}{0.96,0.96,0.96}
\definecolor{YellowGreen}{rgb}{0.60,0.80,0.20}
\definecolor{aliceblue}{rgb}{0.94,0.97,1.00}
\definecolor{antiquewhite}{rgb}{0.98,0.92,0.84}
\definecolor{aquamarine1}{rgb}{0.50,1.00,0.83}
\definecolor{aquamarine2}{rgb}{0.46,0.93,0.78}
\definecolor{aquamarine3}{rgb}{0.40,0.80,0.67}
\definecolor{aquamarine4}{rgb}{0.27,0.55,0.45}
\definecolor{aquamarine}{rgb}{0.50,1.00,0.83}
\definecolor{azure1}{rgb}{0.94,1.00,1.00}
\definecolor{azure2}{rgb}{0.88,0.93,0.93}
\definecolor{azure3}{rgb}{0.76,0.80,0.80}
\definecolor{azure4}{rgb}{0.51,0.55,0.55}
\definecolor{azure}{rgb}{0.94,1.00,1.00}
\definecolor{beige}{rgb}{0.96,0.96,0.86}
\definecolor{bisque1}{rgb}{1.00,0.89,0.77}
\definecolor{bisque2}{rgb}{0.93,0.84,0.72}
\definecolor{bisque3}{rgb}{0.80,0.72,0.62}
\definecolor{bisque4}{rgb}{0.55,0.49,0.42}
\definecolor{bisque}{rgb}{1.00,0.89,0.77}
\definecolor{black}{rgb}{0.00,0.00,0.00}
\definecolor{blanchedalmond}{rgb}{1.00,0.92,0.80}
\definecolor{blue1}{rgb}{0.00,0.00,1.00}
\definecolor{blue2}{rgb}{0.00,0.00,0.93}
\definecolor{blue3}{rgb}{0.00,0.00,0.80}
\definecolor{blue4}{rgb}{0.00,0.00,0.55}
\definecolor{blueviolet}{rgb}{0.54,0.17,0.89}
\definecolor{blue}{rgb}{0.00,0.00,1.00}
\definecolor{brown1}{rgb}{1.00,0.25,0.25}
\definecolor{brown2}{rgb}{0.93,0.23,0.23}
\definecolor{brown3}{rgb}{0.80,0.20,0.20}
\definecolor{brown4}{rgb}{0.55,0.14,0.14}
\definecolor{brown}{rgb}{0.65,0.16,0.16}
\definecolor{burlywood1}{rgb}{1.00,0.83,0.61}
\definecolor{burlywood2}{rgb}{0.93,0.77,0.57}
\definecolor{burlywood3}{rgb}{0.80,0.67,0.49}
\definecolor{burlywood4}{rgb}{0.55,0.45,0.33}
\definecolor{burlywood}{rgb}{0.87,0.72,0.53}
\definecolor{cadetblue}{rgb}{0.37,0.62,0.63}
\definecolor{chartreuse1}{rgb}{0.50,1.00,0.00}
\definecolor{chartreuse2}{rgb}{0.46,0.93,0.00}
\definecolor{chartreuse3}{rgb}{0.40,0.80,0.00}
\definecolor{chartreuse4}{rgb}{0.27,0.55,0.00}
\definecolor{chartreuse}{rgb}{0.50,1.00,0.00}
\definecolor{chocolate1}{rgb}{1.00,0.50,0.14}
\definecolor{chocolate2}{rgb}{0.93,0.46,0.13}
\definecolor{chocolate3}{rgb}{0.80,0.40,0.11}
\definecolor{chocolate4}{rgb}{0.55,0.27,0.07}
\definecolor{chocolate}{rgb}{0.82,0.41,0.12}
\definecolor{coral1}{rgb}{1.00,0.45,0.34}
\definecolor{coral2}{rgb}{0.93,0.42,0.31}
\definecolor{coral3}{rgb}{0.80,0.36,0.27}
\definecolor{coral4}{rgb}{0.55,0.24,0.18}
\definecolor{coral}{rgb}{1.00,0.50,0.31}
\definecolor{cornflowerblue}{rgb}{0.39,0.58,0.93}
\definecolor{cornsilk1}{rgb}{1.00,0.97,0.86}
\definecolor{cornsilk2}{rgb}{0.93,0.91,0.80}
\definecolor{cornsilk3}{rgb}{0.80,0.78,0.69}
\definecolor{cornsilk4}{rgb}{0.55,0.53,0.47}
\definecolor{cornsilk}{rgb}{1.00,0.97,0.86}
\definecolor{cyan1}{rgb}{0.00,1.00,1.00}
\definecolor{cyan2}{rgb}{0.00,0.93,0.93}
\definecolor{cyan3}{rgb}{0.00,0.80,0.80}
\definecolor{cyan4}{rgb}{0.00,0.55,0.55}
\definecolor{cyan}{rgb}{0.00,1.00,1.00}
\definecolor{darkblue}{rgb}{0.00,0.00,0.55}
\definecolor{darkcyan}{rgb}{0.00,0.55,0.55}
\definecolor{darkgoldenrod}{rgb}{0.72,0.53,0.04}
\definecolor{darkgray}{rgb}{0.66,0.66,0.66}
\definecolor{darkgreen}{rgb}{0.00,0.39,0.00}
\definecolor{darkgrey}{rgb}{0.66,0.66,0.66}
\definecolor{darkkhaki}{rgb}{0.74,0.72,0.42}
\definecolor{darkmagenta}{rgb}{0.55,0.00,0.55}
\definecolor{darkolive}{rgb}{0.33,0.42,0.18}
\definecolor{darkorange}{rgb}{1.00,0.55,0.00}
\definecolor{darkorchid}{rgb}{0.60,0.20,0.80}
\definecolor{darkred}{rgb}{0.55,0.00,0.00}
\definecolor{darksalmon}{rgb}{0.91,0.59,0.48}
\definecolor{darksea}{rgb}{0.56,0.74,0.56}
\definecolor{darkslate}{rgb}{0.18,0.31,0.31}
\definecolor{darkslate}{rgb}{0.18,0.31,0.31}
\definecolor{darkslate}{rgb}{0.28,0.24,0.55}
\definecolor{darkturquoise}{rgb}{0.00,0.81,0.82}
\definecolor{darkviolet}{rgb}{0.58,0.00,0.83}
\definecolor{deeppink}{rgb}{1.00,0.08,0.58}
\definecolor{deepsky}{rgb}{0.00,0.75,1.00}
\definecolor{dimgray}{rgb}{0.41,0.41,0.41}
\definecolor{dimgrey}{rgb}{0.41,0.41,0.41}
\definecolor{dodgerblue}{rgb}{0.12,0.56,1.00}
\definecolor{firebrick1}{rgb}{1.00,0.19,0.19}
\definecolor{firebrick2}{rgb}{0.93,0.17,0.17}
\definecolor{firebrick3}{rgb}{0.80,0.15,0.15}
\definecolor{firebrick4}{rgb}{0.55,0.10,0.10}
\definecolor{firebrick}{rgb}{0.70,0.13,0.13}
\definecolor{floralwhite}{rgb}{1.00,0.98,0.94}
\definecolor{forestgreen}{rgb}{0.13,0.55,0.13}
\definecolor{gainsboro}{rgb}{0.86,0.86,0.86}
\definecolor{ghostwhite}{rgb}{0.97,0.97,1.00}
\definecolor{gold1}{rgb}{1.00,0.84,0.00}
\definecolor{gold2}{rgb}{0.93,0.79,0.00}
\definecolor{gold3}{rgb}{0.80,0.68,0.00}
\definecolor{gold4}{rgb}{0.55,0.46,0.00}
\definecolor{goldenrod1}{rgb}{1.00,0.76,0.15}
\definecolor{goldenrod2}{rgb}{0.93,0.71,0.13}
\definecolor{goldenrod3}{rgb}{0.80,0.61,0.11}
\definecolor{goldenrod4}{rgb}{0.55,0.41,0.08}
\definecolor{goldenrod}{rgb}{0.85,0.65,0.13}
\definecolor{gold}{rgb}{1.00,0.84,0.00}
\definecolor{gray0}{rgb}{0.00,0.00,0.00}
\definecolor{gray100}{rgb}{1.00,1.00,1.00}
\definecolor{gray10}{rgb}{0.10,0.10,0.10}
\definecolor{gray11}{rgb}{0.11,0.11,0.11}
\definecolor{gray12}{rgb}{0.12,0.12,0.12}
\definecolor{gray13}{rgb}{0.13,0.13,0.13}
\definecolor{gray14}{rgb}{0.14,0.14,0.14}
\definecolor{gray15}{rgb}{0.15,0.15,0.15}
\definecolor{gray16}{rgb}{0.16,0.16,0.16}
\definecolor{gray17}{rgb}{0.17,0.17,0.17}
\definecolor{gray18}{rgb}{0.18,0.18,0.18}
\definecolor{gray19}{rgb}{0.19,0.19,0.19}
\definecolor{gray1}{rgb}{0.01,0.01,0.01}
\definecolor{gray20}{rgb}{0.20,0.20,0.20}
\definecolor{gray21}{rgb}{0.21,0.21,0.21}
\definecolor{gray22}{rgb}{0.22,0.22,0.22}
\definecolor{gray23}{rgb}{0.23,0.23,0.23}
\definecolor{gray24}{rgb}{0.24,0.24,0.24}
\definecolor{gray25}{rgb}{0.25,0.25,0.25}
\definecolor{gray26}{rgb}{0.26,0.26,0.26}
\definecolor{gray27}{rgb}{0.27,0.27,0.27}
\definecolor{gray28}{rgb}{0.28,0.28,0.28}
\definecolor{gray29}{rgb}{0.29,0.29,0.29}
\definecolor{gray2}{rgb}{0.02,0.02,0.02}
\definecolor{gray30}{rgb}{0.30,0.30,0.30}
\definecolor{gray31}{rgb}{0.31,0.31,0.31}
\definecolor{gray32}{rgb}{0.32,0.32,0.32}
\definecolor{gray33}{rgb}{0.33,0.33,0.33}
\definecolor{gray34}{rgb}{0.34,0.34,0.34}
\definecolor{gray35}{rgb}{0.35,0.35,0.35}
\definecolor{gray36}{rgb}{0.36,0.36,0.36}
\definecolor{gray37}{rgb}{0.37,0.37,0.37}
\definecolor{gray38}{rgb}{0.38,0.38,0.38}
\definecolor{gray39}{rgb}{0.39,0.39,0.39}
\definecolor{gray3}{rgb}{0.03,0.03,0.03}
\definecolor{gray40}{rgb}{0.40,0.40,0.40}
\definecolor{gray41}{rgb}{0.41,0.41,0.41}
\definecolor{gray42}{rgb}{0.42,0.42,0.42}
\definecolor{gray43}{rgb}{0.43,0.43,0.43}
\definecolor{gray44}{rgb}{0.44,0.44,0.44}
\definecolor{gray45}{rgb}{0.45,0.45,0.45}
\definecolor{gray46}{rgb}{0.46,0.46,0.46}
\definecolor{gray47}{rgb}{0.47,0.47,0.47}
\definecolor{gray48}{rgb}{0.48,0.48,0.48}
\definecolor{gray49}{rgb}{0.49,0.49,0.49}
\definecolor{gray4}{rgb}{0.04,0.04,0.04}
\definecolor{gray50}{rgb}{0.50,0.50,0.50}
\definecolor{gray51}{rgb}{0.51,0.51,0.51}
\definecolor{gray52}{rgb}{0.52,0.52,0.52}
\definecolor{gray53}{rgb}{0.53,0.53,0.53}
\definecolor{gray54}{rgb}{0.54,0.54,0.54}
\definecolor{gray55}{rgb}{0.55,0.55,0.55}
\definecolor{gray56}{rgb}{0.56,0.56,0.56}
\definecolor{gray57}{rgb}{0.57,0.57,0.57}
\definecolor{gray58}{rgb}{0.58,0.58,0.58}
\definecolor{gray59}{rgb}{0.59,0.59,0.59}
\definecolor{gray5}{rgb}{0.05,0.05,0.05}
\definecolor{gray60}{rgb}{0.60,0.60,0.60}
\definecolor{gray61}{rgb}{0.61,0.61,0.61}
\definecolor{gray62}{rgb}{0.62,0.62,0.62}
\definecolor{gray63}{rgb}{0.63,0.63,0.63}
\definecolor{gray64}{rgb}{0.64,0.64,0.64}
\definecolor{gray65}{rgb}{0.65,0.65,0.65}
\definecolor{gray66}{rgb}{0.66,0.66,0.66}
\definecolor{gray67}{rgb}{0.67,0.67,0.67}
\definecolor{gray68}{rgb}{0.68,0.68,0.68}
\definecolor{gray69}{rgb}{0.69,0.69,0.69}
\definecolor{gray6}{rgb}{0.06,0.06,0.06}
\definecolor{gray70}{rgb}{0.70,0.70,0.70}
\definecolor{gray71}{rgb}{0.71,0.71,0.71}
\definecolor{gray72}{rgb}{0.72,0.72,0.72}
\definecolor{gray73}{rgb}{0.73,0.73,0.73}
\definecolor{gray74}{rgb}{0.74,0.74,0.74}
\definecolor{gray75}{rgb}{0.75,0.75,0.75}
\definecolor{gray76}{rgb}{0.76,0.76,0.76}
\definecolor{gray77}{rgb}{0.77,0.77,0.77}
\definecolor{gray78}{rgb}{0.78,0.78,0.78}
\definecolor{gray79}{rgb}{0.79,0.79,0.79}
\definecolor{gray7}{rgb}{0.07,0.07,0.07}
\definecolor{gray80}{rgb}{0.80,0.80,0.80}
\definecolor{gray81}{rgb}{0.81,0.81,0.81}
\definecolor{gray82}{rgb}{0.82,0.82,0.82}
\definecolor{gray83}{rgb}{0.83,0.83,0.83}
\definecolor{gray84}{rgb}{0.84,0.84,0.84}
\definecolor{gray85}{rgb}{0.85,0.85,0.85}
\definecolor{gray86}{rgb}{0.86,0.86,0.86}
\definecolor{gray87}{rgb}{0.87,0.87,0.87}
\definecolor{gray88}{rgb}{0.88,0.88,0.88}
\definecolor{gray89}{rgb}{0.89,0.89,0.89}
\definecolor{gray8}{rgb}{0.08,0.08,0.08}
\definecolor{gray90}{rgb}{0.90,0.90,0.90}
\definecolor{gray91}{rgb}{0.91,0.91,0.91}
\definecolor{gray92}{rgb}{0.92,0.92,0.92}
\definecolor{gray93}{rgb}{0.93,0.93,0.93}
\definecolor{gray94}{rgb}{0.94,0.94,0.94}
\definecolor{gray95}{rgb}{0.95,0.95,0.95}
\definecolor{gray96}{rgb}{0.96,0.96,0.96}
\definecolor{gray97}{rgb}{0.97,0.97,0.97}
\definecolor{gray98}{rgb}{0.98,0.98,0.98}
\definecolor{gray99}{rgb}{0.99,0.99,0.99}
\definecolor{gray9}{rgb}{0.09,0.09,0.09}
\definecolor{gray}{rgb}{0.75,0.75,0.75}
\definecolor{green1}{rgb}{0.00,1.00,0.00}
\definecolor{green2}{rgb}{0.00,0.93,0.00}
\definecolor{green3}{rgb}{0.00,0.80,0.00}
\definecolor{green4}{rgb}{0.00,0.55,0.00}
\definecolor{greenyellow}{rgb}{0.68,1.00,0.18}
\definecolor{green}{rgb}{0.00,1.00,0.00}
\definecolor{grey0}{rgb}{0.00,0.00,0.00}
\definecolor{grey100}{rgb}{1.00,1.00,1.00}
\definecolor{grey10}{rgb}{0.10,0.10,0.10}
\definecolor{grey11}{rgb}{0.11,0.11,0.11}
\definecolor{grey12}{rgb}{0.12,0.12,0.12}
\definecolor{grey13}{rgb}{0.13,0.13,0.13}
\definecolor{grey14}{rgb}{0.14,0.14,0.14}
\definecolor{grey15}{rgb}{0.15,0.15,0.15}
\definecolor{grey16}{rgb}{0.16,0.16,0.16}
\definecolor{grey17}{rgb}{0.17,0.17,0.17}
\definecolor{grey18}{rgb}{0.18,0.18,0.18}
\definecolor{grey19}{rgb}{0.19,0.19,0.19}
\definecolor{grey1}{rgb}{0.01,0.01,0.01}
\definecolor{grey20}{rgb}{0.20,0.20,0.20}
\definecolor{grey21}{rgb}{0.21,0.21,0.21}
\definecolor{grey22}{rgb}{0.22,0.22,0.22}
\definecolor{grey23}{rgb}{0.23,0.23,0.23}
\definecolor{grey24}{rgb}{0.24,0.24,0.24}
\definecolor{grey25}{rgb}{0.25,0.25,0.25}
\definecolor{grey26}{rgb}{0.26,0.26,0.26}
\definecolor{grey27}{rgb}{0.27,0.27,0.27}
\definecolor{grey28}{rgb}{0.28,0.28,0.28}
\definecolor{grey29}{rgb}{0.29,0.29,0.29}
\definecolor{grey2}{rgb}{0.02,0.02,0.02}
\definecolor{grey30}{rgb}{0.30,0.30,0.30}
\definecolor{grey31}{rgb}{0.31,0.31,0.31}
\definecolor{grey32}{rgb}{0.32,0.32,0.32}
\definecolor{grey33}{rgb}{0.33,0.33,0.33}
\definecolor{grey34}{rgb}{0.34,0.34,0.34}
\definecolor{grey35}{rgb}{0.35,0.35,0.35}
\definecolor{grey36}{rgb}{0.36,0.36,0.36}
\definecolor{grey37}{rgb}{0.37,0.37,0.37}
\definecolor{grey38}{rgb}{0.38,0.38,0.38}
\definecolor{grey39}{rgb}{0.39,0.39,0.39}
\definecolor{grey3}{rgb}{0.03,0.03,0.03}
\definecolor{grey40}{rgb}{0.40,0.40,0.40}
\definecolor{grey41}{rgb}{0.41,0.41,0.41}
\definecolor{grey42}{rgb}{0.42,0.42,0.42}
\definecolor{grey43}{rgb}{0.43,0.43,0.43}
\definecolor{grey44}{rgb}{0.44,0.44,0.44}
\definecolor{grey45}{rgb}{0.45,0.45,0.45}
\definecolor{grey46}{rgb}{0.46,0.46,0.46}
\definecolor{grey47}{rgb}{0.47,0.47,0.47}
\definecolor{grey48}{rgb}{0.48,0.48,0.48}
\definecolor{grey49}{rgb}{0.49,0.49,0.49}
\definecolor{grey4}{rgb}{0.04,0.04,0.04}
\definecolor{grey50}{rgb}{0.50,0.50,0.50}
\definecolor{grey51}{rgb}{0.51,0.51,0.51}
\definecolor{grey52}{rgb}{0.52,0.52,0.52}
\definecolor{grey53}{rgb}{0.53,0.53,0.53}
\definecolor{grey54}{rgb}{0.54,0.54,0.54}
\definecolor{grey55}{rgb}{0.55,0.55,0.55}
\definecolor{grey56}{rgb}{0.56,0.56,0.56}
\definecolor{grey57}{rgb}{0.57,0.57,0.57}
\definecolor{grey58}{rgb}{0.58,0.58,0.58}
\definecolor{grey59}{rgb}{0.59,0.59,0.59}
\definecolor{grey5}{rgb}{0.05,0.05,0.05}
\definecolor{grey60}{rgb}{0.60,0.60,0.60}
\definecolor{grey61}{rgb}{0.61,0.61,0.61}
\definecolor{grey62}{rgb}{0.62,0.62,0.62}
\definecolor{grey63}{rgb}{0.63,0.63,0.63}
\definecolor{grey64}{rgb}{0.64,0.64,0.64}
\definecolor{grey65}{rgb}{0.65,0.65,0.65}
\definecolor{grey66}{rgb}{0.66,0.66,0.66}
\definecolor{grey67}{rgb}{0.67,0.67,0.67}
\definecolor{grey68}{rgb}{0.68,0.68,0.68}
\definecolor{grey69}{rgb}{0.69,0.69,0.69}
\definecolor{grey6}{rgb}{0.06,0.06,0.06}
\definecolor{grey70}{rgb}{0.70,0.70,0.70}
\definecolor{grey71}{rgb}{0.71,0.71,0.71}
\definecolor{grey72}{rgb}{0.72,0.72,0.72}
\definecolor{grey73}{rgb}{0.73,0.73,0.73}
\definecolor{grey74}{rgb}{0.74,0.74,0.74}
\definecolor{grey75}{rgb}{0.75,0.75,0.75}
\definecolor{grey76}{rgb}{0.76,0.76,0.76}
\definecolor{grey77}{rgb}{0.77,0.77,0.77}
\definecolor{grey78}{rgb}{0.78,0.78,0.78}
\definecolor{grey79}{rgb}{0.79,0.79,0.79}
\definecolor{grey7}{rgb}{0.07,0.07,0.07}
\definecolor{grey80}{rgb}{0.80,0.80,0.80}
\definecolor{grey81}{rgb}{0.81,0.81,0.81}
\definecolor{grey82}{rgb}{0.82,0.82,0.82}
\definecolor{grey83}{rgb}{0.83,0.83,0.83}
\definecolor{grey84}{rgb}{0.84,0.84,0.84}
\definecolor{grey85}{rgb}{0.85,0.85,0.85}
\definecolor{grey86}{rgb}{0.86,0.86,0.86}
\definecolor{grey87}{rgb}{0.87,0.87,0.87}
\definecolor{grey88}{rgb}{0.88,0.88,0.88}
\definecolor{grey89}{rgb}{0.89,0.89,0.89}
\definecolor{grey8}{rgb}{0.08,0.08,0.08}
\definecolor{grey90}{rgb}{0.90,0.90,0.90}
\definecolor{grey91}{rgb}{0.91,0.91,0.91}
\definecolor{grey92}{rgb}{0.92,0.92,0.92}
\definecolor{grey93}{rgb}{0.93,0.93,0.93}
\definecolor{grey94}{rgb}{0.94,0.94,0.94}
\definecolor{grey95}{rgb}{0.95,0.95,0.95}
\definecolor{grey96}{rgb}{0.96,0.96,0.96}
\definecolor{grey97}{rgb}{0.97,0.97,0.97}
\definecolor{grey98}{rgb}{0.98,0.98,0.98}
\definecolor{grey99}{rgb}{0.99,0.99,0.99}
\definecolor{grey9}{rgb}{0.09,0.09,0.09}
\definecolor{grey}{rgb}{0.75,0.75,0.75}
\definecolor{honeydew1}{rgb}{0.94,1.00,0.94}
\definecolor{honeydew2}{rgb}{0.88,0.93,0.88}
\definecolor{honeydew3}{rgb}{0.76,0.80,0.76}
\definecolor{honeydew4}{rgb}{0.51,0.55,0.51}
\definecolor{honeydew}{rgb}{0.94,1.00,0.94}
\definecolor{hotpink}{rgb}{1.00,0.41,0.71}
\definecolor{indianred}{rgb}{0.80,0.36,0.36}
\definecolor{ivory1}{rgb}{1.00,1.00,0.94}
\definecolor{ivory2}{rgb}{0.93,0.93,0.88}
\definecolor{ivory3}{rgb}{0.80,0.80,0.76}
\definecolor{ivory4}{rgb}{0.55,0.55,0.51}
\definecolor{ivory}{rgb}{1.00,1.00,0.94}
\definecolor{khaki1}{rgb}{1.00,0.96,0.56}
\definecolor{khaki2}{rgb}{0.93,0.90,0.52}
\definecolor{khaki3}{rgb}{0.80,0.78,0.45}
\definecolor{khaki4}{rgb}{0.55,0.53,0.31}
\definecolor{khaki}{rgb}{0.94,0.90,0.55}
\definecolor{lavenderblush}{rgb}{1.00,0.94,0.96}
\definecolor{lavender}{rgb}{0.90,0.90,0.98}
\definecolor{lawngreen}{rgb}{0.49,0.99,0.00}
\definecolor{lemonchiffon}{rgb}{1.00,0.98,0.80}
\definecolor{lightblue}{rgb}{0.68,0.85,0.90}
\definecolor{lightcoral}{rgb}{0.94,0.50,0.50}
\definecolor{lightcyan}{rgb}{0.88,1.00,1.00}
\definecolor{lightgoldenrod}{rgb}{0.93,0.87,0.51}
\definecolor{lightgoldenrod}{rgb}{0.98,0.98,0.82}
\definecolor{lightgray}{rgb}{0.83,0.83,0.83}
\definecolor{lightgreen}{rgb}{0.56,0.93,0.56}
\definecolor{lightgrey}{rgb}{0.83,0.83,0.83}
\definecolor{lightpink}{rgb}{1.00,0.71,0.76}
\definecolor{lightsalmon}{rgb}{1.00,0.63,0.48}
\definecolor{lightsea}{rgb}{0.13,0.70,0.67}
\definecolor{lightsky}{rgb}{0.53,0.81,0.98}
\definecolor{lightslate}{rgb}{0.47,0.53,0.60}
\definecolor{lightslate}{rgb}{0.47,0.53,0.60}
\definecolor{lightslate}{rgb}{0.52,0.44,1.00}
\definecolor{lightsteel}{rgb}{0.69,0.77,0.87}
\definecolor{lightyellow}{rgb}{1.00,1.00,0.88}
\definecolor{limegreen}{rgb}{0.20,0.80,0.20}
\definecolor{linen}{rgb}{0.98,0.94,0.90}
\definecolor{magenta1}{rgb}{1.00,0.00,1.00}
\definecolor{magenta2}{rgb}{0.93,0.00,0.93}
\definecolor{magenta3}{rgb}{0.80,0.00,0.80}
\definecolor{magenta4}{rgb}{0.55,0.00,0.55}
\definecolor{magenta}{rgb}{1.00,0.00,1.00}
\definecolor{maroon1}{rgb}{1.00,0.20,0.70}
\definecolor{maroon2}{rgb}{0.93,0.19,0.65}
\definecolor{maroon3}{rgb}{0.80,0.16,0.56}
\definecolor{maroon4}{rgb}{0.55,0.11,0.38}
\definecolor{maroon}{rgb}{0.69,0.19,0.38}
\definecolor{mediumaquamarine}{rgb}{0.40,0.80,0.67}
\definecolor{mediumblue}{rgb}{0.00,0.00,0.80}
\definecolor{mediumorchid}{rgb}{0.73,0.33,0.83}
\definecolor{mediumpurple}{rgb}{0.58,0.44,0.86}
\definecolor{mediumsea}{rgb}{0.24,0.70,0.44}
\definecolor{mediumslate}{rgb}{0.48,0.41,0.93}
\definecolor{mediumspring}{rgb}{0.00,0.98,0.60}
\definecolor{mediumturquoise}{rgb}{0.28,0.82,0.80}
\definecolor{mediumviolet}{rgb}{0.78,0.08,0.52}
\definecolor{midnightblue}{rgb}{0.10,0.10,0.44}
\definecolor{mintcream}{rgb}{0.96,1.00,0.98}
\definecolor{mistyrose}{rgb}{1.00,0.89,0.88}
\definecolor{moccasin}{rgb}{1.00,0.89,0.71}
\definecolor{navajowhite}{rgb}{1.00,0.87,0.68}
\definecolor{navyblue}{rgb}{0.00,0.00,0.50}
\definecolor{navy}{rgb}{0.00,0.00,0.50}
\definecolor{oldlace}{rgb}{0.99,0.96,0.90}
\definecolor{olivedrab}{rgb}{0.42,0.56,0.14}
\definecolor{orange1}{rgb}{1.00,0.65,0.00}
\definecolor{orange2}{rgb}{0.93,0.60,0.00}
\definecolor{orange3}{rgb}{0.80,0.52,0.00}
\definecolor{orange4}{rgb}{0.55,0.35,0.00}
\definecolor{orangered}{rgb}{1.00,0.27,0.00}
\definecolor{orange}{rgb}{1.00,0.65,0.00}
\definecolor{orchid1}{rgb}{1.00,0.51,0.98}
\definecolor{orchid2}{rgb}{0.93,0.48,0.91}
\definecolor{orchid3}{rgb}{0.80,0.41,0.79}
\definecolor{orchid4}{rgb}{0.55,0.28,0.54}
\definecolor{orchid}{rgb}{0.85,0.44,0.84}
\definecolor{palegoldenrod}{rgb}{0.93,0.91,0.67}
\definecolor{palegreen}{rgb}{0.60,0.98,0.60}
\definecolor{paleturquoise}{rgb}{0.69,0.93,0.93}
\definecolor{paleviolet}{rgb}{0.86,0.44,0.58}
\definecolor{papayawhip}{rgb}{1.00,0.94,0.84}
\definecolor{peachpuff}{rgb}{1.00,0.85,0.73}
\definecolor{peru}{rgb}{0.80,0.52,0.25}
\definecolor{pink1}{rgb}{1.00,0.71,0.77}
\definecolor{pink2}{rgb}{0.93,0.66,0.72}
\definecolor{pink3}{rgb}{0.80,0.57,0.62}
\definecolor{pink4}{rgb}{0.55,0.39,0.42}
\definecolor{pink}{rgb}{1.00,0.75,0.80}
\definecolor{plum1}{rgb}{1.00,0.73,1.00}
\definecolor{plum2}{rgb}{0.93,0.68,0.93}
\definecolor{plum3}{rgb}{0.80,0.59,0.80}
\definecolor{plum4}{rgb}{0.55,0.40,0.55}
\definecolor{plum}{rgb}{0.87,0.63,0.87}
\definecolor{powderblue}{rgb}{0.69,0.88,0.90}
\definecolor{purple1}{rgb}{0.61,0.19,1.00}
\definecolor{purple2}{rgb}{0.57,0.17,0.93}
\definecolor{purple3}{rgb}{0.49,0.15,0.80}
\definecolor{purple4}{rgb}{0.33,0.10,0.55}
\definecolor{purple}{rgb}{0.63,0.13,0.94}
\definecolor{red1}{rgb}{1.00,0.00,0.00}
\definecolor{red2}{rgb}{0.93,0.00,0.00}
\definecolor{red3}{rgb}{0.80,0.00,0.00}
\definecolor{red4}{rgb}{0.55,0.00,0.00}
\definecolor{red}{rgb}{1.00,0.00,0.00}
\definecolor{rosybrown}{rgb}{0.74,0.56,0.56}
\definecolor{royalblue}{rgb}{0.25,0.41,0.88}
\definecolor{saddlebrown}{rgb}{0.55,0.27,0.07}
\definecolor{salmon1}{rgb}{1.00,0.55,0.41}
\definecolor{salmon2}{rgb}{0.93,0.51,0.38}
\definecolor{salmon3}{rgb}{0.80,0.44,0.33}
\definecolor{salmon4}{rgb}{0.55,0.30,0.22}
\definecolor{salmon}{rgb}{0.98,0.50,0.45}
\definecolor{sandybrown}{rgb}{0.96,0.64,0.38}
\definecolor{seagreen}{rgb}{0.18,0.55,0.34}
\definecolor{seashell1}{rgb}{1.00,0.96,0.93}
\definecolor{seashell2}{rgb}{0.93,0.90,0.87}
\definecolor{seashell3}{rgb}{0.80,0.77,0.75}
\definecolor{seashell4}{rgb}{0.55,0.53,0.51}
\definecolor{seashell}{rgb}{1.00,0.96,0.93}
\definecolor{sienna1}{rgb}{1.00,0.51,0.28}
\definecolor{sienna2}{rgb}{0.93,0.47,0.26}
\definecolor{sienna3}{rgb}{0.80,0.41,0.22}
\definecolor{sienna4}{rgb}{0.55,0.28,0.15}
\definecolor{sienna}{rgb}{0.63,0.32,0.18}
\definecolor{skyblue}{rgb}{0.53,0.81,0.92}
\definecolor{slateblue}{rgb}{0.42,0.35,0.80}
\definecolor{slategray}{rgb}{0.44,0.50,0.56}
\definecolor{slategrey}{rgb}{0.44,0.50,0.56}
\definecolor{snow1}{rgb}{1.00,0.98,0.98}
\definecolor{snow2}{rgb}{0.93,0.91,0.91}
\definecolor{snow3}{rgb}{0.80,0.79,0.79}
\definecolor{snow4}{rgb}{0.55,0.54,0.54}
\definecolor{snow}{rgb}{1.00,0.98,0.98}
\definecolor{springgreen}{rgb}{0.00,1.00,0.50}
\definecolor{steelblue}{rgb}{0.27,0.51,0.71}
\definecolor{tan1}{rgb}{1.00,0.65,0.31}
\definecolor{tan2}{rgb}{0.93,0.60,0.29}
\definecolor{tan3}{rgb}{0.80,0.52,0.25}
\definecolor{tan4}{rgb}{0.55,0.35,0.17}
\definecolor{tan}{rgb}{0.82,0.71,0.55}
\definecolor{thistle1}{rgb}{1.00,0.88,1.00}
\definecolor{thistle2}{rgb}{0.93,0.82,0.93}
\definecolor{thistle3}{rgb}{0.80,0.71,0.80}
\definecolor{thistle4}{rgb}{0.55,0.48,0.55}
\definecolor{thistle}{rgb}{0.85,0.75,0.85}
\definecolor{tomato1}{rgb}{1.00,0.39,0.28}
\definecolor{tomato2}{rgb}{0.93,0.36,0.26}
\definecolor{tomato3}{rgb}{0.80,0.31,0.22}
\definecolor{tomato4}{rgb}{0.55,0.21,0.15}
\definecolor{tomato}{rgb}{1.00,0.39,0.28}
\definecolor{turquoise1}{rgb}{0.00,0.96,1.00}
\definecolor{turquoise2}{rgb}{0.00,0.90,0.93}
\definecolor{turquoise3}{rgb}{0.00,0.77,0.80}
\definecolor{turquoise4}{rgb}{0.00,0.53,0.55}
\definecolor{turquoise}{rgb}{0.25,0.88,0.82}
\definecolor{violetred}{rgb}{0.82,0.13,0.56}
\definecolor{violet}{rgb}{0.93,0.51,0.93}
\definecolor{wheat1}{rgb}{1.00,0.91,0.73}
\definecolor{wheat2}{rgb}{0.93,0.85,0.68}
\definecolor{wheat3}{rgb}{0.80,0.73,0.59}
\definecolor{wheat4}{rgb}{0.55,0.49,0.40}
\definecolor{wheat}{rgb}{0.96,0.87,0.70}
\definecolor{whitesmoke}{rgb}{0.96,0.96,0.96}
\definecolor{white}{rgb}{1.00,1.00,1.00}
\definecolor{yellow1}{rgb}{1.00,1.00,0.00}
\definecolor{yellow2}{rgb}{0.93,0.93,0.00}
\definecolor{yellow3}{rgb}{0.80,0.80,0.00}
\definecolor{yellow4}{rgb}{0.55,0.55,0.00}
\definecolor{yellowgreen}{rgb}{0.60,0.80,0.20}
\definecolor{yellow}{rgb}{1.00,1.00,0.00}

\usepackage{txfonts}
\usepackage[breaklinks, colorlinks, citecolor=blue]{hyperref}
\usepackage{natbib}
\usepackage{amssymb}
\usepackage{bm}

\begin{document}

\author{\small
Planck and AMI Collaborations:
N.~Aghanim\inst{55}
\and
M.~Arnaud\inst{70}
\and
M.~Ashdown\inst{67, 5}
\and
J.~Aumont\inst{55}
\and
C.~Baccigalupi\inst{79}
\and
A.~Balbi\inst{35}
\and
A.~J.~Banday\inst{88, 8}
\and
R.~B.~Barreiro\inst{63}
\and
E.~Battaner\inst{90}
\and
R.~Battye\inst{66}
% ...
}
\institute{\small
APC, AstroParticule et Cosmologie, Universit\'{e} Paris Diderot, CNRS/IN2P3, CEA/lrfu, Observatoire de Paris, Sorbonne Paris Cit\'{e}, 10, rue Alice Domon et L\'{e}onie Duquet, 75205 Paris Cedex 13, France\\
\and
Aalto University Mets\"{a}hovi Radio Observatory, Mets\"{a}hovintie 114, FIN-02540 Kylm\"{a}l\"{a}, Finland\\
\and
Academy of Sciences of Tatarstan, Bauman Str., 20, Kazan, 420111, Republic of Tatarstan, Russia\\
\and
Agenzia Spaziale Italiana Science Data Center, c/o ESRIN, via Galileo Galilei, Frascati, Italy\\
\and
Astrophysics Group, Cavendish Laboratory, University of Cambridge, J J Thomson Avenue, Cambridge CB3 0HE, U.K.\\
\and
Atacama Large Millimeter/submillimeter Array, ALMA Santiago Central Offices, Alonso de Cordova 3107, Vitacura, Casilla 763 0355, Santiago, Chile\\
\and
CITA, University of Toronto, 60 St. George St., Toronto, ON M5S 3H8, Canada\\
\and
CNRS, IRAP, 9 Av. colonel Roche, BP 44346, F-31028 Toulouse cedex 4, France\\
\and
California Institute of Technology, Pasadena, California, U.S.A.\\
\and
Centre of Mathematics for Applications, University of Oslo, Blindern, Oslo, Norway\\
\and
Centro de Astrof\'{\i}sica, Universidade do Porto, Rua das Estrelas, 4150-762 Porto, Portugal\\
\and
Centro de Estudios de F\'{i}sica del Cosmos de Arag\'{o}n (CEFCA), Plaza San Juan, 1, planta 2, E-44001, Teruel, Spain\\
\and
Computational Cosmology Center, Lawrence Berkeley National Laboratory, Berkeley, California, U.S.A.\\
\and
Consejo Superior de Investigaciones Cient\'{\i}ficas (CSIC), Madrid, Spain\\
\and
DSM/Irfu/SPP, CEA-Saclay, F-91191 Gif-sur-Yvette Cedex, France\\
\and
DTU Space, National Space Institute, Juliane Mariesvej 30, Copenhagen, Denmark\\
\and
D\'{e}partement de Physique Th\'{e}orique, Universit\'{e} de Gen\`{e}ve, 24, Quai E. Ansermet,1211 Gen\`{e}ve 4, Switzerland\\
\and
Departamento de F\'{\i}sica, Universidad de Oviedo, Avda. Calvo Sotelo s/n, Oviedo, Spain\\
\and
Department of Astronomy and Astrophysics, University of Toronto, 50 Saint George Street, Toronto, Ontario, Canada\\
\and
Department of Astronomy and Geodesy, Kazan Federal University,  Kremlevskaya Str., 18, Kazan, 420008, Russia\\
\and
Department of Astrophysics, IMAPP, Radboud University, P.O. Box 9010, 6500 GL Nijmegen,  The Netherlands\\
\and
Department of Physics \& Astronomy, University of British Columbia, 6224 Agricultural Road, Vancouver, British Columbia, Canada\\
\and
Department of Physics and Astronomy, University of Southern California, Los Angeles, California, U.S.A.\\
\and
Department of Physics and Astronomy, University of Sussex, Brighton BN1 9QH, U.K.\\
\and
Department of Physics, Gustaf H\"{a}llstr\"{o}min katu 2a, University of Helsinki, Helsinki, Finland\\
\and
Department of Physics, Purdue University, 525 Northwestern Avenue, West Lafayette, Indiana, U.S.A.\\
\and
Department of Physics, University of California, Berkeley, California, U.S.A.\\
\and
Department of Physics, University of California, Santa Barbara, California, U.S.A.\\
\and
Department of Physics, University of Illinois at Urbana-Champaign, 1110 West Green Street, Urbana, Illinois, U.S.A.\\
\and
Dipartimento di Fisica e Astronomia G. Galilei, Universit\`{a} degli Studi di Padova, via Marzolo 8, 35131 Padova, Italy\\
\and
Dipartimento di Fisica, Universit\`{a} La Sapienza, P. le A. Moro 2, Roma, Italy\\
\and
Dipartimento di Fisica, Universit\`{a} degli Studi di Milano, Via Celoria, 16, Milano, Italy\\
\and
Dipartimento di Fisica, Universit\`{a} degli Studi di Trieste, via A. Valerio 2, Trieste, Italy\\
\and
Dipartimento di Fisica, Universit\`{a} di Ferrara, Via Saragat 1, 44122 Ferrara, Italy\\
\and
Dipartimento di Fisica, Universit\`{a} di Roma Tor Vergata, Via della Ricerca Scientifica, 1, Roma, Italy\\
\and
Dipartimento di Matematica, Universit\`{a} di Roma Tor Vergata, Via della Ricerca Scientifica, 1, Roma, Italy\\
\and
Discovery Center, Niels Bohr Institute, Blegdamsvej 17, Copenhagen, Denmark\\
\and
Dpto. Astrof\'{i}sica, Universidad de La Laguna (ULL), E-38206 La Laguna, Tenerife, Spain\\
\and
European Southern Observatory, ESO Vitacura, Alonso de Cordova 3107, Vitacura, Casilla 19001, Santiago, Chile\\
\and
European Space Agency, ESAC, Planck Science Office, Camino bajo del Castillo, s/n, Urbanizaci\'{o}n Villafranca del Castillo, Villanueva de la Ca\~{n}ada, Madrid, Spain\\
\and
European Space Agency, ESTEC, Keplerlaan 1, 2201 AZ Noordwijk, The Netherlands\\
\and
Helsinki Institute of Physics, Gustaf H\"{a}llstr\"{o}min katu 2, University of Helsinki, Helsinki, Finland\\
\and
INAF - Osservatorio Astronomico di Padova, Vicolo dell'Osservatorio 5, Padova, Italy\\
\and
INAF - Osservatorio Astronomico di Roma, via di Frascati 33, Monte Porzio Catone, Italy\\
\and
INAF - Osservatorio Astronomico di Trieste, Via G.B. Tiepolo 11, Trieste, Italy\\
\and
INAF Istituto di Radioastronomia, Via P. Gobetti 101, 40129 Bologna, Italy\\
\and
INAF/IASF Bologna, Via Gobetti 101, Bologna, Italy\\
\and
INAF/IASF Milano, Via E. Bassini 15, Milano, Italy\\
\and
INFN, Sezione di Roma 1, Universit`{a} di Roma Sapienza, Piazzale Aldo Moro 2, 00185, Roma, Italy\\
\and
INRIA, Laboratoire de Recherche en Informatique, Universit\'{e} Paris-Sud 11, B\^{a}timent 490, 91405 Orsay Cedex, France\\
\and
IUCAA, Post Bag 4, Ganeshkhind, Pune University Campus, Pune 411 007, India\\
\and
Imperial College London, Astrophysics group, Blackett Laboratory, Prince Consort Road, London, SW7 2AZ, U.K.\\
\and
Infrared Processing and Analysis Center, California Institute of Technology, Pasadena, CA 91125, U.S.A.\\
\and
Institut N\'{e}el, CNRS, Universit\'{e} Joseph Fourier Grenoble I, 25 rue des Martyrs, Grenoble, France\\
\and
Institut d'Astrophysique Spatiale, CNRS (UMR8617) Universit\'{e} Paris-Sud 11, B\^{a}timent 121, Orsay, France\\
\and
Institut d'Astrophysique de Paris, CNRS (UMR7095), 98 bis Boulevard Arago, F-75014, Paris, France\\
\and
Institut de Ci\`{e}ncies de l'Espai, CSIC/IEEC, Facultat de Ci\`{e}ncies, Campus UAB, Torre C5 par-2, Bellaterra 08193, Spain\\
\and
Institute for Space Sciences, Bucharest-Magurale, Romania\\
\and
Institute of Astronomy and Astrophysics, Academia Sinica, Taipei, Taiwan\\
\and
Institute of Astronomy, University of Cambridge, Madingley Road, Cambridge CB3 0HA, U.K.\\
\and
Institute of Theoretical Astrophysics, University of Oslo, Blindern, Oslo, Norway\\
\and
Instituto de Astrof\'{\i}sica de Canarias, C/V\'{\i}a L\'{a}ctea s/n, La Laguna, Tenerife, Spain\\
\and
Instituto de F\'{\i}sica de Cantabria (CSIC-Universidad de Cantabria), Avda. de los Castros s/n, Santander, Spain\\
\and
Istituto di Fisica del Plasma, CNR-ENEA-EURATOM Association, Via R. Cozzi 53, Milano, Italy\\
\and
Jet Propulsion Laboratory, California Institute of Technology, 4800 Oak Grove Drive, Pasadena, California, U.S.A.\\
\and
Jodrell Bank Centre for Astrophysics, Alan Turing Building, School of Physics and Astronomy, The University of Manchester, Oxford Road, Manchester, M13 9PL, U.K.\\
\and
Kavli Institute for Cosmology Cambridge, Madingley Road, Cambridge, CB3 0HA, U.K.\\
\and
LAL, Universit\'{e} Paris-Sud, CNRS/IN2P3, Orsay, France\\
\and
LERMA, CNRS, Observatoire de Paris, 61 Avenue de l'Observatoire, Paris, France\\
\and
Laboratoire AIM, IRFU/Service d'Astrophysique - CEA/DSM - CNRS - Universit\'{e} Paris Diderot, B\^{a}t. 709, CEA-Saclay, F-91191 Gif-sur-Yvette Cedex, France\\
\and
Laboratoire de Physique Subatomique et de Cosmologie, Universit\'{e} Joseph Fourier Grenoble I, CNRS/IN2P3, Institut National Polytechnique de Grenoble, 53 rue des Martyrs, 38026 Grenoble cedex, France\\
\and
Laboratoire de Physique Th\'{e}orique, Universit\'{e} Paris-Sud 11 \& CNRS, B\^{a}timent 210, 91405 Orsay, France\\
\and
Lawrence Berkeley National Laboratory, Berkeley, California, U.S.A.\\
\and
Max-Planck-Institut f\"{u}r Astrophysik, Karl-Schwarzschild-Str. 1, 85741 Garching, Germany\\
\and
Max-Planck-Institut f\"{u}r Extraterrestrische Physik, Giessenbachstra{\ss}e, 85748 Garching, Germany\\
\and
Niels Bohr Institute, Blegdamsvej 17, Copenhagen, Denmark\\
\and
Observational Cosmology, Mail Stop 367-17, California Institute of Technology, Pasadena, CA, 91125, U.S.A.\\
\and
Optical Science Laboratory, University College London, Gower Street, London, U.K.\\
\and
SISSA, Astrophysics Sector, via Bonomea 265, 34136, Trieste, Italy\\
\and
School of Physics and Astronomy, Cardiff University, Queens Buildings, The Parade, Cardiff, CF24 3AA, U.K.\\
\and
Space Research Institute (IKI), Profsoyuznaya 84/32, Moscow, Russia\\
\and
Space Research Institute (IKI), Russian Academy of Sciences, Profsoyuznaya Str, 84/32, Moscow, 117997, Russia\\
\and
Space Sciences Laboratory, University of California, Berkeley, California, U.S.A.\\
\and
Special Astrophysical Observatory, Russian Academy of Sciences, Nizhnij Arkhyz, Zelenchukskiy region, Karachai-Cherkessian Republic, 369167, Russia\\
\and
Stanford University, Dept of Physics, Varian Physics Bldg, 382 Via Pueblo Mall, Stanford, California, U.S.A.\\
\and
T\"{U}B\.{I}TAK National Observatory, Akdeniz University Campus, 07058, Antalya, Turkey\\
\and
UPMC Univ Paris 06, UMR7095, 98 bis Boulevard Arago, F-75014, Paris, France\\
\and
Universit\'{e} de Toulouse, UPS-OMP, IRAP, F-31028 Toulouse cedex 4, France\\
\and
University Observatory, Ludwig Maximilian University of Munich, Scheinerstrasse 1, 81679 Munich, Germany\\
\and
University of Granada, Departamento de F\'{\i}sica Te\'{o}rica y del Cosmos, Facultad de Ciencias, Granada, Spain\\
% ...
}

\title{\textit{Planck} Intermediate Results II: Comparison of
  Sunyaev--Zeldovich measurements from \textit{Planck} and from the
  Arcminute Microkelvin Imager for 11 galaxy clusters} 

\date{Received 2 April 2012; Accepted } 

\abstract{A comparison is presented of Sunyaev--Zeldovich measurements
  for 11 galaxy clusters as obtained by \Planck\ and by the
  ground-based interferometer, the Arcminute Microkelvin
  Imager. Assuming a universal spherically-symmetric Generalised
  Navarro, Frenk \& White (GNFW) model for the cluster gas pressure
  profile, we jointly constrain the integrated Compton-$Y$ parameter
  ($Y_{500}$) and the scale radius ($\theta_{500}$) of each
  cluster. Our resulting constraints in the $Y_{500}-\theta_{500}$ 2D
  parameter space derived from the two instruments overlap
  significantly for eight of the clusters, although, overall, there is
  a tendency for AMI to find the Sunyaev--Zeldovich signal to be
  smaller in angular size and fainter than \Planck. Significant
  discrepancies exist for the three remaining clusters in the sample,
  namely A1413, A1914, and the newly-discovered \Planck\ cluster
  PLCKESZ G139.59+24.18. The robustness of the analysis of both the
  \Planck\ and AMI data is demonstrated through the use of detailed
  simulations, which also discount confusion from residual point
  (radio) sources and from diffuse astrophysical foregrounds as
  possible explanations for the discrepancies found. For a subset of
  our cluster sample, we have investigated the dependence of our
  results on the assumed pressure profile by repeating the analysis
  adopting the best-fitting GNFW profile shape which best matches
  X-ray observations.  Adopting the best-fitting profile shape from
  the X-ray data does not, in general, resolve the discrepancies found
  in this subset of five clusters. Though based on a small sample, our
  results suggest that the adopted GNFW model may not be sufficiently
  flexible to describe clusters universally.}

\keywords{Cosmology: observations $-$ Galaxies: cluster: general $-$
  Galaxies: clusters: intracluster medium $-$ Cosmic background
  radiation, X-rays: galaxies: clusters}

\authorrunning{Planck and AMI Collaborations} 

\titlerunning{Comparison of \Planck\ and AMI Sunyaev--Zeldovich
  measurements for 11 galaxy clusters} \maketitle
 

\section{Introduction}
Clusters of galaxies are the most massive gravitationally bound
objects in the Universe and as such are critical tracers of the
formation of large-scale structure. The size and formation history of
massive clusters is such that the ratio of cluster gas mass to total
mass is expected to be representative of the universal ratio, once the
relatively small amount of baryonic matter in the cluster galaxies is
taken into account (e.g., \citealt{white93}). Moreover, the
comoving number density of clusters as a function of mass and redshift
is expected to be particularly sensitive to the cosmological
parameters $\sigma_8$ and $\Omega_{\rm m}$ (e.g., \citealt{battye03}).

The Sunyaev--Zeldovich (SZ) effect (see \citealt{birkinshaw99};
\citealt{carlstrom02} for reviews) produces secondary anisotropies in
the cosmic microwave background (CMB) radiation through
inverse-Compton scattering from the electrons in the hot intracluster
gas (which also radiates via thermal Bremsstrahlung in the X-ray
waveband) and the transfer of some of the energy of the electrons to
the low-energy photons. Moreover, the surface brightness of an SZ
signal does not depend on the redshift $z$ of the cluster. Hence an
SZ-effect flux-density-limited survey can provide a complete catalogue
of galaxy clusters above a limiting mass (e.g., \citealt{bartlett94},
\citealt{kneissl01}, \citealt{kosowsky03}, \citealt{ruhl04}).

Analyses of observations of galaxy clusters via their SZ effect, X-ray 
emission or gravitational lensing are often based on some
spherically-symmetric cluster model in which one assumes parameterised
functional forms for the radial distribution of some cluster
properties, such as electron density and temperature
\citep{sanderson03, vikhlinin05, vikhlinin06, laroque06, feroz09b,
  zwart11, rodriguez-gonzalvez11, hurley-walker11, shimwell10},
electron pressure and density \citep{nagai07, mroczkowski09, arnaud10,
plagge10, planck2011-5.1a}, or electron pressure and entropy
\citep{olamaie10, allison11}.

The motivation for this paper is to augment SZ measurements obtained
with \Planck\ \footnote{\Planck\ (\url{http://www.esa.int/Planck}) is a
  project of the European Space Agency (ESA) with instruments provided
by two scientific consortia funded by ESA member states (in
particular the lead countries France and Italy), with contributions
form NASA (USA) and telescope reflectors provided by a collaboration
between ESA and a scientific consortium led and funded by Denmark.} 
for a sample of 11 galaxy clusters with refined
higher-resolution SZ measurements obtained with the Arcminute
Microkelvin Imager (AMI) interferometer. Such a combination is an
interesting and potentially very powerful way to pin down the gas
pressure profile of individual galaxy clusters as it relies on a
single well-understood astrophysical effect. In addition, \Planck\ and
AMI SZ measurements exploit very different aspects of the SZ
signature: \Planck\ effectively uses its wide frequency coverage to
identify the characteristic frequency spectrum of the SZ effect while
AMI exploits its higher angular resolution to perform spatial
filtering to identify SZ clusters and constrain their parameters.
Combining measurements by these two instruments not only provides a
powerful consistency check on both sets of observations, but may also
break, or at least reduce, the observed parameter degeneracy between
the derived SZ Compton-$Y$ parameter and the cluster angular size
which often results due to the finite resolution of SZ telescopes
\citep{planck2011-5.1a}.

The paper is organised as follows. In Section~\ref
{sec:cluster_sample}, we outline how we selected our sample of 11
galaxy clusters for this comparison work. In
Sections~\ref{sec:planckdata} and \ref{sec:amidata}, we describe the
\Planck\ and AMI observations of our cluster sample, respectively.  In
Section~\ref{sec:SZsignal} we present the pressure profile that we
have used to model the clusters and constrain parameters. The analysis
of the real data from both experiments is also described in
Section~\ref{sec:SZsignal}. We follow this by validating our analysis
methodology and investigating the effects of diffuse foreground
emission on the \Planck\ constraints with simulations in
Section~\ref{sec:planck_sims}. Section~\ref{sec:ami_sims} presents a
similar simulations-based investigation of the effects of residual
point sources and analysis methodology on the constraints derived from
the AMI interferometric data. With a view to explaining some of the
discrepancies we observe, in Section~\ref{sec:implications}, we
investigate the possibility of relaxing the assumptions regarding the
universal pressure profile adopted for our cluster sample. Here we
also examine the consistency of the \Planck\ and AMI SZ results with
complementary constraints from high-quality X-ray observations for a
subset of our cluster sample. We conclude with a discussion in
Section~\ref{sec:discussion}.


\section{Selection of the cluster sample}
\label{sec:cluster_sample}
An original sample of $26$ clusters was defined at the beginning of
this study. $24$ of the clusters were identified as members of the
sample by virtue of the fact that they were both present in the
\Planck\ Early Sunyaev--Zeldovich (ESZ) cluster catalogue
\citep{planck2011-5.1a}, and had also already been observed and
detected with AMI during the course of its normal observing
programme. Note that these $24$ clusters had been observed by AMI as
part of differing scientific programmes and while each programme had a
well-defined sample, the resulting set of clusters used in this paper
does not constitute a well-defined or complete sample.

To this sample of $24$, two newly-discovered \Planck\ clusters were
added, for which AMI made follow-up observations. The complete list of
the original cluster sample, their coordinates and redshifts is
presented in Table~\ref{tab:clsample0}. The sample was then screened
to include only clusters that had (i) a firm SZ detection by AMI
($\geq 3\sigma$ before source subtraction and $\geq 5\sigma$ after
source subtraction) and (ii) a benign environment in terms of radio
point sources (we discard clusters with total integrated source flux
densities greater than $5 \, \rm{mJy}$ within a radius of $3\arcm$ or
greater than $10 \, \rm{mJy}$ within a radius of $10\arcmin$ from the
phase centre). This reduced the sample to $11$ clusters spanning a
wide range in redshift, $0.11 < z < 0.55$ . The clusters in the new
sample are A2034, A1413, A990, A2409, A1914, A2218, A773, MACS
J1149+2223, RXJ0748+5941, PLCKESZ G139.59+24.18, and PLCKESZ
G121.11+57.01. The sample includes two cool-core clusters, A1413 and
A2034 \citep{pratt02, kempner03, vikhlinin05, govoni09} and two
newly-discovered \Planck\ clusters, PLCKESZ G139.59+24.18 and PLCKESZ
G121.11+57.01 \citep{planck2011-5.1a}. The last two have been
observered in the optical with the RTT-150
telescope\footnote{\url{http://hea.iki.rssi.ru/rtt150/en/}} as part of 
the \Planck\ follow-up programme. The resulting spectroscopic
redshifts measured for the brightest cluster galaxies within these two
clusters are given in Table~\ref{tab:clsample0}.

\begin{table*}
\centering

\caption{Original sample of 26 clusters. References for cluster
  information are: (1) {\protect \cite{crawford95}}; (2) {\protect
    \cite{appenzeller98}}; (3) {\protect \cite{ebeling98}}; (4)
  {\protect \cite{bohringer00}}; (5) {\protect \cite{ebeling02}}; (6)
  {\protect \cite{ebeling07}}; (7) {\protect \cite{kocevski07}}; (8)
  {\protect \cite{planck2011-5.1a}}; and (9) \Planck\ RTT
  follow-up programme; see Section~\ref{sec:cluster_sample}. The meaning of the notes in the
  right-most column are: (i) cluster has SZ detection
  smaller than $3\sigma$ before source subtraction and smaller than
  $5\sigma$ after source subtraction; (ii-a) cluster has a total
  integrated source flux density greater than 5 mJy within a radius of
  $3\arcm$; (ii-b) cluster has a total integrated source flux density
  greater than 10 mJy within a radius of $10\arcm$. ``in sample''
  indicates that the cluster is included in the 11-cluster sample
  analysed in this paper. \label{tab:clsample0}}
\begin{tabular}{llcccccl}
\noalign{\doubleline}
ESZ cluster name & Alternative cluster name & Right Ascension & Declination & RA and Dec &Redshift& Redshift & Rejection\\
                 &                          &   (J2000)       &  (J2000)    & reference  &        & reference &reason\\
\noalign{\vskip 3pt\hrule\vskip 5pt}
PLCKESZ G139.59+24.18 & $\cdots$             &06:21:58.00  &+74:42:15.7&  8   & 0.27 & 9 &   in sample     \\
PLCKESZ G167.65+17.64 &ZwCl 0634.1+4750      &06:38:04.00  &+47:47:53.0& 1,7    &0.17&1,7& i\\
PLCKESZ G056.80+36.30 &MACS J0717.5+3745     &07:17:30.93  &+37:45:29.7&  6  &0.55&6&ii-b\\
PLCKESZ G157.43+30.33 &RXJ0748+5941          &07:48:45.60  &+59:41:41.0&  2   & 0.55&  2&in sample         \\
PLCKESZ G149.70+34.70 &A665                  &08:30:58.50  &+65:51:03.7&  3    &0.18&3&ii-b\\
PLCKESZ G186.40+37.30 &A697                  &08:42:57.56  &+36:21:59.3&  3    &0.28&3&ii-b\\
PLCKESZ G166.13+43.39 &A773                  &09:17:52.97  &+51:43:55.5&  3   &0.22& 3&in sample         \\
PLCKESZ G195.60+44.10 &A781                  &09:20:26.08  &+30:30:54.0&  3   &0.30&3&ii-a\\
PLCKESZ G163.70+53.50 &A980                  &10:22:28.10  &+50:07:15.6&  3  &0.16&3&ii-b\\
PLCKESZ G165.08+54.11 &A990                  &10:23:41.83  &+49:08:38.2&  3  & 0.14& 3 &in sample         \\
PLCKESZ G228.15+75.19 &MACS J1149+2223       &11:49:34.30  &+22:23:42.5&  6   & 0.55&   6& in sample    \\
PLCKESZ G226.24+76.76 &A1413                 &11:55:18.24  &+23:24:28.6&  3   & 0.14& 3& in sample         \\
PLCKESZ G180.60+76.70 &A1423                 &11:57:22.10  &+33:37:55.2&  3   &0.21&3&i\\
PLCKESZ G229.60+78.00 &A1443                 &12:01:17.00  &+23:06:18.0&  4    &0.27&4& ii-a,b\\
PLCKESZ G125.70+53.90 &A1576                 &12:36:58.96  &+63:11:26.5&  3  &0.30&3&ii-a\\
PLCKESZ G121.11+57.01 & $\cdots$             &12:59:23.80  &+60:05:24.8&  8   & 0.34 & 9 & in sample  \\
PLCKESZ G118.40+39.30 &RXCJ1354.6+7715       &13:54:37.80  &+77:15:34.6&  4    &0.40&4& i\\
PLCKESZ G067.23+67.46 &A1914                 &14:26:02.15  &+37:50:05.8&  3  & 0.17 & 3 & in sample       \\
PLCKESZ G053.52+59.54 &A2034                 &15:10:10.80  &+33:30:21.6&  3   & 0.11 & 3& in sample       \\	
PLCKESZ G044.20+48.70 &A2142                 &15:58:22.10  &+27:13:58.8&  3   &0.09&3&ii-b\\	
PLCKESZ G097.73+38.11 &A2218                 &16:35:52.80  &+66:12:50.0&  3   & 0.17 &3 & in sample       \\
PLCKESZ G086.50+15.30 &A2219                 &16:40:22.56  &+46:42:32.4&  3  &0.23	&3&ii-a,b\\
PLCKESZ G056.80+36.30 &A2244                 &17:02:42.87  &+34:03:42.8&  3  &0.10&3&ii-a\\		 
PLCKESZ G055.60+31.90 &A2261                 &17:22:27.09  &+32:08:01.7&  3   &0.22&3&ii-a,b\\	 
PLCKESZ G086.50+15.30 &CIZA J1938.3+5409     &19:38:18.60  &+54:09:33.0&  5   &0.26& 5 & i        \\
PLCKESZ G077.90$-$26.64 &A2409                 &22:00:53.03  &+20:57:38.3&  3  & 0.15 & 3 &in sample       \\
\noalign{\vskip 5pt\hrule\vskip 3pt}
\end{tabular}
\footnotetext[1]{table footnote 1}
\end{table*} 
\section{Description of \textit{Planck} data}
\label{sec:planckdata}

\Planck\ \citep{tauber2010a, Planck2011-1.1} is the third generation
space mission to measure the anisotropy of the cosmic microwave
background (CMB).  It observes the sky in nine frequency bands
covering 30--857\,GHz with high sensitivity and angular resolution
from 31\arcm\ to 5\arcm.  The Low Frequency Instrument (LFI;
\citealt{Mandolesi2010, Bersanelli2010, Planck2011-1.4}) covers the 30,
44, and 70\,GHz bands with amplifiers cooled to 20\,\hbox{K}.  The
High Frequency Instrument (HFI; \citealt{Lamarre2010,
Planck2011-1.5}) covers the 100, 143, 217, 353, 545, and 857\,GHz
bands with bolometers cooled to 0.1\,\hbox{K}.  Polarisation is
measured in all but the highest two bands \citep{Leahy2010,
Rosset2010}.  A combination of radiative cooling and three mechanical
coolers produces the temperatures needed for the detectors and optics
\citep{Planck2011-1.3}.  Two Data Processing Centers (DPCs) check and
calibrate the data and make maps of the sky \citep{Planck2011-1.7,
Planck2011-1.6}.  \Planck's sensitivity, angular resolution, and
frequency coverage make it a powerful instrument for Galactic and
extragalactic astrophysics as well as cosmology.  Early astrophysics
results are given in Planck Collaboration VIII--XXVI 2011, based on data
taken between 13~August 2009 and 7~June 2010.  Intermediate
astrophysics results are now being presented in a series of papers
based on data taken between 13~August 2009 and 27~November 2010.

We note that the \Planck\ maps used for the analysis in this paper are
not the same as those used in the Early \Planck\ results papers. In
particular, we stress that both the data and the analysis techniques
employed for this study are not the same as was used to construct the
\Planck\ ESZ catalogue \citep{planck2011-5.1a}. A later version of the
data has been used for the analysis here and we use different analysis
techniques for reasons which will be explained in
Section~\ref{sec:planck_data_analysis}. However, as part of our suite
of internal tests, we have repeated our analysis on the older version
of the \Planck\ data which was used to derive the ESZ catalogue and we
find excellent agreement.

\begin{figure*}
\centerline{\includegraphics[width=5.55cm,clip=,angle=-90.]{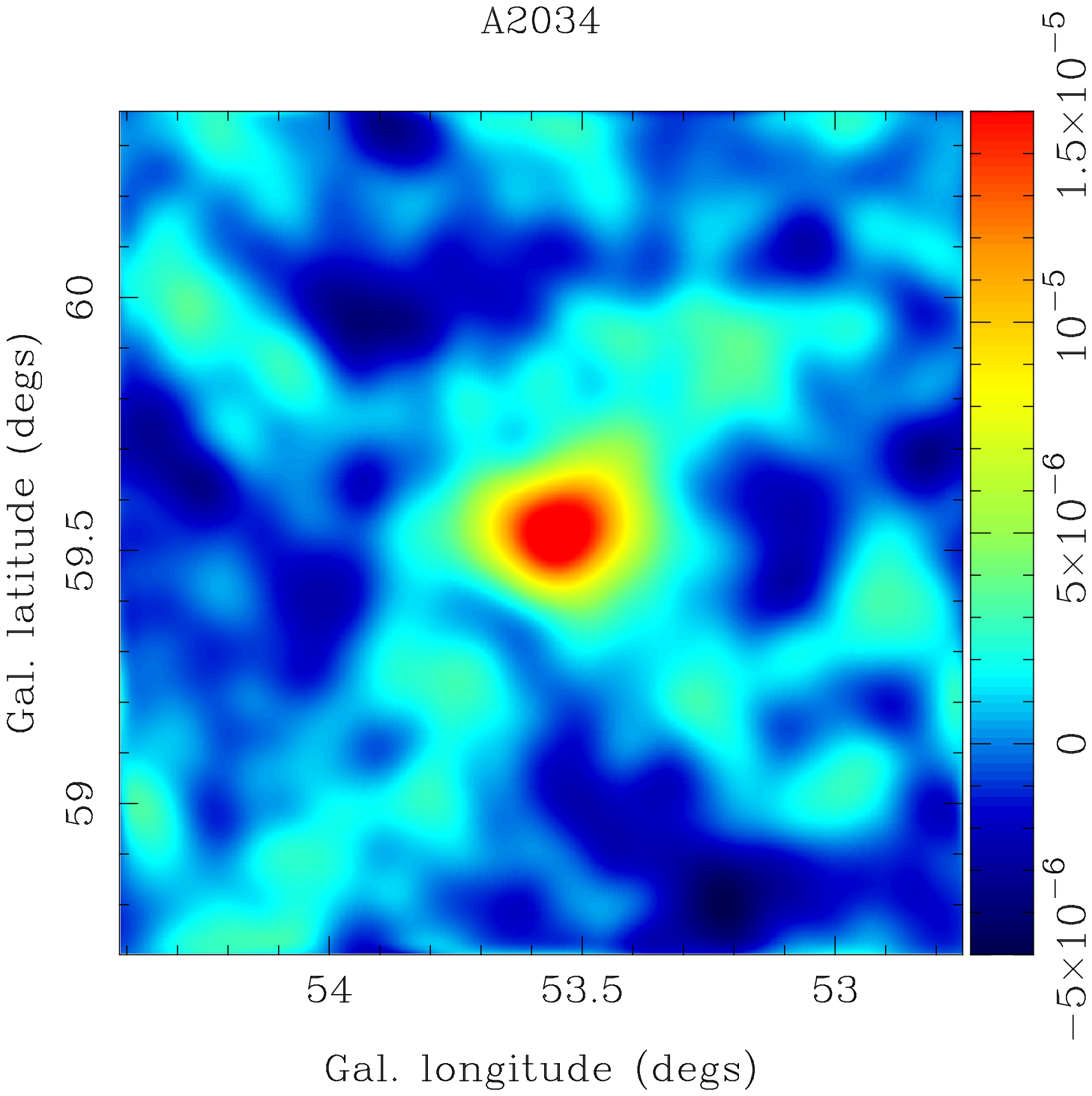}
\qquad\includegraphics[width=5.55cm,clip=,angle=-90.]{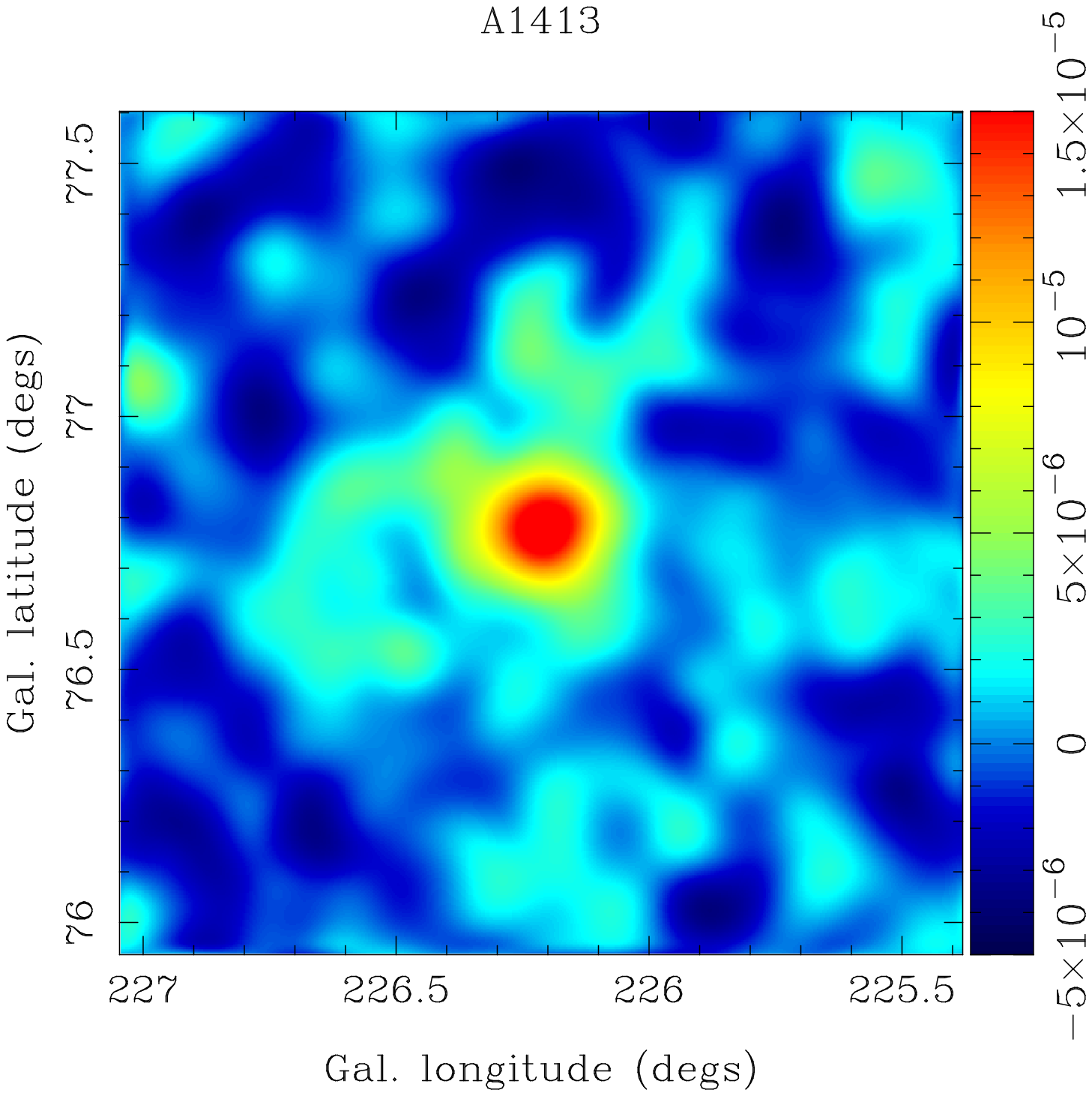}\qquad
\includegraphics[width=5.55cm,clip=,angle=-90.]{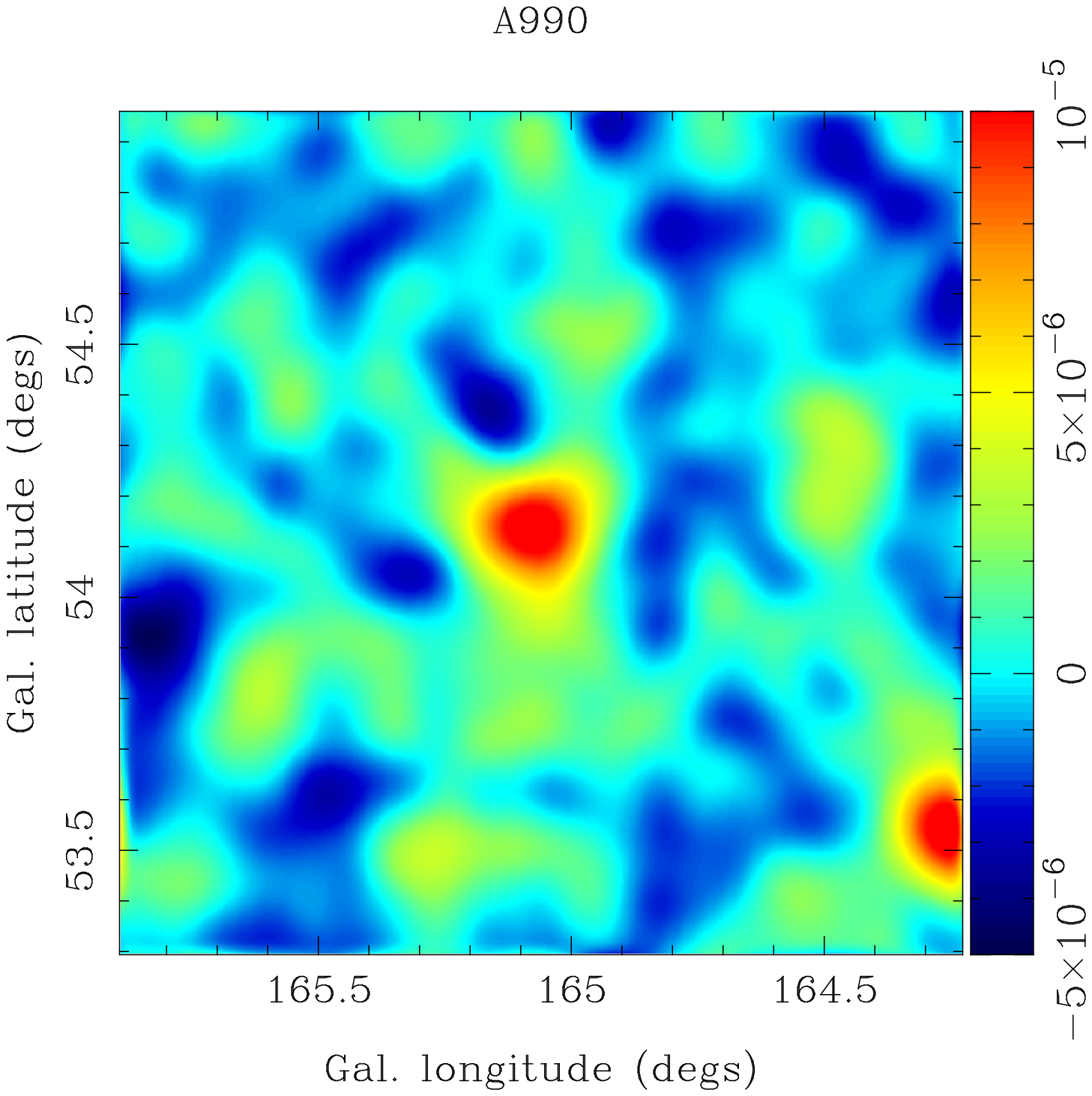}}\vspace{0.45cm}
\centerline{\includegraphics[width=5.55cm,clip=,angle=-90.]{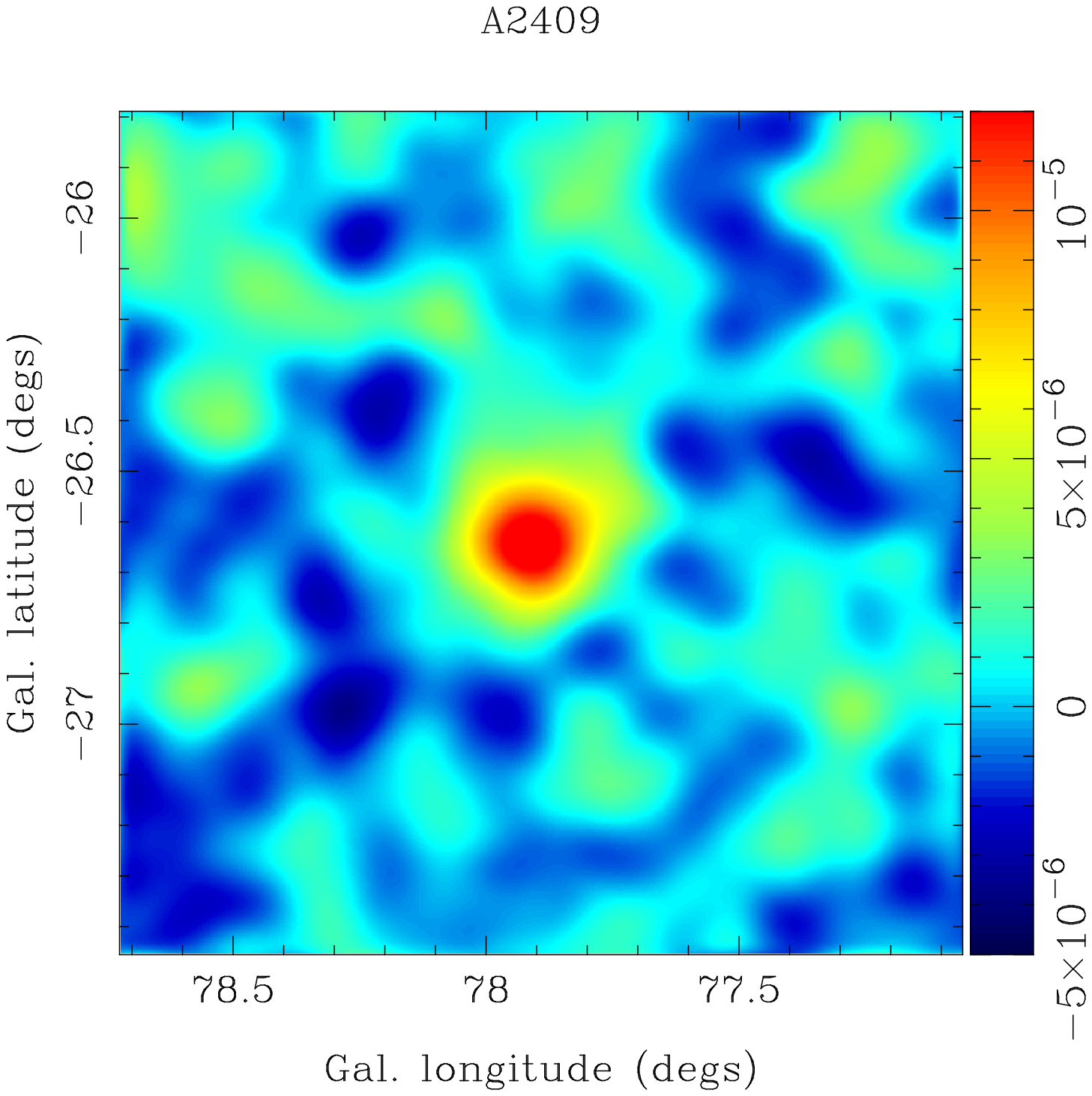}
\qquad\includegraphics[width=5.55cm,clip=,angle=-90.]{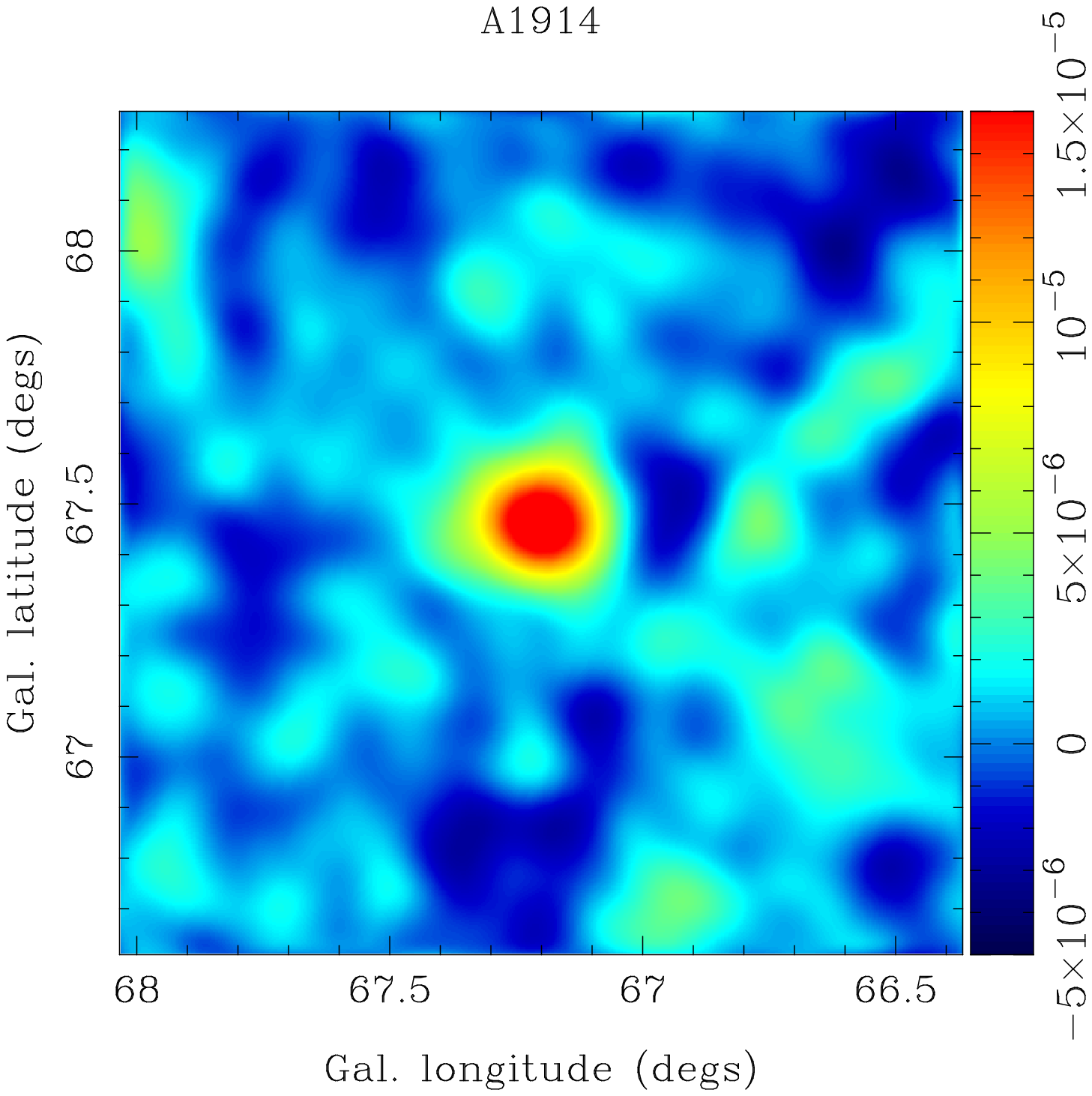}\qquad
\includegraphics[width=5.55cm,clip=,angle=-90.]{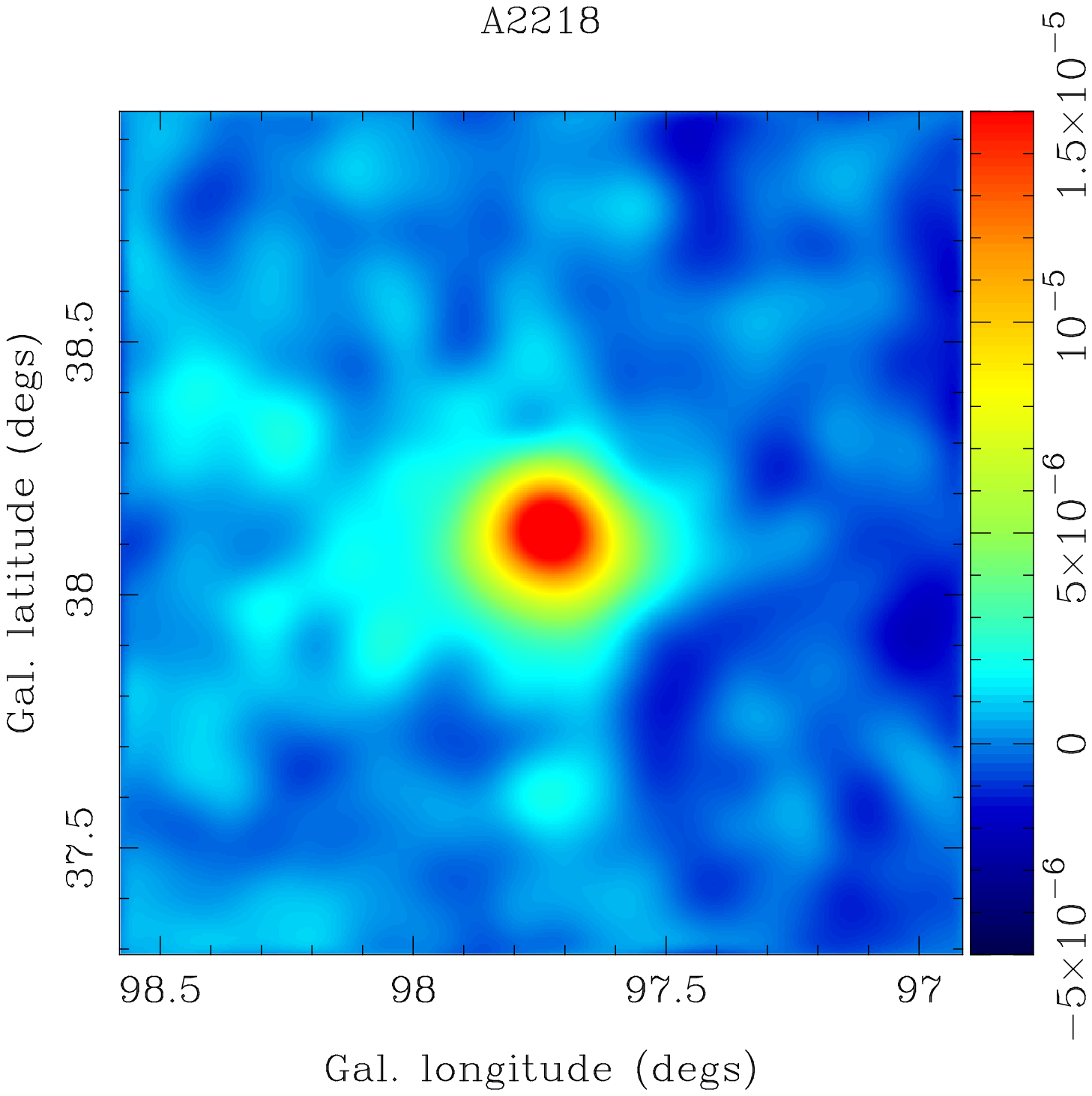}}\vspace{0.45cm}
\centerline{\includegraphics[width=5.55cm,clip=,angle=-90.]{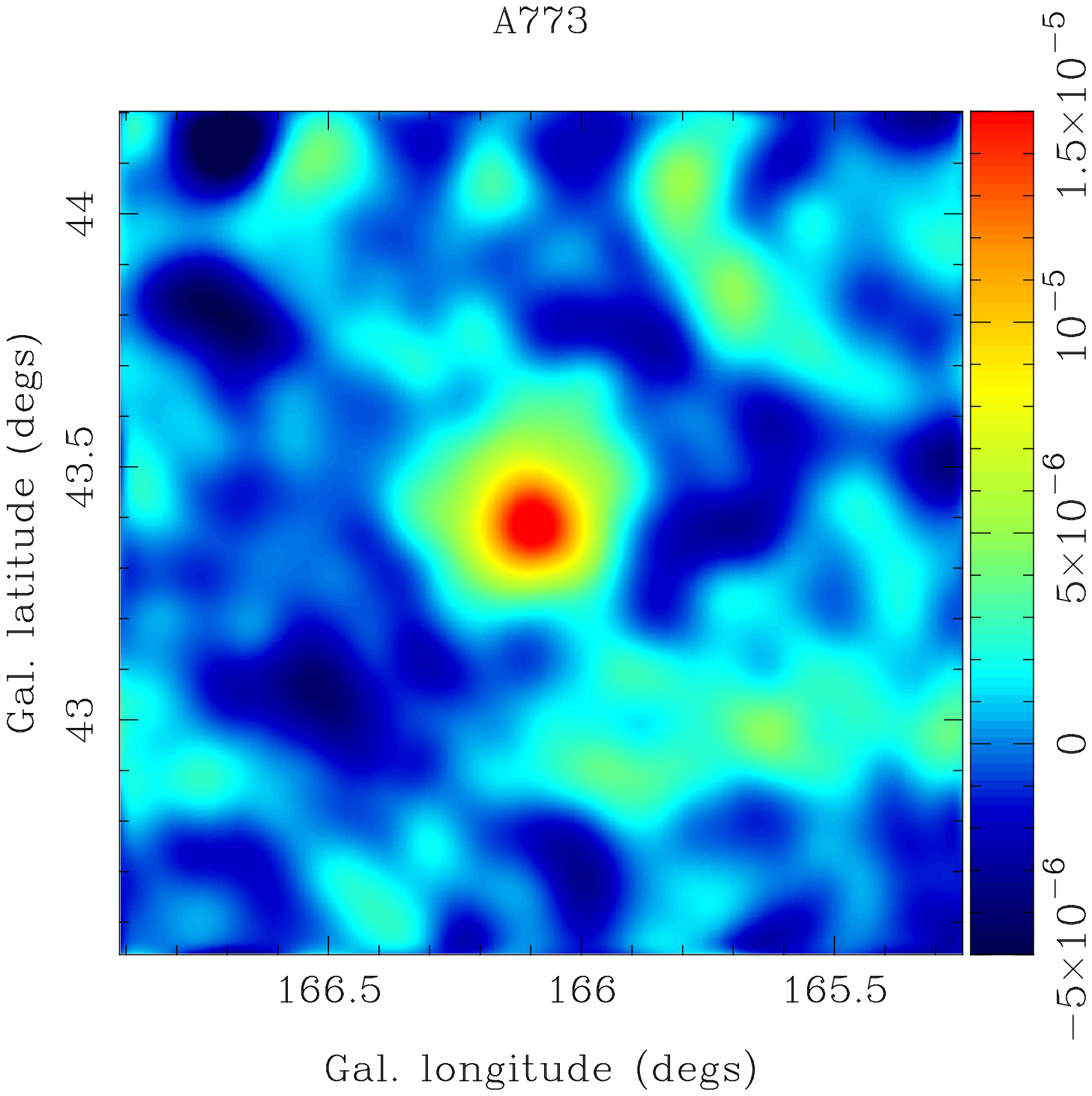}
\qquad\includegraphics[width=5.55cm,clip=,angle=-90.]{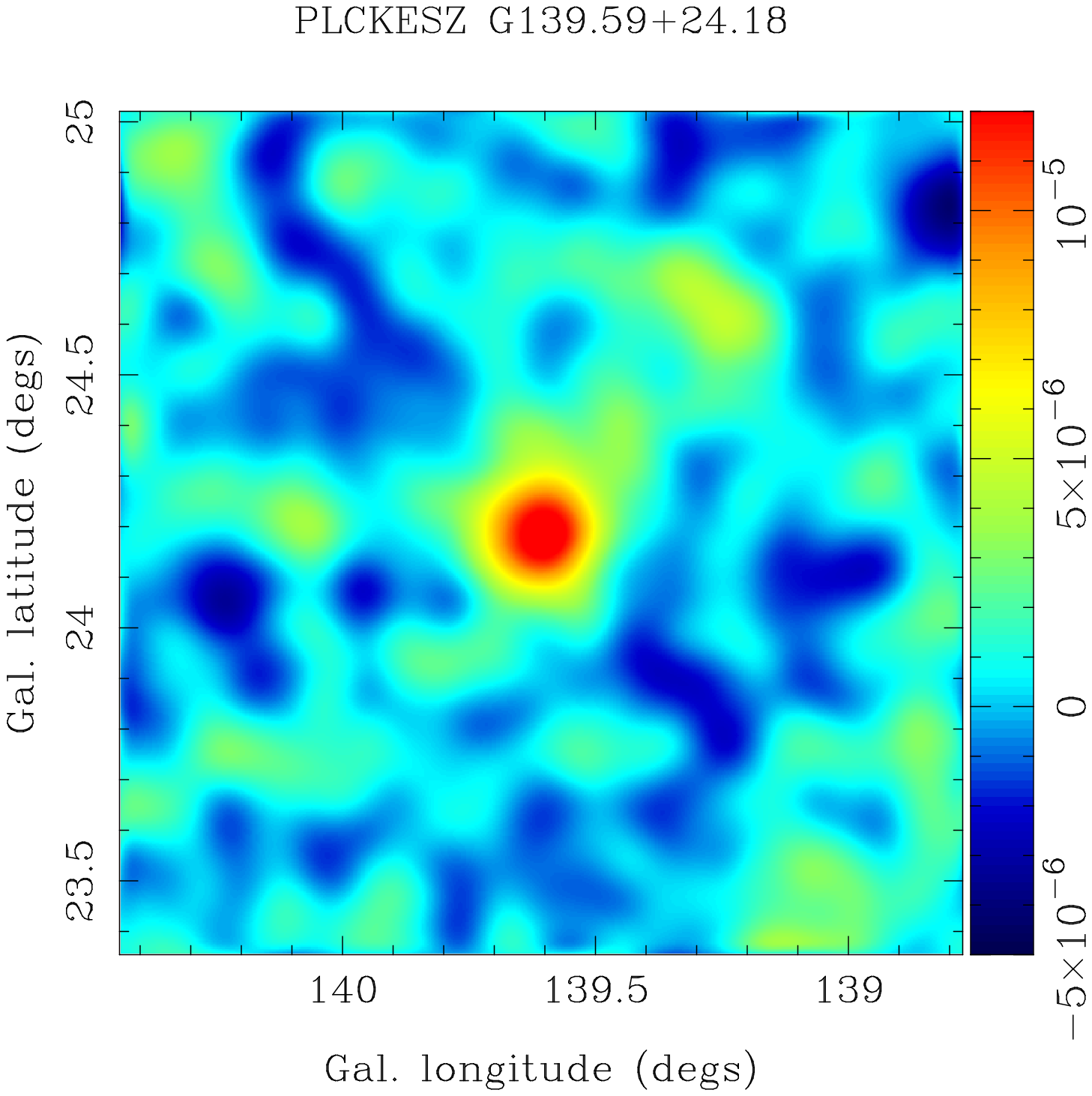}
\qquad\includegraphics[width=5.55cm,clip=,angle=-90.]{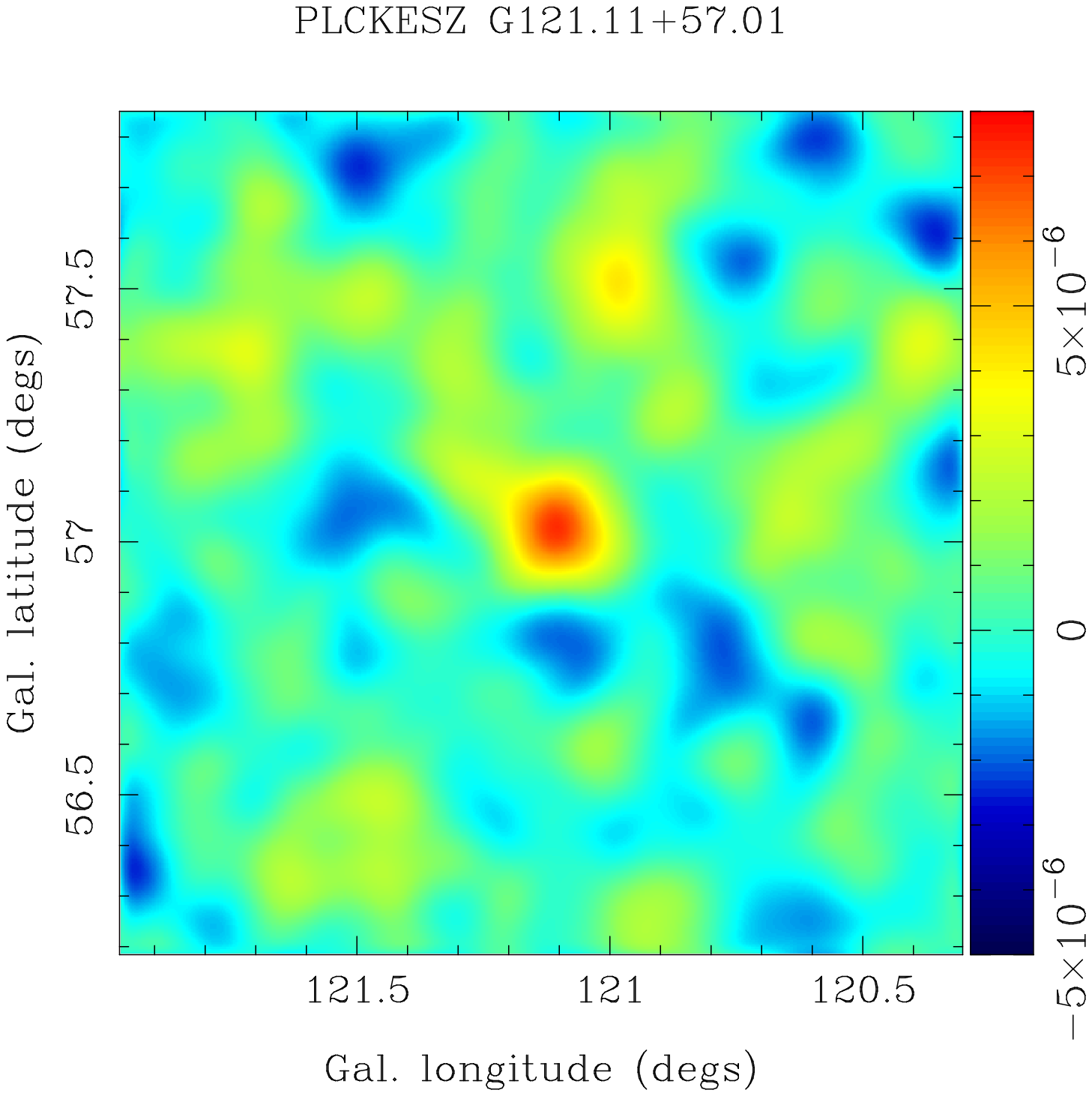}}\vspace{0.45cm}
\centerline{\includegraphics[width=5.55cm,clip=,angle=-90.]
{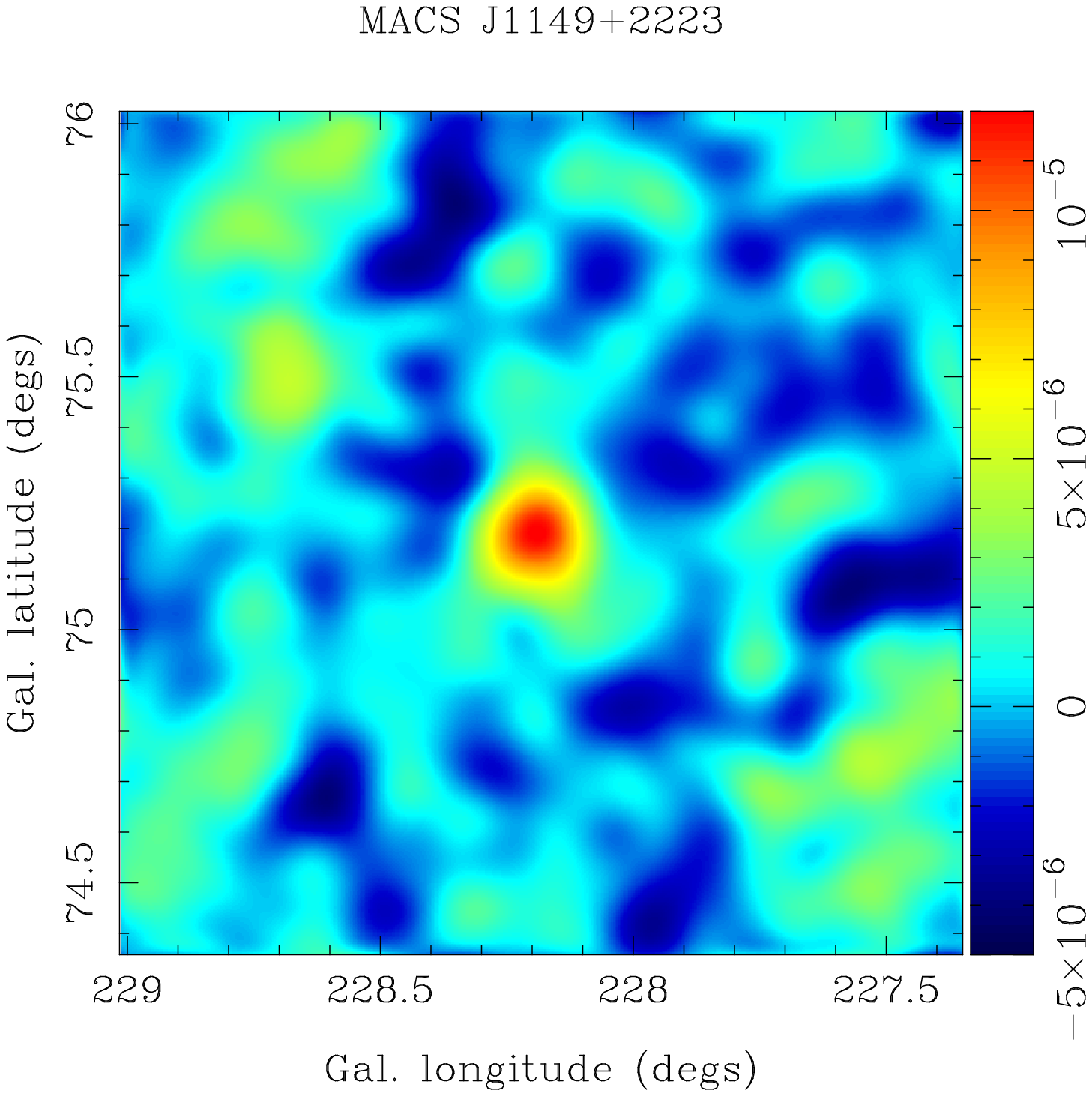}\qquad\includegraphics[width=5.55cm,clip=,angle=-90.]
{RXJ0748+5941pm.eps}}
\caption{Maps of the dimensionless Compton-$y$ parameter
  (equation~\ref{eq:y_obs}) as estimated from the  
  \Planck\ observations using the MILCA algorithm. The maps have an
  effective resolution of 10~arcmin. The clusters are ordered in terms of
increasing redshift, from top left to bottom right. Each panel shows a $100^\prime \times
100^\prime$ region. \label{fig:planckmap}} 
\end{figure*}

In Fig.~\ref{fig:planckmap} we present maps of the dimensionless
Compton-$y$ parameter for each of the clusters as estimated from the
\Planck\ data. The $y$-parameter is related to the observed brightness
as a function of frequency ($\nu$) by
\begin{equation}\label{eq:y_obs}
\frac{\DeltaT_{\rm tSZ}}{T_{\rm CMB}} (\nu) = y \cdot g(\nu),
\end{equation}
where $\DeltaT_{\rm tSZ}$ is the brightness fluctuation due to the
thermal SZ (tSZ) effect and $T_{\rm CMB}$ is the temperature of the
CMB, which we take to be $2.7255\pm0.0006$ \citep{fixsen09}. The
function, $g(\nu)$ is the frequency dependence of the SZ effect
\citep{sunyaev72}. For the HFI channel frequencies ($100, 143, 217,
353, 545$ and $857$ GHz), $g(\nu)$ takes the values ($-4.03, -2.78,
0.19, 6.19, 14.47$ and $26.36$ $K_{\rm CMB} / y$).  The maps of
Fig.~\ref{fig:planckmap} were estimated from HFI channel data taken
between 13 August 2009 and 27 November 2010, corresponding to slightly
more than 2.5 full-sky scans. The measured noise levels on the
\Planck\ frequency channel maps are listed in
Table~\ref{tab:plancknoise} (Planck HFI Core Team, in prep.). The tSZ
signal reconstructions were performed using the MILCA method
(\citealt{hurier10} and references therein) on the six
\Planck\ all-sky maps from 100~GHz to 857~GHz. MILCA (Modified
Internal Linear Combination Algorithm) is a component separation
approach aimed at extracting a chosen component (here the tSZ signal)
from a multi-channel set of input maps. It is based on the well known
ILC approach (e.g., \citealt{eriksen04}), which searches for the
linear combination of the input maps that minimises the variance of
the final reconstructed map, while imposing spectral constraints. For our
cluster SZ reconstructions, we applied MILCA using two constraints,
the first one to preserve the tSZ signal and the second one to remove
CMB contamination in the final tSZ $y$-map. In addition, to compute
the weights of the linear combination, we have used the extra degrees
of freedom to minimise residuals from other components (two degrees) and
from the noise (two degrees). The noise covariance matrix was estimated
from jack-knife maps. The final $y$-maps have an effective resolution
of 10~arcmin.  Note that, in general, the properties of the foreground
emission depend on both the position on the sky and on the frequency
of observation. We have therefore allowed the weights to vary as a
function of both position and frequency. We have confirmed using
simulations that such an approach maximises the signal-to-noise and
minimises the bias in the extraction of the tSZ signal.  We emphasise
that the MILCA SZ reconstructions presented in
Fig.~\ref{fig:planckmap} are intended for visual examination and
qualitative assessment of the cluster signals only. Our quantitative
analysis of the \Planck\ data, which we use to compare with the AMI
results, is based mainly on the PowellSnakes \citep{carvalho11} SZ
extraction algorithm (see Section~\ref{sec:SZsignal}).

\begin{table}
\centering
\caption{Noise levels per $N_{\rm side} = 2048$ pixel for each \Planck\ frequency band.
\label{tab:plancknoise}}
\begin{tabular}{lrrrrrr}
\noalign{\doubleline}
Freq. (GHz)     & $100$ & $143$ & $217$ & $353$ & $545$ & $857$  \\
Noise ($\mu$K)  & 80    &  34 &     56  & 167 & 1414  & 170200 \\
\noalign{\vskip 5pt\hrule\vskip 3pt}
\end{tabular}
\end{table}
\section{Description of AMI data}
\label{sec:amidata}
AMI comprises two arrays, the Small Array (SA) and the Large Array
(LA), located at the Mullard Radio Astronomy Observatory near
Cambridge. The SA consists of ten 3.7-m diameter equatorially--mounted
antennas, with a baseline range of $ \simeq5$--20\,m and synthesised
beam (resolution) of around $3\arcm$. The LA consists of eight 13-m
diameter antennas with a baseline range of $\simeq20$--100\,m and
synthesised beam of around $30\arcsec$. Both arrays observe Stokes $I+Q$ in
the band 13--18\,GHz, each with system temperatures of about
25~K. Note that AMI defines Stokes $Q$ and $U$ with respect to celestial
north. The backends are analogue Fourier transform spectrometers,
from which the complex signals in each of eight channels of 750-MHz
bandwidth are synthesised, and the signals in the adjacent channels
are correlated at the $\simeq$10\,\% level. Further details of
the instrument are given in \cite{zwart08}.

SA pointed observations of our cluster sample were taken during
$2007$--$2011 $. The observation lengths per cluster before any
flagging of the data are presented in Table~\ref{tab:amiobs1}; the
noise levels on the SA maps reflect the actual observation time
used. The SA observations were made with single pointings interspersed
with a phase calibration source, while the LA observations were made
in a 61+19-point raster mode configuration with $4\arcm$ spacing.
This consisted of 61 pointings arranged in a hexagonal grid, with grid
points separated by $4\arcm$ with further observations of the central
19 pointings designed to increase the sensitivity at the centre of the
field. In this mode the integration time on the area $\leq7.5\arcm$
from the cluster centre is twice as long as the area $>7.5\arcm$ away,
so as to provide a better match to the primary beam sensitivity of the
AMI~SA. Phase calibrators were chosen from the Jodrell Bank VLA 
Astrometric Survey (JVAS, \citealt{patnaik92}) on the basis of
proximity ($\leq2^\circ$ for the AMI~LA, $\leq8^\circ$ for the AMI~SA)
and 15\,GHz flux density ($\geq0.2$\,Jy for the AMI~LA, $\geq0.7$\,Jy
for the AMI~SA). The JVAS is based on observations made with
the VLA in ``A" configuration \citep{condon85, condon86, condon89,
  white92}.

The reduction of the AMI data was performed using a dedicated software
tool \textsc{reduce}. This is used to apply path-compensator and
path-delay corrections, to flag interference, shadowing and hardware
errors, to apply phase and amplitude calibrations and to Fourier
transform the correlator data readout to synthesise the frequency
channels, before outputting to disk.

Flux calibration was performed using short observations of 3C48 and
3C286 near the beginning and end of each run. The assumed $I+Q$ flux
densities for these sources in the AMI channels are listed in
Table~\ref{tab:Fluxes-of-3C286} and are consistent with
\cite{baars77}. As Baars et~al.  measure $I$ and AMI measures $I+Q$,
these flux densities include corrections for the polarisation of the
sources. An amplitude correction is also made for the intervening air
mass during the observation. Flux calibration is expected to be
accurate to ${\simeq}\,3$\% for the AMI~SA and ${\simeq}\,5$\% for the
AMI~LA.  After phase calibration, the phase of both arrays over one
hour is generally stable to $5^{\circ}$ for channels 4--7, and to
$10^{\circ}$ for channels 3 and 8. (Channels 1 and 2 are generally not
used for science analysis as they tend to suffer from interference
problems.)
\begin{table}
\centering
\caption{Assumed $I+Q$ flux densities of 3C286 and 3C48 over the 
commonly-used AMI band, and the full width at half maximum of the LA
primary beam (approximate field of view, $\Theta_{\rm LA}$) for each
channel.\label{tab:Fluxes-of-3C286}}
\begin{tabular}{cccccc}
\noalign{\doubleline}
 Channel & $\nu$/GHz & $S^{{\rm {3C286}}}$/Jy & $S^{{\rm {3C48}}}$/Jy &
 $\Theta_{\rm LA}$/ arcmin \\ 
\noalign{\vskip 3pt\hrule\vskip 5pt}
 3 & 13.9 & 3.74 & 1.89 & 6.08\\
 4 & 14.6 & 3.60 & 1.78 & 5.89\\
 5 & 15.3 & 3.47 & 1.68 & 5.70\\
 6 & 16.1 & 3.35 & 1.60 & 5.53\\
 7 & 16.9 & 3.24 & 1.52 & 5.39\\
 8 & 17.6 & 3.14 & 1.45 & 5.25\\ 
\noalign{\vskip 5pt\hrule\vskip 3pt}
\end{tabular}
\end{table}

\begin{figure}
\centering
\includegraphics[angle=-90,width=75mm]{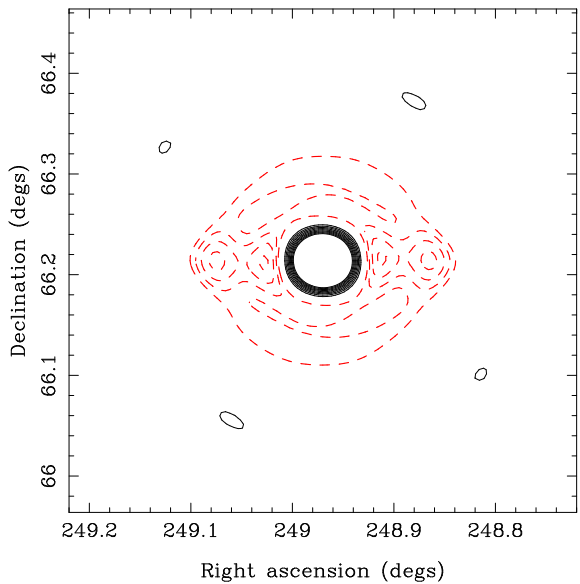}
\caption{SA synthesised beam for the A2218
observations. Contours start at $6\,\%$ and increase linearly by $3\,\%$ 
per contour. Contours drawn as red dashed lines are negative. The 
synthesised beams for the other cluster observations are qualitatively
similar.} \label{fig:beamA2218}
\end{figure}

Maps were made using the Astronomical Image Processing System
(\textsc{aips}, \citealt{greisen03}) from each channel of the AMI~SA and
LA; however here we present only the combined-channel maps of the SA
and LA observations. The \textsc{aips} task \texttt{imean} was used on
the LA individual maps to attach the map noise to the map
header. \texttt{imean} fits a Gaussian to the histogram of the map
pixels (ignoring extreme pixels that might be due to sources) and uses
the standard deviation of the fitted Gaussian as a measure of the
random noise in the data. The \textsc{aips} task \texttt{flatn} was
then used to form a mosaiced image from the multiple pointings. Data
from the pointings were primary beam corrected using parameters listed
in Table~\ref{tab:Fluxes-of-3C286} and weighted accordingly when
combined. The AMI~SA combined-channel map noise and the LA map noise
are given in Table~\ref{tab:amiobs2}. The raw $uv$ data for all good
observations were concatenated together to make a visibility data file
for each channel. All maps were made using natural $uv$ weighting and
all images were \textsc{cleaned} to three times the thermal noise with
a single clean box encompassing the entire map. The data were also
binned into bins of width $40\lambda$. This reduced the size of the
data to a manageable level without adversely affecting the subsequent
inference of cluster properties. Fig.~\ref{fig:beamA2218} shows a
typical example of the SA synthesised beam, in this case for the
observations of A2218.

\begin{figure*}
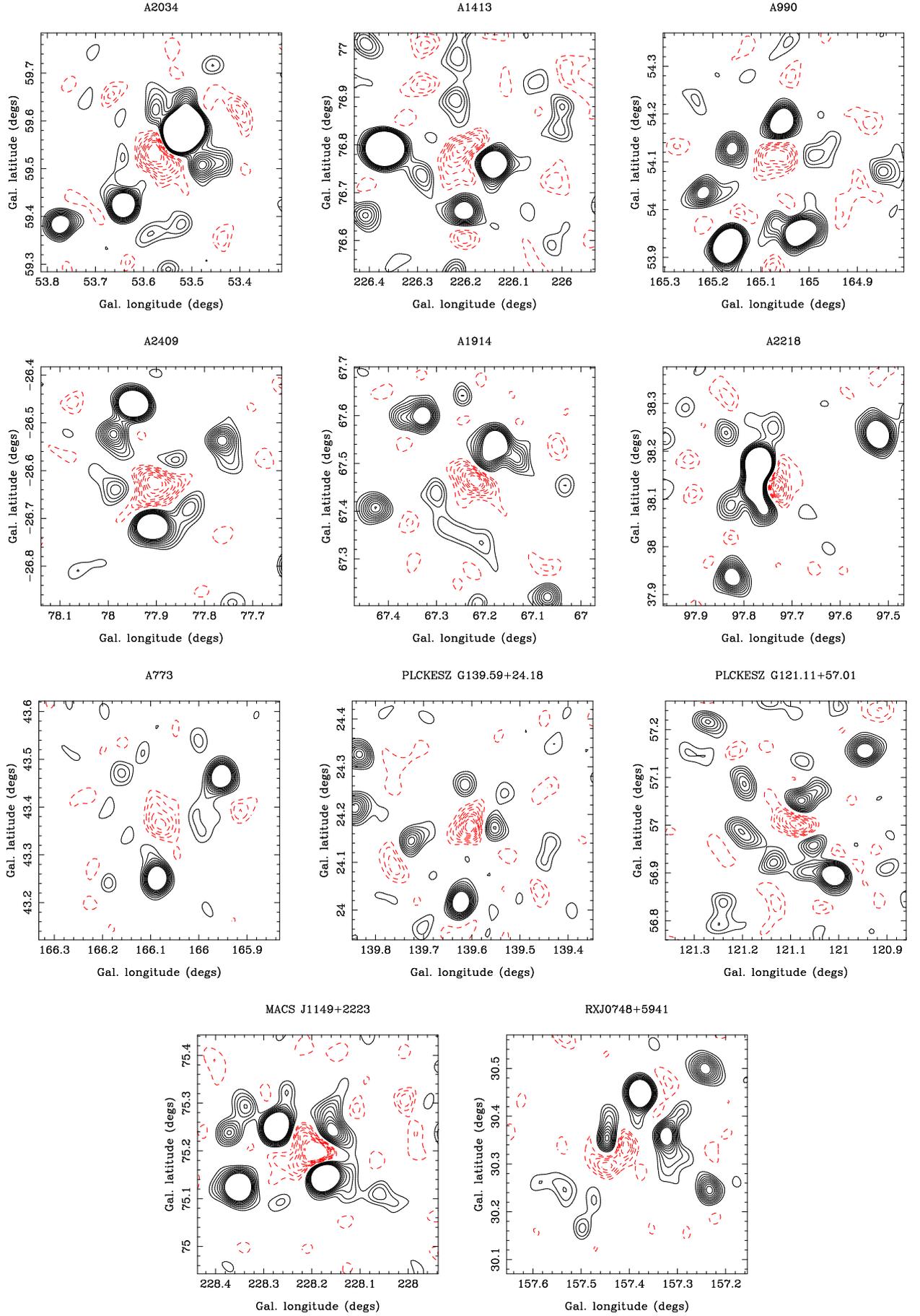

\centerline{\includegraphics[width=5.65cm,clip=,angle=-90.]{A2034am_bsub.ps}
\qquad\includegraphics[width=5.65cm,clip=,angle=-90.]{A1413am_bsub.ps}\qquad
\includegraphics[width=5.65cm,clip=,angle=-90.]{A990am_bsub.ps}}\vspace{0.45cm}
\centerline{\includegraphics[width=5.65cm,clip=,angle=-90.]{A2409am_bsub.ps}
\qquad\includegraphics[width=5.65cm,clip=,angle=-90.]{A1914am_bsub.ps}\qquad
\includegraphics[width=5.65cm,clip=,angle=-90.]{A2218am_bsub.ps}}\vspace{0.45cm}
\centerline{\includegraphics[width=5.65cm,clip=,angle=-90.]{A773am_bsub.ps}
\qquad\includegraphics[width=5.65cm,clip=,angle=-90.]{PLJ0621+7442am_bsub.ps}
\qquad\includegraphics[width=5.65cm,clip=,angle=-90.]{PLJ1259+6005am_bsub.ps}}\vspace{0.45cm}
\centerline{\includegraphics[width=5.65cm,clip=,angle=-90.]
{MAJ1149+2223am_bsub.ps}\qquad\includegraphics[width=5.65cm,clip=,angle=-90.]
{RXJ0748+5941am_bsub.ps}}
\caption{AMI maps before source subtraction. The clusters are ordered
  as in Fig.~\ref{fig:planckmap}, in terms of increasing
  redshift. Black solid lines represent positive contours and red
  dashed lines indicate negative contours. The contours increase
  linearly from $\pm2\sigma_{\rm SA}$ to $\pm10\sigma_{\rm SA}$ where
  $\sigma_{\rm SA}$ is listed in Table~\ref{tab:amiobs2} for each
  cluster. Each map covers a region approximately $30^\prime \times
  30^\prime$ and the resolution is around $3^\prime$. \label{fig:amimap_bsub}}
 \end{figure*}

\begin{figure*}
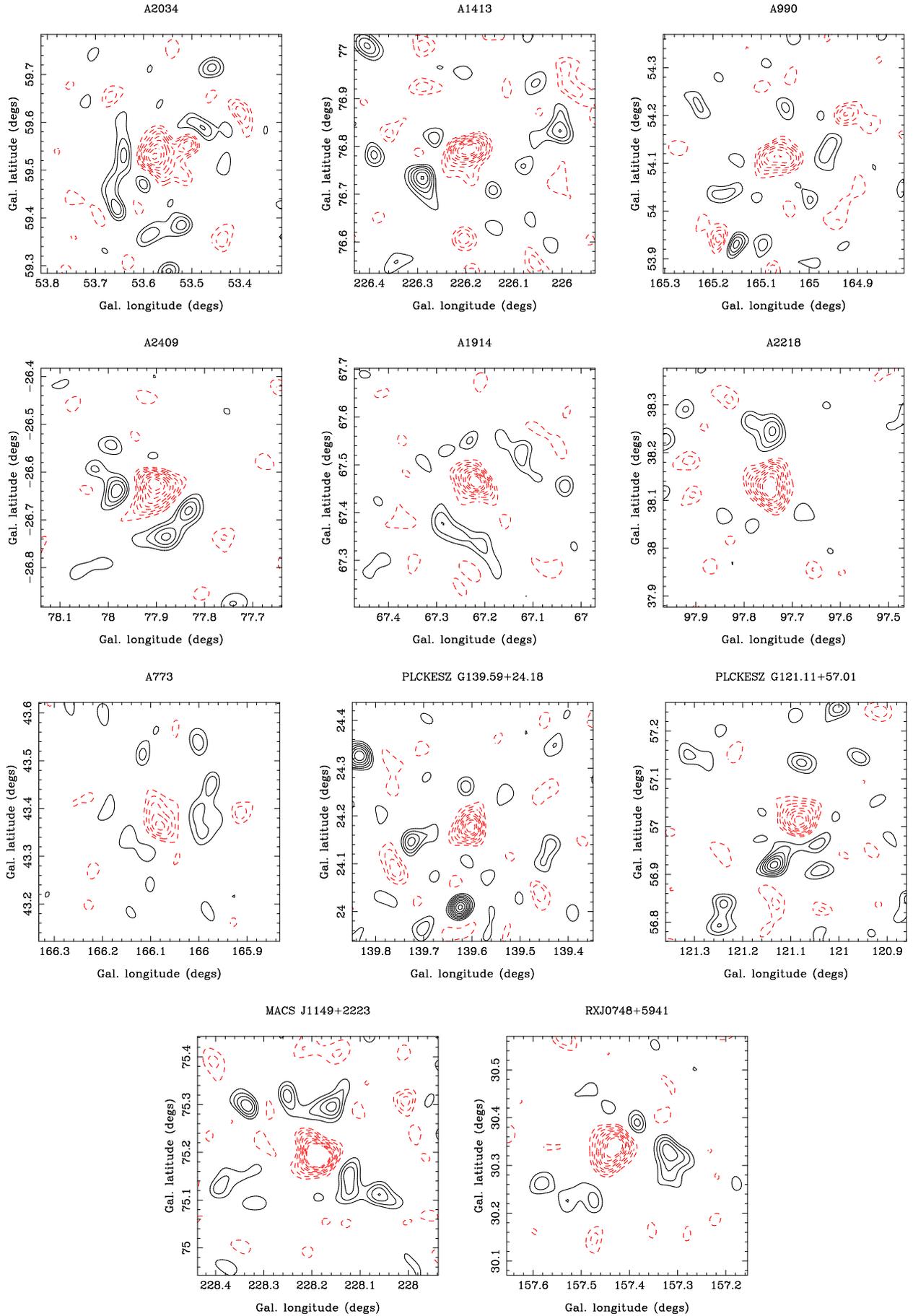

\centerline{\includegraphics[width=5.65cm,clip=,angle=-90.]{A2034am_asub.ps}
\qquad\includegraphics[width=5.65cm,clip=,angle=-90.]{A1413am_asub.ps}\qquad
\includegraphics[width=5.65cm,clip=,angle=-90.]{A990am_asub.ps}}\vspace{0.45cm}
\centerline{\includegraphics[width=5.65cm,clip=,angle=-90.]{A2409am_asub.ps}
\qquad\includegraphics[width=5.65cm,clip=,angle=-90.]{A1914am_asub.ps}\qquad
\includegraphics[width=5.65cm,clip=,angle=-90.]{A2218am_asub.ps}}\vspace{0.45cm}
\centerline{\includegraphics[width=5.65cm,clip=,angle=-90.]{A773am_asub.ps}
\qquad\includegraphics[width=5.65cm,clip=,angle=-90.]{PLJ0621+7442am_asub.ps}
\qquad\includegraphics[width=5.65cm,clip=,angle=-90.]{PLJ1259+6005am_asub.ps}}\vspace{0.45cm}
\centerline{\includegraphics[width=5.65cm,clip=,angle=-90.]
{MAJ1149+2223am_asub.ps}\qquad\includegraphics[width=5.65cm,clip=,angle=-90.]
{RXJ0748+5941am_asub.ps}}
\caption{AMI maps after subtraction of radio point sources. The solid
  black lines represent positive contours and the dashed red lines
  indicate negative contours. The cluster ordering, contour levels and
  resolution are the same as in
  Fig.~\ref{fig:amimap_bsub}. \label{fig:amimap_asub}} 
\end{figure*}

As contamination from radio sources at $16$ GHz tends to be 
significant, removing or modelling this emission accurately can often 
be essential to recover SZ decrements from AMI maps.
To address this issue, we use the AMI-developed source extraction
software, \textsc{sourcefind} \citep{waldram03, franzen11} to
determine the position, flux density and spectral index of the radio
sources with flux density $\geq 3.5\sigma_{\rm{LA}}$ on the
\textsc{cleaned} LA continuum maps, where $\sigma_{\rm{LA}}$ is the LA
thermal noise. Spectral indices were fit with a Markov Chain Monte
Carlo (MCMC) method using LA maps for all six channels and assuming that
source flux densities follow a power-law relation of $S\propto\nu^{-\alpha}$
for the AMI frequencies. These source parameter estimates are
subsequently used as priors in our Bayesian analysis of the cluster SZ
signals (Section~\ref {sec:AMI_data_analysis}).
Tables~\ref{tab:amiobs1} and \ref{tab:amiobs2} summarise the
observational details of our cluster sample and
Figs.~\ref{fig:amimap_bsub} and \ref {fig:amimap_asub} present the
maps of the AMI observations of these clusters before and after source
subtraction, respectively.
Once again, as with the reconstructed \Planck\ maps presented in the
previous section, the AMI maps presented in
Figs.~\ref{fig:amimap_bsub} and \ref{fig:amimap_asub} are intended for
visual examination and qualitative assessment of the cluster signals
only. Our quantitative analysis of the AMI data is described later in
Section~\ref{sec:AMI_data_analysis}.

\begin{table*} 
\centering
\caption{Details of the AMI observations of our cluster sample.\label{tab:amiobs1}}
\begin{tabular}{llcccc}
\noalign{\doubleline}
Cluster     & Observing dates & AMI~SA observing & LA observing & AMI~SA phase & LA phase   \\
            &                 & time (hours) & time (hours) & calibrator & calibrator \\
\noalign{\vskip 3pt\hrule\vskip 5pt}
A2034       &2010 Jan, Mar, Dec& 38 & 23 & J1506+3730 & J1504+3249 \\
A1413       &2007 Mar, Nov; 2010 Mar& 40 & 19 & J1159+2914 & J1150+2417 \\
A990        &2007 Feb, Nov; 2009 Apr& 31 & 17 & J0958+4725 & J1015+4926 \\
A2409       &2007 Mar, Apr, May; 2009 Dec& 38 & 21 & J2225+2118 & J2200+2137 \\
A1914       &2008 May; 2009 Jan, Jun& 32 & 35 & J1419+3821 & J1419+3821 \\
A2218       &2008 Jan; 2009 Feb, Jun& 30 & 12 & J1642+6856 & J1623+6624 \\
A773        &2007 Sep, Oct; 2009 May& 40 & 19 & J0903+4651 & J0929+5013 \\
MACS J1149+2223&2010 Apr, May, Nov, Dec& 38 & 11 & J1150+2417 & J1150+2417 \\
RXJ0748+5941&2011 Feb& 45 & 35 & J0753+5352 & J0737+5941  \\
PLCKESZ G139.59+24.18 & 2010 Nov& 52 & 37 & J0639+7324 & J0639+7324 \\
PLCKESZ G121.11+57.01 & 2011 Jan& 40 & 67 & J1302+5748 & J1302+5748 \\
\noalign{\vskip 5pt\hrule\vskip 3pt}
\end{tabular}
\end{table*}
\begin{table*} 
\centering
\caption{Details of radio point sources and the thermal noise levels
  for the AMI observations of the cluster sample. Here $\sigma_{\rm{SA}}$
  and $\sigma_{\rm{LA}}$ refer to the thermal noise levels reached in
  the LA and SA maps, respectively. The S/N values are calculated by
  dividing the peak flux values by $\sigma_{\rm{SA}}$.\label{tab:amiobs2}}
\begin{tabular}{lccccr}
\noalign{\doubleline}
Cluster&Number of LA                        &total source flux     &$\sigma_{\rm {SA}}$&$\sigma_{\rm {LA}}$&S/N\\
       &3.5\,$\sigma_{\mathrm{LA}}$ sources &on the map (${\rm mJy}\,{\rm beam}^{-1}$) & (${\rm mJy}\,{\rm beam}^{-1}$) & (${\rm mJy}\,{\rm beam}^{-1}$)   &  \\
\noalign{\vskip 3pt\hrule\vskip 5pt}
A2034  &13                                  &32.5                 &0.08                 &0.08             &10      \\
A1413  &20                                  &47.6                 &0.09                 &0.12             &12 \\
A990   &20                                  &56.3                 &0.11                 &0.10             &8 \\
A2409  &14                                  &49.2                 &0.11                 &0.10             &11 \\
A1914  &14                                  &31.5                 &0.12                 &0.10             &11 \\
A2218  &\,\,\,9                             &31.5                 &0.11                 &0.15             &12 \\
A773   &\,\,\,6                             &12.8                 &0.12                 &0.12             &6   \\
MACS J1149+2223&13                          &44.6                 &0.09                 &0.14             &17   \\
RXJ0748+5941&15                             &17.1                 &0.07                 &0.06             &12    \\
PLCKESZ G139.59+24.18 &13                   &13.4                 &0.09                 &0.06             &11     \\
PLCKESZ G121.11+57.01 &19                   &25.4                 &0.07                 &0.06             &10   \\
\noalign{\vskip 5pt\hrule\vskip 3pt}
\end{tabular}
\end{table*} 

\section{Analysing the SZ signal}
\label{sec:SZsignal}
The SZ surface brightness (the Compton-$y$ parameter, cf.,
equation~\ref{eq:y_obs}) is proportional to the line of sight
integral of the electron pressure, 
\begin{equation}\label{eq:y}
y=\frac{\sigma_{\rm T}}{m_{\rm e}c^2} \int_{-\infty}^{+\infty}{P_{\rm e}(r){\rm d}
l},
\end{equation}
where $P_{\rm e}(r)$ is the electron pressure at radius $r$, $\sigma_
{\rm T}$ is the Thomson scattering cross-section, $m_{\rm e}$ is the
electron mass, $c$ is the speed of light and $dl$ is the line element
along the line of sight. In this context, \cite{nagai07} analysed the
pressure profiles of a series of simulated clusters \citep{kravtsov05}
as well as a sample of relaxed real clusters presented in
\cite{vikhlinin05, vikhlinin06}. They found that the pressure profiles
of all of these clusters could be described by a generalisation of the
Navarro, Frenk, and White (NFW, \citealt{navarro97}) model used to
describe the dark matter halos of simulated clusters. Assuming
spherical geometry, the GNFW pressure profile \citep{nagai07} reads
\begin{equation}\label{eq:GNFW}
 P_{\rm e}(r) = P_{\rm {0}}\left(\frac{r}{r_{\rm s}}\right)^{-\gamma}
\left[1+\left(\frac{r}{r_{\rm s}}\right)^{\alpha}\right]^{\,(\gamma-\beta)/\alpha},
\end{equation}
where $P_{\rm {0}}$ is the overall normalisation coefficient of the
pressure profile and $r_{\rm s}$ is the scale radius. It is common to
define the latter in terms of $r_{\rm 500}$, the radius at which the
mean density is 500 times the critical density at the cluster
redshift, and to define the gas concentration parameter, $c_{\rm 500}=r_{\rm
  500}/r_{\rm s}$. The parameters $(\alpha, \beta, \gamma)$ describe
the slopes of the pressure profile at $r\simeq r_ {\rm s}$, $r>
r_{\rm s}$, and $r \ll r_{\rm s}$ respectively. In order to retain
consistency between the \Planck\ and AMI analysis pipelines, we follow
\cite{arnaud10} (see also \citealt{planck2011-5.1a, planck2011-5.2a})
and fix the values of the gas concentration parameter and the slopes to be
$(c_{\rm 500},\alpha,\beta,\gamma) =(1.156,1.0620, 5.4807, 0.3292)$.
These values describe the ``universal pressure profile'', derived from
{\it XMM-Newton} observations of the REXCESS cluster sample
\citep{bohringer07}, and from three different sets of detailed numerical
simulations by \cite{borgani04}, \cite{piffaretti08}, and
\cite{nagai07}, which take into account radiative cooling, star
formation, and energy feedback from supernova explosions. In Section
\ref{sec:implications}, we will relax these restrictions for a subset
of our cluster sample and will include information from X-ray
observations of individual clusters in our analysis. We note that the
profile of equation~(\ref{eq:GNFW}) has recently been used to analyse SZ 
data from the South Pole Telescope (SPT, \citealt{plagge10}) in 
addition to the \Planck\ survey data \citep{planck2011-5.1a,
  planck2011-5.2a}.

The integral of the $y$ parameter over the solid angle $\Omega$
subtended by the cluster is denoted by $Y_{\rm SZ}$, and is
proportional to the volume integral of the gas pressure. It is thus a
good indicator of the total thermal energy content of the cluster and
its mass (e.g., \citealt{bartlett94}). The determination of the
normalisation and the slope of the $Y_{\rm SZ}-M$ relation has
therefore been a major goal of studies of the SZ effect
\citep{daSilva04, nagai06, kravtsov06, plagge10, arnaud10,
  andersson11, planck2011-5.1a, planck2011-5.1b, planck2011-5.2a,
  planck2011-5.2b, planck2011-5.2c}. In particular, \cite{andersson11}
investigated the $Y_{\rm SZ}-Y_ {\rm X}$ scaling relation within a
sample of 15 clusters observed by SPT, {\it Chandra} and {\it
  XMM-Newton} and found a slope of close to unity ($0.96 \pm
0.18$). Similar studies were carried out by \cite{planck2011-5.2b}
using a sample of 62 nearby ($z < 0.5$) clusters observed by both
\Planck\ and by {\it XMM-Newton}. The results are consistent with
predictions from X-ray studies \citep{arnaud10, andersson11}. These
studies at low redshifts, where data are available from both X-ray and
SZ observations of galaxy clusters, are crucial to calibrate the
$Y_{\rm SZ}-M$ relation, as such a relation can then be scaled and
used to determine masses of SZ selected clusters at high redshifts in
order to constrain cosmology.

The integrated $y$ parameter ($Y_{\rm SZ}$) adopting a spherical
geometry $Y_{\rm sph}$, is given by 
\begin{equation}\label{eq:volumey}
Y_{\rm sph}(r)= \frac{\sigma_{\rm T}}{m_{\rm e}c^2}\int_{0}^{r}{P_{\rm e}(r')4 \pi {r'}^{2}{\rm d}r'}.
\end{equation}
Following \cite{arnaud10}, we consider the radius of $5r_{500}$ 
as the boundary of the cluster where the pressure profile flattens, and 
we use this boundary to define the total volume integrated SZ signal,
$Y_{\rm tot}$.

In the simplest case, where $\alpha$, $\beta$, $\gamma$, and $c_{500}$
in equation~(\ref {eq:GNFW}) have fixed values, our cluster model
depends only on four parameters: $x_{\rm c}$ and $y_{\rm c}$ which define the 
projected cluster position on the sky and ${P_{\rm {0}}}$ and $r_{\rm 
s}$ in the pressure profile (equation~\ref{eq:GNFW}). In this paper, we
define clusters in terms of the parameter set $\mbox{\boldmath$\Theta$}
_{\rm c}\equiv (x_{\rm c}\,,\, y_{\rm c}\,, \,\theta_{\rm s}=
r_{\rm s}/D_{\rm A}\, , \,Y_{\theta}=Y_{\rm tot}/D^2_{\rm A})$, where
$D_{\rm A}$ is the angular-diameter distance to the cluster
\citep{planck2011-5.1a}, and we determine the model parameter 
${P_{\rm {0}}}$ by evaluating equation (\ref{eq:volumey}) at
$r = 5r_{500}$. To calculate $D_{\rm A}$ we assume a flat Universe
with matter density $\Omega_m = 0.27$ and Hubble constant $H_0 =
70.4 \, {\rm km}\,{\rm s}^{-1} {\rm Mpc}^{-1}$ \citep{komatsu2010}.

We adopt an exponential prior for $\theta_{\rm s}$ and a power-law
prior for $Y_{\theta}$ to analyse both the \Planck\ and AMI data
\citep{carvalho11}. The prior on $\theta_{\rm s}$ is 
$\lambda e^{-\lambda \theta_{\rm s}}$ for 
$1.3\arcmin<\theta_{\rm s}<45\arcmin$ and 
zero outside this range, with $\lambda =0.2$. The prior on 
$Y_{\theta}$ is $Y^{-\alpha}_{\theta}$
for 
$5.0 \times 10^{-4} \, \rm{arcmin}^2 < {\it Y}_{\theta} < 0.2 \, \rm{arcmin}^2$  
and zero outside this range, with $\alpha = 1.6$.
These priors have already been used in \Planck\ detection and 
extraction algorithms to identify and characterise compact objects 
buried in a diffuse background \citep{planck2011-5.1a}. For the 
cluster position, however, in order to ensure that we are comparing 
integrated SZ fluxes centred on identical positions on the sky, we 
performed an initial analysis of the AMI data using a Gaussian prior 
centred on the cluster phase centre and with a standard deviation of 
1\arcm\ in order to find the best-fitting cluster coordinates. We then 
fixed the cluster position to these best-fitting coordinates in the 
subsequent analysis of the \Planck\ data and also in a subsequent 
re-analysis of the AMI data. 
\subsection{Analysis of \Planck\ data}
\label{sec:planck_data_analysis}
The analysis of the \Planck\ data was performed using
\textsc{PowellSnakes} (PwS), which is a Bayesian package for discrete
object detection, photometry and astrometry, as described in
\cite{carvalho2009, carvalho11}. PwS is part of the \Planck\ HFI
pipeline and is regularly used to produce catalogues of objects
\citep{planck2011-1.10} and to measure and characterise the SZ signal
\citep{planck2011-5.1a}. Note that we have chosen to use PwS as our
primary SZ extraction algorithm for the \Planck\ analysis in this
study as PwS naturally returns the full posterior distribution in the
$Y_{500} - \theta_{500}$ 2D parameter space. It is thus naturally
suited for combining with the AMI results to produce joint
constraints. We will also present a comparison with results obtained
using the Matched Multi-Filter algorithm (MMF3; \citealt{melin2006})
which was the reference algorithm adopted for the production of the
\Planck\ ESZ catalogue \citep{planck2011-5.1a}.  However, in its
current implementation, the MMF3 algorithm does not produce any
information on the correlation between the two cluster parameters of
interest ($Y_{500}$ and $\theta_{500}$) and so producing joint
constraints as obtained from AMI and \Planck\ via MMF3 is currently
not possible.

The analysis of the \Planck\ data using the PwS algorithm proceeded as
follows. For each cluster, flat patches
($14.7^\circ\times14.7^\circ$;
 $512 \times 512$ pixels) were created
using a gnomonic projection,
 centred on the targeted cluster, for
each of the \Planck\ HFI
 channels. By operating on such a large
patch of sky enough statistics are collected in order to produce a
smooth cross-channel covariance matrix. 
The position of each cluster was assumed to be known precisely (we
adopted delta-function priors at the AMI-determined position) as
described in Section~\ref{sec:SZsignal}. 

The data model for a single isolated cluster located in the centre of
the patch is then described by
\begin{equation}
\label{ed:SourcesModel2}
{\bf d}({\bf x}) = Y \, {\bm f} \, \Gamma(\theta_{\rm s},{\bf x}) 
+ {\bf n}({\bf x}),
\end{equation}
where ${\bf x}$ is sky position, ${\bf d}({\bf x})$ is a vector
containing the data, ${\bf n}({\bf x})$ is the background composed of
instrumental noise plus all other astronomical components except the
SZ signal, ${\bm f}$ is a vector containing the SZ surface brightness
at each frequency, $Y$ is the total integrated Comptonisation
parameter, $\theta_{\rm s}$ is a parameter controlling the cluster
radial scale and $\Gamma(\theta_{\rm s},{\bf x})$ is the convolution
of the canonical GNFW model integrated along the line of sight with
the \Planck\ beam at that channel. It is assumed that the background
is a realisation of a stationary Gaussian random field.

A direct computation of the likelihood is very expensive. Therefore,
PwS instead computes the likelihood ratio of two competing models
describing the data: a cluster is present ($H_1$); or no cluster is
present ($H_0$). Note that the latter hypothesis does not contain any
parameters and therefore only multiplies the target likelihood in
$H_1$ by a constant. The representation of the likelihood ratio in
real space reads
\begin{eqnarray}
\label{eq:LikeFilterComplete}
\ln\left[\frac{\mathcal{L}_{H}({\bf \Theta})}{\mathcal{L}_{H_0}({\bf \Theta})}
\right] &=& Y \mathcal{F}^{-1}
\left[\mathcal{P}_j({\bm \eta}) 
\widetilde{\tau}(-{\bm \eta};\theta_{\rm s})\right]_{{\bf x}={\bf 0}} - \nonumber\\
&  &\frac{1}{2}Y^2 \sum_{{\bm \eta}}
\mathcal{Q}_{jj}({\bm \eta})
|\widetilde{\tau}({\bm \eta};\theta_{\rm s})|^2,
\end{eqnarray}
where ${\bm \eta}$ is the spatial frequency (the conjugate variable to
${\bf x}$) and $\mathcal{F}^{-1}[\ldots]_{\bf x}$ denotes the inverse Fourier
transform of the quantity in brackets, evaluated at the point
${\bf x}$. We have also defined the quantities $\mathcal{P}_j({\bm
\eta}) \equiv \widetilde{{\bf d}}^t({\bm \eta})
{\bf \mathcal{N}}^{-1}({\bm \eta}) {\bm\psi}({\bm\eta})$ and
$\mathcal{Q}_{ij}({\bm \eta}) \equiv
\widetilde{{\bm\psi}}_i^t({\bm\eta})
{\bf\mathcal{N}}^{-1}({\bm\eta}) {\bm\psi}_j({\bm\eta})$, in
which the vector $\bm \psi_i({\bm\eta})$ has the components
$({\bm\psi}_{i})_\nu = \widetilde{B}_\nu({\bm\eta})
({\bf f}_{i})_\nu$, with $\nu$ labeling frequency channels and $\widetilde{B}_\nu({\bm\eta})$ 
is the beam transfer function. The quantity $\widetilde{\tau}(-{\bm \eta};\theta_{\rm s})$ is the Fourier 
transform of $\Gamma(\theta_{\rm s},{\bf x})$ and the
matrix ${\bf\mathcal{N}}({\bm\eta})$ contains the generalised
noise cross-power spectra. We refer the interested reader to
\cite{carvalho2009, carvalho11} for further technical details on the
PwS algorithm. 

The cross-channel covariance matrix is computed directly from the
pixel data, by averaging the Fourier modes in concentric annuli. This
operation is only possible because of the assumed isotropy of the
background. To reduce the contamination of the background by the
signal itself, the estimation of the covariance matrix is performed
iteratively. After an initial estimate, all detected clusters in the
patch are subtracted from the data using their best fit values and
the covariance matrix is re-estimated. To enforce our assumption of a
single source in the centre of the patch, PwS removes from the data
all other detections with SNRs higher than our target cluster to
reduce possible contamination of the signal by power leakage from
nearby objects. Bright point sources are masked or subtracted from the
maps as part of a pre-processing routine run prior to the production of
the flat patches.

To construct the joint posterior distributions of $(Y, \theta_{\rm s})$, we 
have used the set of priors as described in Section \ref{sec:SZsignal}. To 
draw the posterior distribution manifold, PwS grids the parameter 
space using a uniformly spaced lattice of $(256 \times 256)$ cells, 
appropriately chosen to enclose all posterior regions significantly 
different from zero. 

Since the LFI channels of \Planck\ have relatively coarse resolution,
the use of LFI bands in current implementations of the extraction
algorithms results in beam dilution of the SZ signal and thus
decreases the S/N for the detected clusters
\citep{planck2011-5.1a}. This can potentially be improved in the
future with modifications to the algorithms but for the purposes of
the present study, we use only the HFI data.
\subsection{Analysis of AMI data}
\label{sec:AMI_data_analysis}
An interferometer like AMI operating at a frequency $\nu$ measures samples
from the complex visibility plane $\tilde{I}_\nu({\bf u})$. These are given by
a weighted Fourier transform of the surface brightness $I_\nu({\bf x})$, namely
\begin{equation}\label{eq:Isky}
 \tilde{I}_\nu({\bf u})=\int{A_\nu({\bf x})I_\nu({\bf x})\exp(2\pi i{\bf u\cdot x})
{\rm d}{\bf x}},
\end{equation}
where ${\bf x}$ is the position relative to the phase centre, $A_\nu({\bf x})$
is the (power) primary beam of the antennas at observing frequency $\nu$
(normalised to unity at its peak) and ${\bf u}$ is the baseline vector in units
of wavelength. In our model, the measured visibilities are defined as
\begin{equation}\label{eq:vis}
 V_\nu({\bf u})=S_\nu({\bf u}) + N_\nu({\bf u}),
\end{equation}
where the signal component, $S_\nu({\bf u})$, contains the contributions
from the SZ cluster and identified radio point sources, whereas the generalised
noise part, $N_\nu({\bf u})$, contains contributions from a background of 
unsubtracted 
radio point sources, primary CMB anisotropies and instrumental noise.

We assume a Gaussian distribution for the generalised noise. This then 
defines the likelihood function for the data
\begin{equation}\label{eq:like}
 \mathcal{L}(\mbox{\boldmath $\Theta$})=\frac{1}{Z_{\rm N}}\exp\left(-\frac{1}{2}\chi^2\right),
\end{equation}
where $\chi^2$ is the standard statistic quantifying the misfit between the
observed data $\mbox{\boldmath $D$}$ and the predicted data
$\mbox{\boldmath $D$}^p(\mbox{\boldmath$\Theta$})$,
\begin{equation}\label{eq:chi}
 \chi^2=\sum_{\nu , \nu' }(\mbox{\boldmath $D$}_\nu -\mbox{\boldmath $D$}_\nu^p)^T
(\mbox{\boldmath $C$}_
{\nu , \nu'})^{-1}(\mbox{\boldmath $D$}_{\nu'}-\mbox{\boldmath $D$}_
{\nu'}^p),
\end{equation}
where $\nu$ and $\nu'$ are channel frequencies. Here $\mbox{\boldmath $C$}$ is the 
generalised noise covariance matrix
\begin{equation}\label{eq:covmat}
 \mbox{\boldmath $C$}=\mbox{\boldmath $C$}^{\rm {rec}}_{\nu , \nu'} +\mbox
{\boldmath $C$}^{\rm {CMB}}_{\nu , \nu'} +\mbox{ \boldmath $C$}^{\rm {conf}}_{\nu , \nu'}.
\end{equation}
The first term on the right hand side of equation~(\ref{eq:covmat}) is
a diagonal matrix with elements
$\sigma^2_{\nu,i}\,\delta_{ij}\delta_{\nu \nu^{\prime}}$, where
$\sigma_{\nu,i}$ is the rms Johnson (receiver) noise on the $i$th
element of the data vector $\bf D_{\nu}$ at frequency $\nu$. The
second term denotes the noise due to primordial CMB anisotropies and
contains significant off-diagonal elements, both between visibility
positions and between frequencies. This term can be calculated from a
given primary CMB power spectrum $C^{\rm {CMB}}_{l}(\nu)$ following
\cite{hobson02}; note that in intensity units the CMB power
spectrum is a function of frequency. To calculate this term, we adopt
the best-fitting CMB power spectrum to the {\it WMAP} 7-year data
\citep{komatsu2010}. The third term on the right hand side of
equation~(\ref{eq:covmat}) is the source confusion noise, which
accounts for remaining unresolved radio sources with flux densities
less than the flux limit ($S_{\rm {lim}}$) of the AMI observations and
which remain after high resolution observation and subtraction. We
estimate this term assuming that sources are randomly distributed on
the sky, in which case we can describe the source confusion noise
with a power spectrum calculated as
\begin{equation}
C^{\rm conf}_\ell(\nu) = \int_0^{S_{\rm lim}} S^2 n_\nu(S) \, {\rm d}S, 
\end{equation}
where $n_\nu(S) = dN_\nu(>\!S) / dS$ is the differential source count at
frequency $\nu$ as a function of flux density $S$. We use the source counts as
measured by the 10C survey \citep{davies11} for our calculation. The
limiting flux density for the integration ($S_{\rm lim}$) is determined from
the noise in the LA maps and is different for each cluster, but is
typically in the range 0.2--0.5 mJy. 

The normalisation factor $Z_{\rm N}$ in equation~(\ref{eq:like}) is given by
\begin{equation}\label{eq:norm}
 Z_{\rm N}=(2\pi)^{(2N_{\rm {vis}})/2}|\mbox{\boldmath $C$}|^{1/2},
\end{equation}
where $N_{vis}$ is the total number of visibilities. Further details 
on our Bayesian methodology, generalised noise model, likelihood 
function and resolved radio point-source models are given in
\cite{feroz08} and \cite{feroz09a, feroz09b}. 

Radio sources detected in the LA maps were modeled by four source
parameters, $\mbox {\boldmath$\Theta$}_{\rm s}\equiv (x_{\rm s}\, ,\,
y_{\rm s}\,,\, S_0\, ,\,\alpha)$, where $x_{\rm s}$ and $y_{\rm s}$
refer to the right ascension and declination of radio sources,
respectively, while $S_0$ and $\alpha$ are the flux density and spectral
index of the radio source at the central frequency, $\nu_0$. As
mentioned in Section \ref{sec:amidata}, this modelling is necessary
because of source variability and some difficulty with inter-array
calibration. Therefore, the properties of point sources detected at
$>3.5\,\sigma_{\mathrm{LA}}$ by the LA were used as priors
when modelling the SA data. We used a delta-function prior on the
position of the source since the resolution of the LA is around three
times that of the SA. We used Gaussian priors on the source flux
densities, with LA (integrated, where applicable) flux densities
generating the peak of the prior, and the Gaussian $\sigma$s were set
to 40\,\% of the source flux densities. Spectral index ($\alpha$)
priors were also set as Gaussians, with $\sigma$ equal to the error on
the spectral index fit. This is because for sources with high
signal-to-noise ratio, the determination of the spectral index is
dominated by the AMI frequency channel mean and the error on $\alpha$
is Gaussian. For sources with low signal-to-noise ratio which just
meet our continuum detection threshold, the spectral index probability
distribution is dominated by the prior, which is determined from the
10C survey \citep{davies11}.

\subsection{Results}
\label{sec:resultsreal}

\begin{figure*}
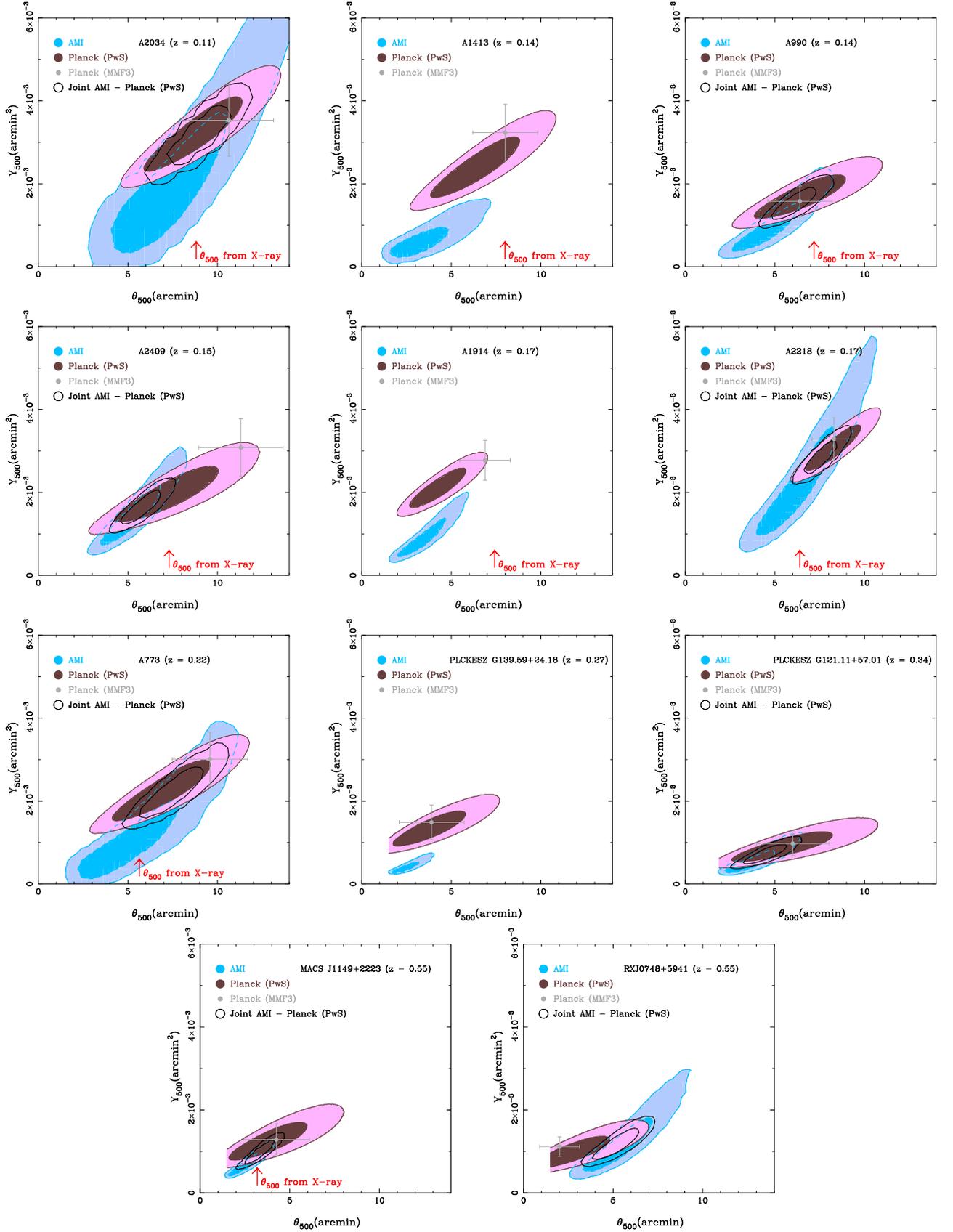

\centerline{\includegraphics[width=5.5cm,clip=,angle=-90]{a2034.eps}
      \qquad\includegraphics[width=5.5cm,clip=,angle=-90]{a1413.eps}
      \qquad\includegraphics[width=5.5cm,clip=,angle=-90]{a990.eps}}\vspace{0.1cm}
\centerline{\includegraphics[width=5.5cm,clip=,angle=-90]{a2409.eps}
      \qquad\includegraphics[width=5.5cm,clip=,angle=-90]{a1914.eps}
      \qquad\includegraphics[width=5.5cm,clip=,angle=-90]{a2218.eps}}\vspace{0.1cm}
\centerline{\includegraphics[width=5.5cm,clip=,angle=-90]{a773.eps}
      \qquad\includegraphics[width=5.5cm,clip=,angle=-90]{plj0621+7442.eps}
      \qquad\includegraphics[width=5.5cm,clip=,angle=-90]{plj1259+6005.eps}}\vspace{0.1cm}
\centerline{\includegraphics[width=5.5cm,clip=,angle=-90]{maj1149+2223.eps}
      \qquad\includegraphics[width=5.5cm,clip=,angle=-90]{rxj0748+5941.eps}}
\caption{Recovered \Planck\ and AMI 2D posterior distributions in 
the $Y_{500}-\theta_{500}$ plane. Blue contour 
plots are the results from the AMI analysis and purple contours
show the \Planck\ results (specifically using the PwS method). Red arrows show the values of 
$\theta_{500}$ as determined from X-ray measurements of these clusters
where available. The grey points with error bars show the MMF3
\Planck\ results. The inner and outer contours in each set indicate
the areas enclosing 68\,\% and 95\,\% of the probability distribution,
while the MMF3 error-bars indicate the $1\sigma$ uncertainties. Where
the recovered AMI and PwS \Planck\ constraints are consistent, the
joint constraints are also indicated by the heavy black contours. In
cases where the contours do not close at the lower ends of the
parameter ranges, the corresponding constraints represent upper limits
only. \label{fig:realpos1}} 
\end{figure*}

Fig.~\ref{fig:realpos1} presents the 2D marginalised posterior
distributions in the $Y_{500}-\theta_{500}$ plane and
Table~\ref{tab:realplanckamino1} summarises the mean and the
dispersion of these two parameters for each cluster, as estimated from
the \Planck\ and AMI data, respectively. Note that in
Fig.~\ref{fig:realpos1} the inner and outer contours show the areas
enclosing $68\,\%$ and $95\,\%$ of the probability
distributions. Estimates of $\theta_{500}$ as derived from X-ray
observations, and which were included in the \Planck\ ESZ catalogue
\citep{planck2011-5.1a}, are also indicated in the figures for
comparison.

\begin{table*} 
\centering
\caption{Recovered mean and dispersion values for $\theta_{500}$ and
  $Y_{500}$ for the $11$ clusters. Where consistency is found between
  the \Planck\ and AMI measurements, the joint constraints are also
  given. The cluster redshift and signal-to-noise of the PwS
  detections are also listed.\label{tab:realplanckamino1}}
\begin{tabular}{lcrcccccc}
\noalign{\doubleline}
             & $z$& S/N & \Planck\ $\theta_{500}$ & \Planck\ $Y_{500}$ & AMI $\theta_{500}$ & AMI $Y_{500}$ & Joint $\theta_{500}$ & Joint $Y_{500}$ \\  
Cluster      &    & PwS & (arcmin) & ($10^{-4}$ arcmin$^2$)& (arcmin) & ($10^{-4}$ arcmin$^2$) & (arcmin) & ($10^{-4}$ arcmin$^2$) \\         
\noalign{\vskip 3pt\hrule\vskip 5pt}
A2034                 & 0.11 &$13$ & $9.0\pm1.9$ & $33\pm6$ &$8.0\pm2.5$ & $24\pm14$ & $9.1\pm1.5$ & $32\pm4$ \\
A1413                 & 0.14 &$11$ & $6.7\pm1.7$ & $25\pm5$ &$3.6\pm1.3$ & $\,\,\,7\pm \,\,\,3$ & $\cdots$          & $\cdots$            \\
A990                  & 0.14 &$10$  & $6.8\pm1.7$ & $17\pm4$ &$4.9\pm1.3$ & $10\pm \,\,\,4$ & $6.2\pm0.8$ & $15\pm2$ \\
A2409                 & 0.15 &$9$  & $7.6\pm1.9$ & $20\pm4$ &$5.5\pm1.2$ & $16\pm \,\,\,5$ & $5.8\pm0.8$ & $17\pm3$ \\
A1914                 & 0.17 &$14$ & $4.4\pm1.0$ & $21\pm3$ &$3.6\pm0.9$ & $10\pm \,\,\,3$ & $\cdots$       & $\cdots$         \\
A2218                 & 0.17 &$20$ & $8.4\pm1.0$ & $31\pm4$ &$6.7\pm1.5$ & $25\pm \,\,\,9$ & $7.6\pm0.7$ & $29\pm3$ \\
A773                  & 0.22 &$11$ & $7.3\pm1.9$ & $23\pm5$ &$5.7\pm2.1$ & $13\pm \,\,\,7$ & $7.7\pm1.2$ & $23\pm4$ \\
MACS J1149+2223       & 0.55 &$8$  & $4.2\pm1.5$ & $13\pm3$ &$2.7\pm0.7$ & $\,\,\,7\pm \,\,\,2$ & $3.4\pm0.6$ & $10\pm2$  \\
RXJ0748+5941          & 0.55 &$8$  & $3.4\pm1.4$ & $11\pm2$ &$5.7\pm1.4$ & $13\pm \,\,\,5$ & $5.3\pm0.9$ & $12\pm3$ \\
PLCKESZ G139.59+24.18 & 0.27 &$9$  & $4.2\pm1.4$ & $14\pm3$&$2.6\pm0.5$ & $\,\,\,4\pm \,\,\,1$ & $\cdots$ & $\cdots$   \\
PLCKESZ G121.11+57.01 & 0.34 &$7$  & $5.9\pm1.9$ & $\,\,\,9\pm2$ &$3.9\pm1.0$ & $\,\,\,6\pm \,\,\,2$ & $4.6\pm0.8$ & $\,\,\,7\pm2$ \\
\noalign{\vskip 5pt\hrule\vskip 3pt}
\end{tabular}
\end{table*} 

Recall that in this figure the clusters are ordered in terms of
increasing redshift. The constraints from both \Planck\ and AMI
demonstrate a strong cluster size--integrated Compton parameter
($\theta_{500}-Y_{500}$) degeneracy/correlation in all cases. It is
important to account for such effects when attempting to use the SZ
signal to estimate cluster masses \citep{daSilva04, arnaud07}. We also
note that the \Planck\ constraints appear to be weaker for high
redshift clusters, which can generally be understood as a resolution
effect -- \Planck's relatively poor resolution (e.g., as compared with
AMI) means it has difficulty resolving and thus estimating the
parameters of clusters with small angular extent -- and high-redshift
clusters are likely to be smaller in angular size. AMI's increased
resolution, on the other hand, means that it can still constrain the
sizes of these high-redshift, small-angular size clusters.

For three clusters (A1413, A1914, and PLCKESZ G139.59+24.18),
the AMI and \Planck\ constraints are clearly discrepant. 
On the other hand there is significant overlap in the posterior
distributions for the remaining eight clusters. However, taking our
cluster sample as an ensemble, there is some evidence that the cluster
parameter estimates derived from the AMI data are systematically lower
than those derived from the \Planck\ data (i.e., AMI is finding the
clusters to be fainter and smaller in angular extent compared to what
the \Planck\ data indicate). 

In addition, the $Y_{500} - \theta_{500}$ degeneracies
are significantly different for the \Planck\ and AMI constraints, with
the degeneracies of the AMI constraints being generally steeper than
the \Planck\ ones. This arises because of the interplay between the
angular size--redshift relation and the differing angular scales that
AMI and \Planck\ are sensitive to, as well as the very different
observational techniques and frequencies used by the two instruments.

In the cases where the \Planck\ and AMI-derived constraints are
compatible with one another, we also overplot the joint constraints
obtained from multiplying the \Planck\ and AMI posteriors. In many
cases, the resulting joint constraints are far tighter than either
analysis alone which is a direct result of the differing parameter
degeneracies for the two instruments, as described above. The
marginalised constraints from this combined analysis are also
presented in Table~\ref{tab:realplanckamino1}.
 
In Fig.~\ref{fig:realpos1}, we have also over-plotted the constraints
as obtained from the \Planck\ data using the MMF3 algorithm. Comparing
these results with the PwS \Planck\ results, we see good agreement in
most cases, although there may be a tendency for the MMF3 estimates to
be systematically brighter and larger than the PwS results. However,
it is clear that our broad conclusions regarding the general levels of
agreement between the \Planck\ and AMI results remain unchanged
if we consider the MMF3 results in place of the PwS constraints. 

\begin{figure}
\centering
\includegraphics[angle=-90,width=75mm]{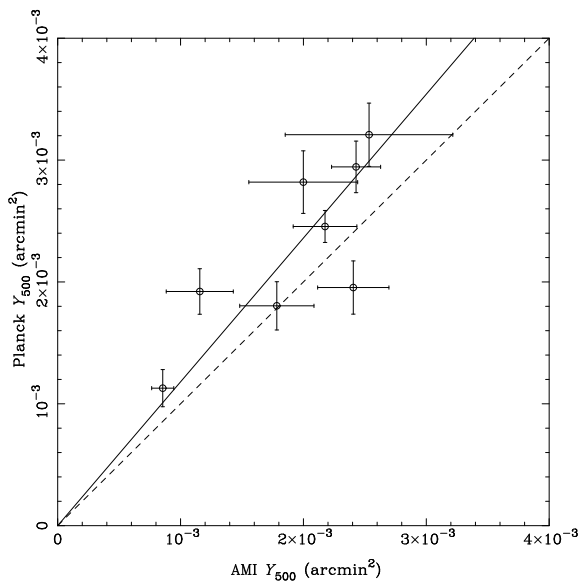}
\caption{Comparison of the integrated Compton-$Y$
    parameters obtained from the \Planck\ and AMI fits, when the cluster
    size is fixed to that determined from X-ray observations. The
    one-to-one relation is denoted with the dashed line. The
    best-fitting linear relation is plotted as the unbroken line. The
    slope of this latter relation is $1.18\pm0.07$ and the
    correlation coefficient is $0.79$. Note that the same general
    behaviour (slope $> 1$) is also observed when we fix the cluster
    size to be that determined from either the \Planck\ or the AMI SZ
    observations.} \label{fig:y_y_corr}
\end{figure}

In Fig.~\ref{fig:y_y_corr}, we plot the \Planck\ determined integrated
Compton-$Y$ parameter versus the Compton-$Y$ parameter as derived from
the AMI data. Note that, for this correlation plot, we have fixed the
cluster scale size to be that determined from X-ray observations (as
indicated by the red arrows in Fig.~\ref{fig:realpos1}). (Three of the
clusters have no reported X-ray size so only eight of the 11 clusters
contribute to this correlation analysis.)  The measured correlation
coefficient is 0.79 and the best-fitting linear relationship has a
slope of $1.18\pm0.07$, again indicating that the \Planck\ SZ
fluxes appear to be systematically larger that the AMI derived
fluxes. We have also repeated this analysis fixing the cluster size to
both the \Planck-determined size and the AMI-determined size. In
both cases we see the same general trend, with the \Planck\ Compton-$Y$
parameter being consistently larger than the AMI-derived value. 

In summary, our results suggest a systematic difference between the
\Planck\ and AMI measurements of the SZ signal coming from our cluster
sample. Such a systematic difference could be an indicator of a
shortcoming in some part of our analysis and could have important
implications for performing cosmological studies with larger samples
of SZ clusters. For example, the observed systematic could indicate
that the way the clusters are being modeled in the analysis (e.g., the
fixed GNFW profile adopted and/or the assumption of spherical
symmetry) is not flexible enough to describe both the \Planck\ and AMI
results simultaneously. If this were to be the source of the
discrepancy then such effects would need to be accounted for in future
cosmological studies. However, before considering such an explanation,
it is important to first consider if possible instrumental and/or 
astrophysical systematic effects could be responsible for the results
we have found. We now turn to simulations to investigate the potential
impact of such effects. 

\section{Simulations}
\label{simulation}
In order to test the SZ signal extraction techniques used and to
investigate whether the systematic discrepancy observed in the real
data is due to unaccounted-for astrophysical foregrounds, instrumental
systematics or data-analysis induced biases, we have conducted
detailed simulations of both the \Planck\ and AMI experimental setups.

For each of the 11 clusters in our sample, to create an input SZ
signal for the simulations, we simulated a cluster SZ signal using the
GNFW pressure profile (equation~\ref{eq:GNFW}), with input parameters based
on the best-fitting $Y_{500}$ and $\theta_{500}$ values from an
analysis based on intermediate \Planck\ maps, which are, in practice,
close to the best-fitting \Planck\ parameters quoted in
Table~\ref{tab:realplanckamino1}.

\subsection{\Planck\ simulations}
\label{sec:planck_sims}
For \Planck, the main worry in terms of astrophysical systematics is
probably thermal emission from dust in the Galaxy. The \Planck\ simulation 
ensemble comprised of CMB and noise realisations and a fixed foreground 
dust template, produced by re-scaling the \Planck\ 857 GHz channel map to the 
other HFI frequencies and reconvolving so as to apply the appropriate beam 
for each channel. The dust template assumed a modified blackbody spectrum with 
emissivity $\beta = 1.8$ and temperature $T=18$~K. The beams were assumed 
to be Gaussian with the appropriate mean FWHM for each channel as 
calculated by the \textsc{FEBeCoP} algorithm \citep{Planck2011-1.7}. 

The noise component of the simulations was generated using the
\textsc{Springtide} destriping pipeline \citep{ashdown2007b}. This pipeline
creates realisations of the noise in the nominal mission time-ordered
data streams, compresses them to rings and destripes the rings to
produce noise maps. It is assumed that the noise is uncorrelated
between rings and that in each ring it is drawn from a power
spectrum. For these simulations, the noise power spectrum used was the
mean of the ring-by-ring spectra. In turn, these were determined by applying the noise
estimation pipeline \citep{Planck2011-1.7} to the exact same version
and time-span of the \Planck\ data that was used for the real SZ
cluster analysis of the previous section.

The simulations were then analysed using the PwS algorithm in exactly
the same manner as was applied during the analysis of the real
data. For each of the 11 clusters, ten simulations were run. The dust
template based on the \Planck\ 857 GHz map was the same for each of
these ten simulations, but the CMB and noise realisations were
different. 

\begin{figure*}
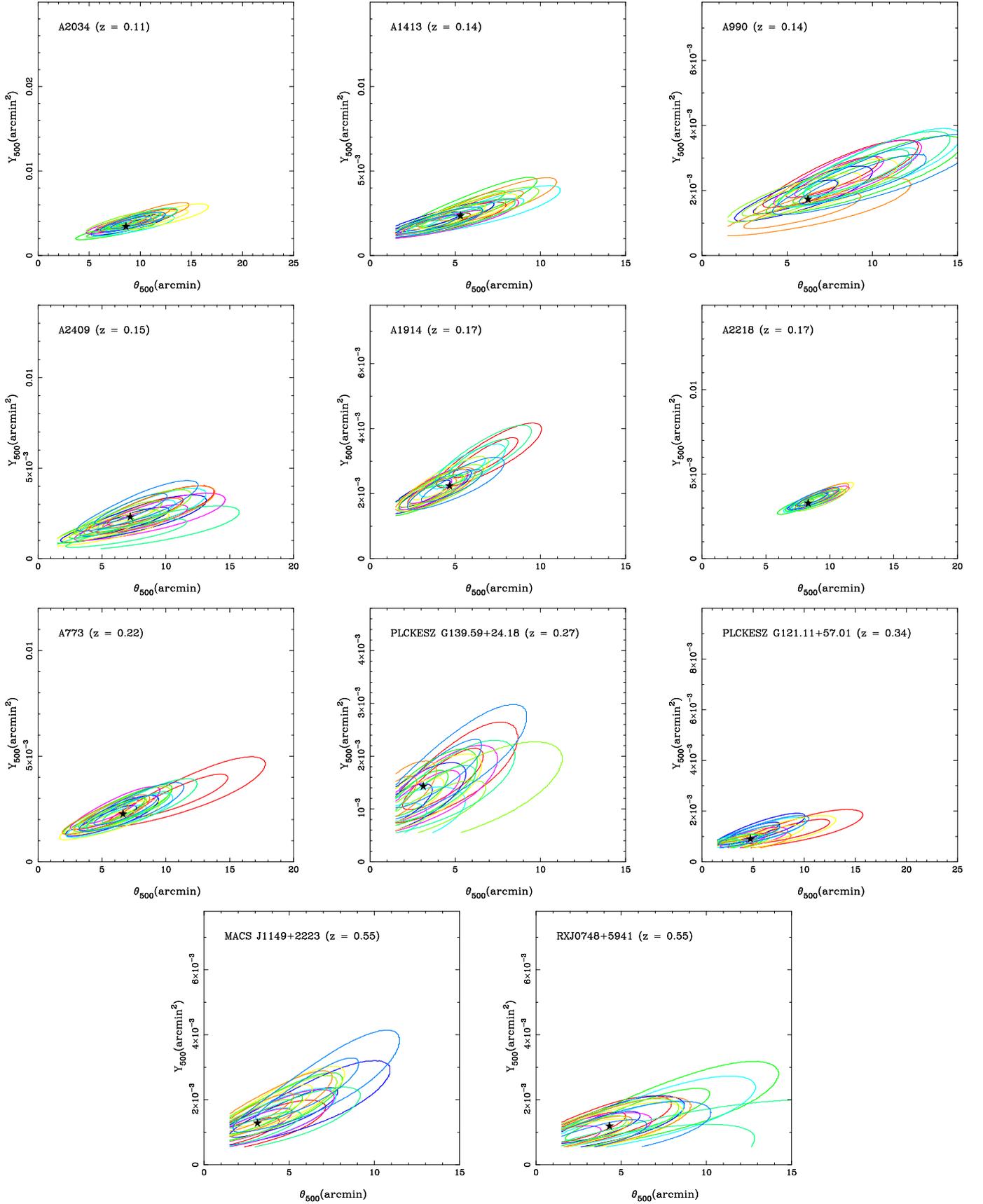

\centerline{\includegraphics[width=5.5cm,clip=,angle=-90.]{A2034plsim.ps}
      \qquad\includegraphics[width=5.5cm,clip=,angle=-90.]{A1413plsim.ps}
      \qquad\includegraphics[width=5.5cm,clip=,angle=-90.]{A990plsim.ps}}\vspace{0.2cm}
\centerline{\includegraphics[width=5.5cm,clip=,angle=-90.]{A2409plsim.ps}
      \qquad\includegraphics[width=5.5cm,clip=,angle=-90.]{A1914plsim.ps}
      \qquad\includegraphics[width=5.5cm,clip=,angle=-90.]{A2218plsim.ps}}\vspace{0.2cm}
\centerline{\includegraphics[width=5.5cm,clip=,angle=-90.]{A773plsim.ps}
      \qquad\includegraphics[width=5.5cm,clip=,angle=-90.]{PLJ0621+7442plsim.ps}
      \qquad\includegraphics[width=5.5cm,clip=,angle=-90.]{PLJ1259+6005plsim.ps}}\vspace{0.2cm}
\centerline{\includegraphics[width=5.5cm,clip=,angle=-90.]{MAJ1149+2223plsim.ps}
      \qquad\includegraphics[width=5.5cm,clip=,angle=-90.]{RXJ0748+5941plsim.ps}}
\caption{Recovery of SZ cluster parameters from
    simulated \Planck\ observations (see Section~\ref{sec:planck_sims}
    for details). Each set of recovered parameter constraints
    (contours with different colours) represents a different
    realisation of the instrument noise and primordial CMB
    fluctuations and the star shows the input parameter
    values. The inner and outer contours in each set indicate the
    areas enclosing 68\,\% and 95\,\% of the probability distribution. Any
    bias in the recovery of the input parameters averaged over
    realisations is negligible compared to the random errors. In cases
    where the contours do not close at the lower ends of the parameter
    ranges, the corresponding constraints represent upper limits
    only.}\label{fig:plancksim1}
\end{figure*}

The results of the simulations are shown in
Fig.~\ref{fig:plancksim1}. In each panel, input parameters are
indicated with a star and the recovered parameter constraints from the
ten different simulations are indicated by ten sets of differently
coloured contours. Comparing with the input parameters, the recovered
constraints for each of our clusters are clearly distributed about the
input model and there is no indication of any significant bias due to
dust contamination, noise bias, the \Planck\ beams or the PwS
extraction technique employed. 

\subsection{AMI simulations}
\label{sec:ami_sims}
As mentioned in Section~\ref{sec:amidata}, contamination from radio
point sources is a significant issue at AMI frequencies
($\sim$16~GHz). Although the AMI~LA observations are used to accurately
find and model sources in the AMI~SA observations, there is the
possibility of contamination from source residuals if this modelling
is not perfect.

In a similar manner to the \Planck\ simulations described in the
previous subsection, we have investigated potential issues associated
with either residual foreground radio sources or with the AMI
data-analysis methodology and instrument response using
simulations. The simulated input clusters were the same as used for
the \Planck\ simulations.

To simulate the interferometric AMI observations, we used the in-house
simulation package, \textsc{profile} \citep{grainge02} to create the
mock visibilities. In addition to the cluster signal, the simulations
included primordial CMB fluctuations and Gaussian noise, the amplitude
of which was chosen to match that measured from the real
observations. The simulation package also mimics the actual $uv$
coverage and synthesised beam of the real observations. The point
sources in each cluster were simulated using the best-fitting values
from the analysis of the real data.
These simulated observations were then analysed in the exact same way
as was used for the real data. Once again, for each of the 11
clusters, ten simulations were performed. Here, the point source
environment was kept the same for these ten simulations but the CMB and
noise realisations were again different.

\begin{figure*}
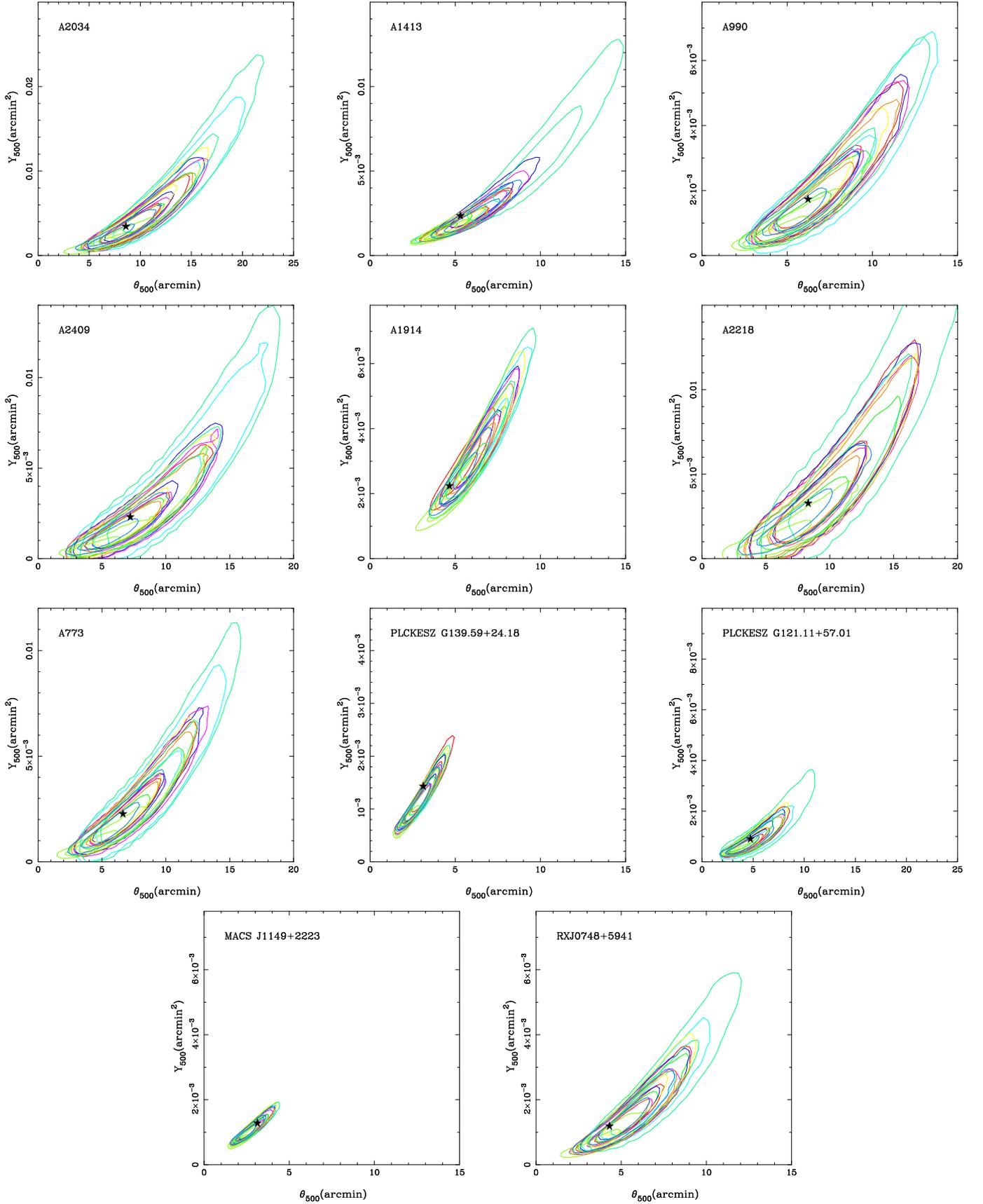

\centerline{\includegraphics[width=5.5cm,clip=,angle=-90.]{A2034amisim.ps}
      \qquad\includegraphics[width=5.5cm,clip=,angle=-90.]{A1413amisim.ps}
      \qquad\includegraphics[width=5.5cm,clip=,angle=-90.]{A990amisim.ps}}\vspace{0.2cm}
\centerline{\includegraphics[width=5.5cm,clip=,angle=-90.]{A2409amisim.ps}
      \qquad\includegraphics[width=5.5cm,clip=,angle=-90.]{A1914amisim.ps}
      \qquad\includegraphics[width=5.5cm,clip=,angle=-90.]{A2218amisim.ps}}\vspace{0.2cm}
\centerline{\includegraphics[width=5.5cm,clip=,angle=-90.]{A773amisim.ps}
      \qquad\includegraphics[width=5.5cm,clip=,angle=-90.]{PLJ0621+7442amisim.ps}
      \qquad\includegraphics[width=5.5cm,clip=,angle=-90.]{PLJ1259+6005amisim.ps}}\vspace{0.2cm}
\centerline{\includegraphics[width=5.5cm,clip=,angle=-90.]{MAJ1149+2223amisim.ps}
      \qquad\includegraphics[width=5.5cm,clip=,angle=-90.]{RXJ0748+5941amisim.ps}}
\caption{Recovery of SZ cluster parameters from the
    simulated AMI observations in the presence of residual point
    source contamination from imperfectly modeled radio sources for
    each cluster in the sample. Also included in the simulations are
    the AMI $uv$ coverage, the instrument beams and realisations of
    the instrument noise and primordial CMB. The different sets of
    contours indicate different CMB and noise realisations and the star
    shows the input parameters used to generate the simulated
    cluster. The inner and outer contours in each set indicate the
    areas enclosing 68\,\% and 95\,\% of the probability
    distribution.}\label{fig:amisim1}
\end{figure*}

The results of the simulations are shown in Fig.~\ref{fig:amisim1}. In
each panel, input parameters are indicated with a star and the
recovered parameter constraints from the ten different simulations are
indicated by ten sets of differently coloured contours. As was the case
with the \Planck\ simulations, the recovered constraints for each of our
clusters are clearly distributed about the input model. Once again, there is no
indication of any significant bias due to residual point source
contamination, noise bias, the AMI $uv$ coverage and resolution effects,
or the extraction technique employed.

\section{Adopting individual pressure profiles as measured from X-ray
  observations}
\label{sec:implications}
The simulations presented in the previous section indicate that the
discrepancies seen in the analysis of the real data cannot be
easily explained by astrophysical contamination, instrumental effects,
or any issues associated with the SZ signal extraction techniques.
It is then interesting to ask whether the discrepancies observed might
be associated with the way in which the clusters have been modeled
using the universal pressure profile \citep{arnaud10}.

For a number of clusters in the sample, we have high-quality
determinations of the clusters' individual pressure profiles, as
estimated from X-ray observations. Rather than adopting the
\cite{arnaud10} profile (which is essentially an average profile taken
over many clusters), one might expect to achieve better consistency on
a case-by-case basis if we use these individual best-fitting X-ray
derived profiles in the SZ analysis.

\begin{figure*}
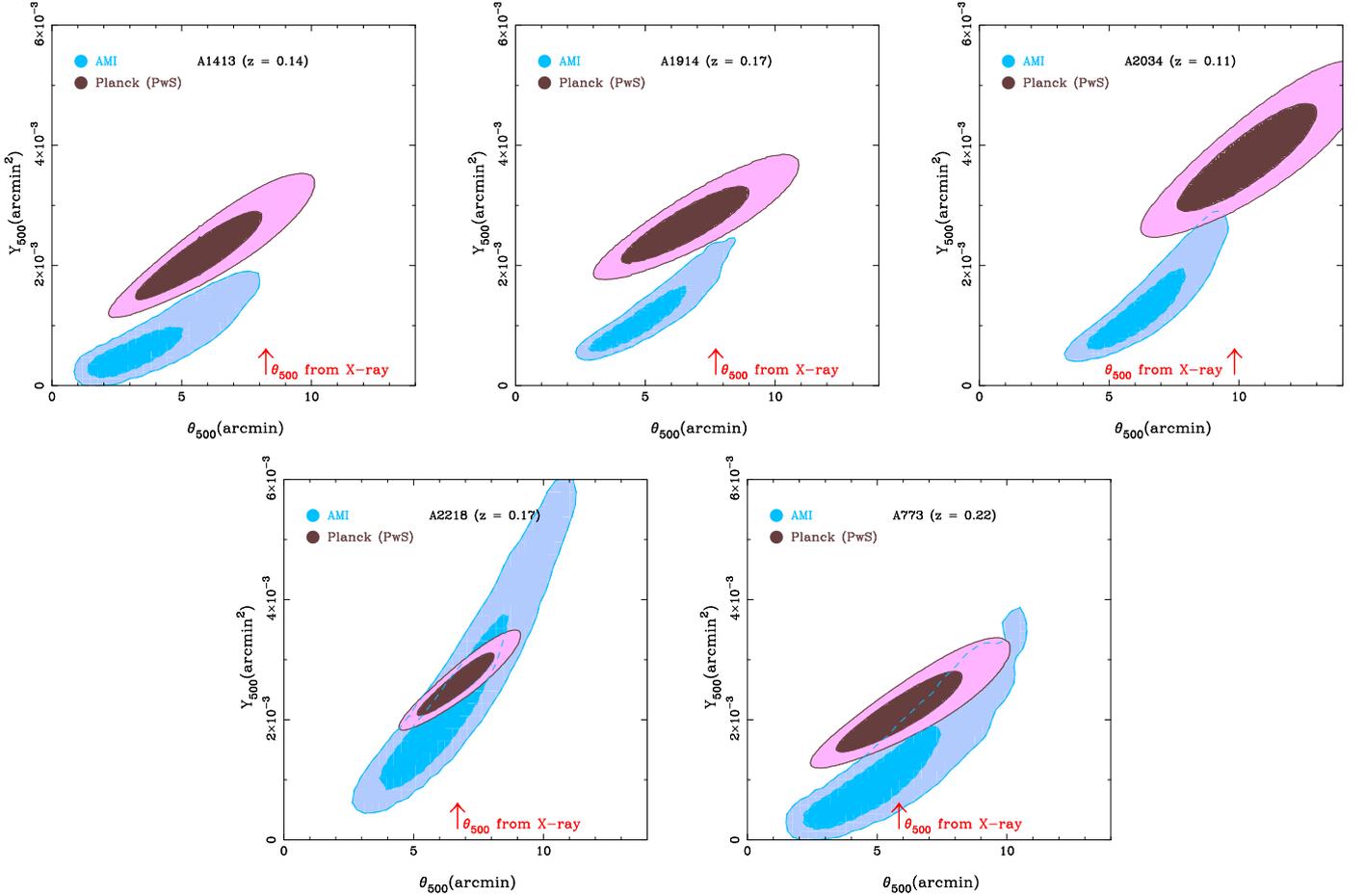

\centerline{\includegraphics[angle=-90,width=5.62cm]{A1413_xray_fits.ps} \qquad
            \includegraphics[angle=-90,width=5.62cm]{A1914_xray_fits.ps}
            \qquad \includegraphics[angle=-90,width=5.62cm]{A2034_xray_fits.ps}}\vspace{0.2cm}
\centerline{\includegraphics[angle=-90,width=5.62cm]{A2218_xray_fits.ps} \qquad
            \includegraphics[angle=-90,width=5.62cm]{A773_xray_fits.ps}}
\caption{Constraints obtained from the re-analysis of
    the \Planck\ and AMI observations for the five clusters for which high-quality
    X-ray observations are available. These re-analyses adopted the
    GNFW shape parameters which best fit the X-ray data as given in
    Table~\ref{tab:xrayfitsGNFW}. Comparison with the corresponding
  panels in Fig.~\ref{fig:realpos1} reveals no obvious improvement in the
  level of agreement between the \Planck\ and AMI constraints. The
  X-ray sizes indicated here are also derived from the GNFW fits to
  the high-quality X-ray data. These are slightly different from the
  X-ray sizes plotted in Fig.~\ref{fig:realpos1}, which were taken from
  the ESZ catalogue.}\label{fig:xraypressureprofiles}
\end{figure*}

We have performed such a re-analysis for five clusters in our sample for
which we have high-quality measured X-ray profiles. The clusters
concerned are A1413, A1914, A2034, A2218, and A773. We fitted a GNFW
pressure profile to the measured X-ray profiles and the results are
presented in Table~\ref{tab:xrayfitsGNFW}. 
We then re-analysed the \Planck\ and AMI SZ data for these five clusters
using the best fitting values of the profile shape parameters $(c_{\rm
  500},\alpha,\beta,\gamma)$ as given in Table~\ref{tab:xrayfitsGNFW}.
The resulting constraints from this re-analysis are shown in
Fig.~\ref{fig:xraypressureprofiles}.

\begin{table*} 
\centering
\caption{Best-fitting GNFW shape and concentration parameters
  (cf.~equation~\ref{eq:GNFW}) derived by fitting the parameterised
  GNFW profile to the measured X-ray pressure profiles of five
  clusters in our 11 cluster sample. Note that $\beta=5.49$ is fixed
  and a prior of $\gamma > 0$ is imposed. This latter constraint is
  enforced to avoid unphysical pressure gradients being allowed by the
  GNFW parameterisation.\label{tab:xrayfitsGNFW}}
\begin{tabular}{ccccccc}
\noalign{\doubleline}
Cluster     &$R_{500}$ (${\rm Mpc}$) & $P_{500}$ & $P_0$ & $c_{500}$ & $\alpha$ & $\gamma$ \\         
\noalign{\vskip 3pt\hrule\vskip 5pt}
A1413       &$1.240$& $3.229$ & $31.08$&$0.90$ &$0.69$&$0.191$       \\
A1914       &$1.348$&$4.045$ & $49.94$&$1.88$&$0.95$ & $0.000$   \\
A2034       & $1.211 $&$ 2.899$ &$9.14$ & $1.84$&$1.72$&$0.000$  \\
A2218       &$1.169$ &$3.039$ &$40.92$ & $1.02$&$0.74$&$0.000$   \\
A773        &$1.232$&$3.724$ & $20.61$& $1.25$&$0.96$&$0.000$  \\
\noalign{\vskip 5pt\hrule\vskip 3pt}
\end{tabular}
\end{table*}
Comparing with the corresponding constraints for these five clusters in
our original analysis, we see that the updated constraints for A2034
have tightened significantly. This appears to be due to the fact that
the previously used \cite{arnaud10} GNFW profile was not a good match to
this particular cluster's pressure profile, particularly in the
central region of the cluster, where AMI is sensitive. The GNFW profile
variant used to produce the updated constraints is a much better match
to the measured X-ray profile and so the AMI data are better able to
constrain the remaining cluster parameters.  

Apart from this single case, comparing with our original results,
there does not appear to be a systematic improvement in the agreement
between the \Planck\ and AMI constraints when we move from the
\cite{arnaud10} profile to the best-fitting GNFW profile as measured
from the individual X-ray observations. This, and similar reasoning
based on an adaptation of these modified profiles to the other
clusters in our sample, suggests that a more significant widening of
the parameter space describing the cluster profiles will be required
in order to simultaneously fit both the \Planck\ and AMI SZ
measurements for the entire cluster sample considered in this paper.

\section{Conclusions}
\label{sec:discussion}
We have studied the $Y_{500}-\theta_{500}$ degeneracy from the SZ
effect for a sample of $11$ clusters ($0.11<z<0.55$) observed with
both \Planck\ and AMI. This is motivated by the fact that such a study
can potentially break the well-known $Y$-size degeneracy which
commonly results from SZ experiments with limited
resolution. Modelling the radial pressure distribution in each cluster
using a universal GNFW profile, we have shown that there is
significant overlap in the 2D posterior distributions for eight of the
clusters. However, overall, AMI finds the SZ signal to be smaller in
angular extent and fainter than \Planck\ finds. The derived 
parameter degeneracies are significantly different for the two
instruments. Hence, where the constraints from the two instruments are
mutually consistent, their combination can be powerful in terms of
reducing the parameter uncertainties. Significant discrepancies are
found between the \Planck\ and AMI parameter constraints for the
remaining three clusters in our sample.

We have investigated the origin of these discrepancies by carrying out
a detailed analysis of a series of simulations assessing the potential
impact of diffuse thermal emission from dust and residual
contamination from imperfectly modeled radio point sources. Our
simulations also include a number of systematic effects associated
with the two instruments in addition to primordial CMB fluctuations
and thermal noise. We find that the results of the simulations of both
the \Planck\ and AMI analyses are unbiased, confirming the accuracy of
the two analysis pipelines and their corresponding methodologies.

We have attempted to reconcile some of the discrepancies seen by
re-analysing the \Planck\ and AMI data adopting individual
best-fitting pressure profiles, as measured from high-quality X-ray
observations for five of the clusters in the sample. However, we do
not observe a systematic improvement in the agreement between the
\Planck\ and AMI parameter constraints when we perform this
re-analysis.

We conclude that: either (i) there remain unaccounted for systematic
effects in one or both of the data sets beyond what are included in
our simulations; or (ii) a further expansion of the parameter space
used to model the SZ cluster signal is required to simultaneously fit
the \Planck\ and AMI SZ data. Such further expansion of the model
parameter space, which we leave for future studies, could potentially
include using the \Planck\ and AMI data in conjunction with X-ray
observations to find a global fit for the GNFW shape parameters, going
beyond the GNFW parameterisation to investigate other cluster profiles,
and/or dropping the assumption of spherical symmetry for the SZ (and
X-ray) emission.

\begin{acknowledgements}
A description of the Planck Collaboration and a list of its members,
indicating which technical or scientific activities they have been
involved in, can be found at http://www.rssd.esa.int/Planck.
The Planck Collaboration acknowledges the support of: ESA; CNES and
CNRS/INSU-IN2P3-INP (France); ASI, CNR, and INAF (Italy); NASA and DoE
(USA); STFC and UKSA (UK); CSIC, MICINN and JA (Spain); Tekes, AoF and
CSC (Finland); DLR and MPG (Germany); CSA (Canada); DTU Space
(Denmark); SER/SSO (Switzerland); RCN (Norway); SFI (Ireland);
FCT/MCTES (Portugal); and DEISA (EU). The AMI telescope is supported
by Cambridge University and the STFC. The AMI data analysis was
carried out on the COSMOS UK National Supercomputer at DAMTP,
University of Cambridge and the AMI Consortium thanks Andrey Kaliazin
for computing assistance.

\end{acknowledgements}

\bibliographystyle{aa}
\bibliography{ms}

\raggedright

\listofobjects
\end{document}